\documentclass[12pt,a4paper]{article}
\pdfoutput=1
\usepackage{jheppub}
\usepackage[T1]{fontenc}
\usepackage{slashed}
\usepackage{caption}
\usepackage{subcaption} 

\usepackage{xcolor}
\usepackage{color,soul}

\usepackage{graphicx}
\usepackage{feynmp}

\def\[{\left[}
\def\]{\right]}
\def\({\left(}
\def\){\right)}

\newcommand{\be}{\beta}

\def\Tr{\mathrm{Tr}}

\def\Tr{\mathrm{Tr}}

\def \be {\begin{equation}}
\def \ee {\end{equation}}
\def \bea {\begin{eqnarray}}
\def \eea {\end{eqnarray}}
\def \beal#1 {\begin{align}#1\end{align}}

\everymath{\displaystyle}

\preprint{TIFR/TH/19-9}

\title{Matter Chern Simons Theories in a Background Magnetic Field}

 \author[a,1]{Indranil
  Halder,\note{indranil.halder@tifr.res.in} }   \author[a,2]{Shiraz
  Minwalla\note{minwalla@theory.tifr.res.in}}

\affiliation[a]{Department of Theoretical Physics, Tata Institute of Fundamental Research, Homi Bhabha Rd, Mumbai 400005, India}

\abstract{We study large $N$ 2+1 dimensional fermions in the fundamental representation of an $SU(N)_k$ Chern Simons gauge group in the presence of a uniform background magnetic field for the $U(1)$ global symmetry of this theory. The magnetic field modifies the Schwinger Dyson equation for the propagator in an interesting way; the product between the self energy and the Greens function is replaced by a Moyal star product. Employing a basis of functions previously used in the study of non-commutative solitons, we are able to exactly solve the Schwinger Dyson equation and so determine the fermion propagator. The propagator has a series of poles (and no other singularities)  whose locations yield a spectrum of single particle energies at arbitrary t' Hooft coupling and chemical potential. 
The usual free fermion Landau levels spectrum is shifted 
and broadened out; we  compute the shifts and widths of these levels at arbitrary t'Hooft coupling. As a check on our results we  independently solve for the propagators of the conjecturally dual theory of Chern Simons gauged large $N$ fundamental  Wilson Fisher bosons also in a background magnetic field but this time only at zero chemical potential. The spectrum of single particle states of the bosonic theory  precisely agrees with those of the fermionic theory under  Bose-Fermi duality.}

\begin{document}

\maketitle

\section{Introduction}

Three dimensional Chern Simons theories coupled to matter 
fields in the fundamental representation are interesting for at least six related but  distinct reasons. 
First because of the (by now compelling) evidence for a 
level rank type Bose Fermi duality between pairs of these these theories \footnote{Motivated by the conjectured dual bulk Vasiliev duals of these theories \cite{Klebanov:2002ja, Sezgin:2002rt,
		Giombi:2009wh, Giombi:2011kc, Chang:2012kt}, 
	the first suggestion that such a duality might exist 
	was made in \cite{Giombi:2011kc} The papers \cite{Giombi:2011kc,Maldacena:2011jn,Maldacena:2012sf} also demonstrated that these theories are effectively `solvable' in the large $N$ limit. 
	The authors of  \cite{Aharony:2012nh,GurAri:2012is} combined the results of \cite{Maldacena:2011jn,Maldacena:2012sf} and 
Schwinger-Dyson techniques developed in \cite{Giombi:2011kc}) to find the first quantitative 
evidence for such a duality and to propose a concrete duality map. Several subsequent large $N$ - and more 
recently finite $N$ - computations have supplied overwhelming evidence for the duality. A  precise version of the duality conjecture was presented in \cite{Aharony:2015mjs,Seiberg:2016gmd,Karch:2016sxi,Murugan:2016zal}. There have been many subsequent studies of finite $N$ versions and generalizations of these dualities; see \cite{hsin:2016blu, aharony:2016jvv, Gomis:2017ixy, Cordova:2017vab, Cordova:2017kue, Metlitski:2016dht} for a very partial list 
of references.}. Second because these theories are solvable in the 
large $N$ limit \cite{Giombi:2011kc}. Third because the 
all orders large 
$N$ solution of various quantities (like S matrices
\cite{Jain:2014nza,Dandekar:2014era,Inbasekar:2015tsa,Yokoyama:2016sbx,Inbasekar:2017ieo,Inbasekar:2017sqp}
and thermal partition functions  \cite{Giombi:2011kc,Jain:2012qi,Yokoyama:2012fa, Aharony:2012ns,
	Jain:2013py,Takimi:2013zca,Yokoyama:2013pxa,Jain:2013gza,Minwalla:2015sca,xyz,Geracie:2015drf,RD:2015,Choudhury:2018iwf, OSM, Dey:2018ykx}),
 apart from supplying overwhelming 
 evidence for duality,  reveal qualitatively novel dynamics. \footnote{See also 
 \cite{Bedhotiya:2015uga, Turiaci:2018dht, Aharony:2018npf} 
 for checks of this duality at the level
 of four point functions of gauge invariant operators.} Fourth because these theories 
 have interesting conjectured bulk dual descriptions in terms  of Vasiliev theory. Fifth because the effective excitations  in massive deformations of these theories are non abelian anyons. Sixth because there exists an intricate and rich set of finite $N$ exact - but physically incompletely understood - results (see  e.g. \cite{ Kim:2009wb, Kapustin:2009kz, Drukker:2011zy, Imamura:2011su}) for supersymmetric observables in supersymmetric versions of this theory. 
 
In this paper we carry forward the study of matter Chern 
Simons theories in the large $N$ limit. We 
focus attention on the simplest and best studied matter 
Chern Simons theories, namely the so called regular fermion 
theory (a theory of a single fundamental multiplet of otherwise free fermions interacting with an $SU(N_F)$ Chern Simons gauge field) and the conjecturally dual so called 
critical boson theory (the theory of a single multiplet of 
fundamental Wilson Fisher bosons interacting with a $U(N_B)$ 
Chern Simons gauge field). Each of these theories enjoys
invariance under a $U(1)$ global symmetry. \footnote{In the 
	case of the fermionic theory the $U(1)$ global symmetry is the obvious 
	global symmetry that `completes' $SU(N_F)$ to $U(N_F)$. 
	In the case of the bosonic theory, the $U(1)$ global symmetry is the so called topological symmetry of the 
	$U(1)$ part of the gauge field.} In previous work each of these theories has intensively been studied (at leading order at large $N$ but at all orders in the t' Hooft coupling) both in the absence and in the presence of a chemical potential coupling to this gauge field. In this
    paper we generalize these older studies in a manner we now describe.
    
    As the theories we study enjoy invariance under a $U(1)$ global symmetry, they can be studied in the presence of a background gauge field $a_\mu$ that couples to global symmetry current.  Turning on a chemical potential $\mu$ for the $U(1)$ global symmetry is equivalent, in Euclidean space, to studying the theory in the background of an (imaginary) complex constant gauge field in the time direction
    \begin{equation}\label{chemp}
    a_3=i \mu
    \end{equation}
    While a generic background field $a_\mu(x^\mu)$ completely breaks the $SO(3)$ rotational symmetry of our theory, the special choice \eqref{chemp} preserves an $SO(2)$. In this paper we study a generalization of \eqref{chemp} that continues to preserve $SO(2)$. The generalized background field studied in this paper is that of a uniform magnetic
    field together with a chemical potential, i.e. 
    \begin{equation}\label{backgf}
    \begin{split}
    & a_3=i \mu \\
    &a_j = -b\frac{\epsilon_{ji} x^i} {2}~~~{\rm i.e}~~~ f_{ij}=b\epsilon_{ij}\\
\end{split}
\end{equation}

Working in the background \eqref{chemp}, it has been demonstrated in several papers (starting with \cite{Giombi:2011kc}) that the fermion propagator $\alpha$ and the fermion self energy $\Sigma$ in the regular fermion theory obeys the gap equations\footnote{\eqref{sdentinto} applies at leading order in the 
large $N$ limit in a lightcone gauge. The equation has been presented in Fourier space in time but in coordinate space in space. $G_{\mu\nu}$ is the 
lightcone gauge gauge boson propagator,  $\hat{\alpha}$ is the equal time fermion propagator and 
$m_F$ is the bare fermion mass).} 
\begin{equation} \label{sdentinto}
\begin{split}
&\Sigma(\omega,{\vec x}-{\vec y})=-\frac{N}{2}   \gamma^\mu {\hat \alpha({\vec x}-{\vec y})} \gamma^\nu G_{\mu\nu}(\omega,{\vec x}-{\vec y})\\
&\left(  i \omega \gamma^3 
+ \gamma^i \partial_i + m_F\right) 
 \alpha(\omega, {\vec x}-{\vec y}) + 
\left[\Sigma(\omega) \alpha(\omega) \right]( {\vec x}-{\vec y}) = - \delta^{2}( {\vec x}-{\vec y})\\
\end{split}
\end{equation}
 where $[AB]$ denotes 
the convolution of the two dimensional functions $A$ and 
$B$ i.e.  
\begin{equation}\label{convol}
\left[ A B \right]=  \int d^2 w A(\vec{x}-\vec{w}) B(\vec{w}).
\end{equation} In this paper 
we demonstrate that the `reduced' fermion propagator $\alpha_R$ and `reduced'\footnote{See 
	the discussion around \eqref{ussmr} for what `reduced' 
means.} self energy $\Sigma_R$
in the magnetic field background \eqref{backgf} obey 
the strikingly similar gap equation
\begin{equation} \label{sdentintt}
\begin{split}
&\Sigma_R(\omega,{\vec x}-{\vec y})=-\frac{N}{2}   \gamma^\mu {\hat \alpha_R({\vec x}-{\vec y})} \gamma^\nu G_{\mu\nu}(\omega,{\vec x}-{\vec y})\\
&\left( i \omega \gamma^3 
+  D^{(x-y)}_i \gamma^i + m_F
\right) \alpha_R(\omega, {\vec x}-{\vec y}) + \left( \Sigma_R(\omega)
*_b \alpha_R(\omega) \right)({\vec x}-{\vec y})= - \delta^{2}({\vec x}-{\vec y})
\end{split}
\end{equation}
While the first lines in \eqref{sdentinto} and \eqref{sdentintt}
are simply identical to each other, the second lines in these
two equations differ in two ways. First, 
the ordinary derivative $\partial_i$ in the first term 
in the second line of \eqref{sdentinto} is replaced by 
$D_i$, the background covariant derivative. Second, that the ordinary convolution $[\Sigma \alpha]$ in the second term of 
the second line of \eqref{sdentinto} turns into the 
twisted convolution defined by 
\begin{equation}\label{stpint}
\left( A *_{b} B \right) (\vec{x})= \int d^2 w
A(\vec{x}-\vec{w}) e^{-i \frac{b}{2} \epsilon_{ij}(x^i-w^i) (w^j) } B(\vec{w})
\end{equation}

We pause to explore the twisted convolution \eqref{stpint}
in more detail. The operation \eqref{stpint} is the Fourier 
transform of the familiar Moyal product in momentum 
space. More explicitly let 
\begin{equation}\label{ab} \begin{split}
&A(\vec{x})=\int \frac{d^2k}{(2 \pi)^2} {\hat A}(\vec{k}) e^{i {\vec k}. {\vec x}}\\
&B(\vec{x})=\int \frac{d^2k}{(2 \pi)^2} {\hat B}(\vec{k}) e^{i {\vec k}. {\vec x}}\\
\end{split}
\end{equation}
Then 
\begin{equation}\label{asb} 
\begin{split}
&\left( A *_{b} B \right)(\vec{x})= \int \frac{d^2k}{(2 \pi)^2} \left( {\hat A} \star_b {\hat B} \right)(\vec{k}) e^{i {\vec k}. {\vec x}}~~~{\rm where}\\
& \left( {\hat A} \star_b {\hat B} \right)(\vec{k}) = \exp  \left[ {i \frac{b}{2} \epsilon^{ij} \partial^A_{k_i} \partial^B_{k_j} } \right] {\hat A} (\vec{k})
{\hat B} (\vec{k})\\
	\end{split}
\end{equation}
In the second line of \eqref{asb} the derivative $\partial^A_{k_i}$ acts only on the 
function ${\hat A}$ while the derivative $\partial^B_{k_j}$ acts only on the function 
${\hat B}$. The RHS of the second line of \eqref{asb} defines the associative but non commutative Moyal star product in momentum space. Note that under this product 
\begin{equation}\label{mstp}
\left[ k_i, k_j\right]_{\star} =
k_i \star_b k_j-k_j \star_b  k_i = ib \epsilon_{ij}
\end{equation} 

The effective non commutativity of momentum expressed by 
\eqref{mstp} is a familiar phenomenon in the presence of a background magnetic field \footnote{We thank N. Seiberg for reminding us of this fact, emphasizing its relevance and for related discussions.}.  Consider the 
theory of a non relativistic spinless particle of unit 
charge in a background magnetic field (see \eqref{Sch}). While canonical momentum $\pi_i$ conjugate to the variable $x^i$ is represented by the (non gauge convariant) operator $-i \partial_i$, the more physical mechanical momentum $p_i=
m  \frac{d x^i}{dt}$  is represented by the (gauge 
covariant) operator $-i D_i=-i(\partial_ i -i a_i)
=-(i\partial_i+a_i)$. It follows that 
\begin{equation}\label{landau}
[p_i,p_j]= \left[ -i D_i, -i D_j \right]= i \left(\partial_i a_j -\partial_j a_i \right)
= i f_{ij}= i b \epsilon_{ij}\end{equation}
where we have used \eqref{backgf} in the last step.
Note that \eqref{landau} agrees exactly with \eqref{mstp}. 
The  familiar and elementary  noncommutativity of mechanical momenta in the presence of a uniform magnetic field \eqref{landau} thus supplies an 
intuitive explanation for the appearance of the momentum space Moyal 
star product in the gap equation \eqref{sdentintt}.

We now return to the gap equations \eqref{sdentintt}. 
The appearance of the Moyal star product 
in the equations \eqref{sdentintt} allows us to borrow 
the technology developed in the study of non-commutative solitons almost twenty years ago in 
in e.g. \cite{Gopakumar:2000zd, Aganagic:2000mh} and many subsequent works. \footnote{The study of noncommutative solitons was motivated by the demonstration \cite{Seiberg:1999vs} that the dynamics of D branes is governed by non commutative field theories in a particular combined low energy and large field scaling limit \cite{Seiberg:1999vs} and also by the discovery of novel IR effects in such theories at the quantum level \cite{Minwalla:1999px}.}
 We are able to simplify \eqref{sdentintt} by expanding the fermion propagator and self energy in a basis 
of functions introduced in \cite{Gopakumar:2000zd}, and rewriting the gap 
equations \eqref{sdentintt} as a set of nonlinear equations 
for the coefficients in such an expansion. Quite remarkably, 
the equations so obtained turn out to be exactly solvable; the solution yields a reasonably explicit solution for the zero temperature fermion fermion propagator in the background 
\eqref{backgf}.  We present our derivation 
for this propagator in section \ref{regfer};  detailed final results for the fermion propagator are summarized in subsection \ref{sumfinp}. In the next few paragraphs we describe some aspects of these results in qualitative terms. 

Of course the fermion fermion  propagator $\alpha_R$ that 
appears in \eqref{sdentintt} - and that we are largely able 
to solve for -  is not gauge invariant and so is not 
completely physical. The unambiguously physical information in this propagator lies in its singularities. At least at values 
of the chemical potential at which we are able to perform 
our computations completely reliably, these singularities 
all turn out to be poles that represent on-shell 
`one particle states' in the background \eqref{backgf}. 
We are able to find an almost completely explicit solution for the positions of the poles of this propagator and so 
are able to completely understand the structure of `single particle excitations' of our theory about the background \eqref{backgf}, as we now explain in some detail.

Let us first consider the limit $\lambda_F=0$. In this 
limit the regular fermion theory is free. The single 
fermion energy spectrum in this limit is simply that of 
the free Dirac equation in a background magnetic field.
As we review in detail below (see around \eqref{Efer}) these energy levels are given by 
\begin{equation}\label{Eferinto}
E^{\eta \nu J}= \eta \sqrt{m_F^2 + 2b \left( \nu+ \frac{1}{2} -\eta \frac{{\rm sgn}(m_F)}{2} \right) }
\end{equation}
In \eqref{Eferinto} $\nu$ is a Landau level label that ranges from $0$ to $\infty$, $J$ is an angular momentum label that 
parameterizes degeneracy within Landau Levels and $\eta=\pm 1$ decides whether the energy levels in question have 
positive or negative energy. \eqref{Eferinto} differs from the 
corresponding formula for free minimally coupled scalar in only 
one way - the term proportional to $\frac{{\rm sgn}(m_F)}{2}$ in \eqref{Eferinto} is missing in 
the simpler scalar case. The reason for this difference 
is easy to understand;  this term in \eqref{Eferinto} has its origins in a $B.S$ coupling (note the spin of a Dirac fermion of mass $m_F$ equals $\frac{{\rm sgn}(m_F)}{2}$ in 2+1 
dimensions) which is obviously absent for a free scalar particle.  

At (absolute value of) energies larger than the (absolute value of) the chemical potential, the exact finite coupling results of this paper for 
the effective spectrum of single particle states   differs from \eqref{Eferinto} in only 
two ways. First the term $m_F^2$ that appears in the 
square root on the RHS of \eqref{Efer} is replaced by 
a coupling constant dependent dynamical mass, whose 
value is given by the solution to an explicit but 
complicated equation. Second, the 
quantity $\frac{{\rm sgn}(m_F)}{2}$, which also appears 
in under the square root on the RHS of \eqref{Efer} is 
replaced by 
\begin{equation}\label{replan}
\frac{{\rm sgn}(m_F)}{2} \rightarrow s(\lambda_F)
\end{equation}
where
\begin{equation}\label{slamint}
s(\lambda_F)= \frac{{\rm sgn} (m_F)-\lambda_F }{2}
\end{equation}
The replacement \eqref{replan} is simply a consequence 
of the fact \cite{Witten:1988hf} (explained and explored in detail in the upcoming 
paper \cite{Shiraz:2019}) that this free spin is
 renormalized to $s(\lambda)$ \footnote{The exact 
 formula for this effective spin is \cite{Witten:1988hf}  
 $$s=\frac{{\rm sgn} (m_F) }{2}-
 \frac{c_2(F)}{\kappa_F}= \frac{{\rm sgn} (m_F) }{2}-\frac{N_F^2-1}{2 N_F\kappa_F}$$
 where $C_2(F)$ is the quadratic Casimir of the 
 fundamental representation of $SU(N)$ (see subsection 
 2.6 of \cite{Jain:2013py} for notation). In the large 
 $N$ limit $s$ reduces to the quantity $s(\lambda_F)$ reported in  \eqref{slamint}.} 
  in the interacting theory. At zero chemical 
potential, in other words, the spectrum of particle energies 
of our theory is essentially that of a `non interacting 
spin $s(\lambda)$ particle' with a coupling constant 
(and magnetic field) dependent mass.

The effective mass that appears in the spectrum 
of `single particle states' of each particle, mentioned in the paragraph above, depends on the chemical potential, 
i.e. on how many Landau levels are filled. There are two particularly interesting values of the chemical potential associated with any given Landau level. First, the lowest chemical potential at which this level is completely 
filled. Second, the highest chemical potential at which this Landau level is completely unfilled. 
Unlike in the free theory these two chemical 
potentials are not equal (roughly speaking this 
happens because the effective 
mass is a function of the chemical potential). We call the difference between these two chemical potentials the width of the 
corresponding Landau Level. This width - which may be either positive or negative - represents the broadening of the the Landau level as it is filled. We are able to explicitly 
compute the width of every Landau level as a function of the 
coupling constant; our explicit results are presented in 
\eqref{energydifffp}, \eqref{energydifffn} at small $|\lambda_F|$, in \eqref{energydiffB} at small $|\lambda_B|$ and in 
the graphs presented in subsection \ref{numint} at arbitrary 
values of the coupling. 

In this paper we are reliably able to compute the Fermion 
Greens function only outside the width of any given Landau 
level. This limitation prevents us from studying the (presumably fascinating) detailed physics within a Landau 
level width in the current paper. As we discuss at some length in the discussion section it may be possible to do better in 
future work.

It is particularly interesting to investigate the behaviour of our single particle energy spectrum in the limit in which the modulus of the fermionic t'Hooft coupling tends to 
unity (recall that our theory is conjectured 
to admit a weakly coupled dual bosonic 
description in this limit). 

When all  Landau levels are unfilled, it turns out that the $\lambda_B \to 1$ (taken with $m_{B}^{{\rm cri}}$ held fixed) is smooth. Focussing on this case we have 
used the 
conjectured dual description of the theory in 
terms of bosonic variables to obtain a detailed 
and precise calculational check of our fermionic
single particle spectrum. In section \ref{regbo} we have performed the bosonic 
Wilson Fisher analogue of the fermionic computations describe above in some detail. 
The final output of this section is, once again, a spectrum of single particle energies, this time 
read off from the poles of the bosonic propagator. 
The final results for this spectrum agree perfectly under duality with the spectrum obtained from 
the poles of the fermionic propagator providing  a rather impressive check of either the conjectured Bose Fermi duality between the regular fermion and critical boson theories or of the computations presented in this paper, according to taste.

The computation 
of the  bosonic propagator is technically 
complicated in a situation in which some Landau levels are filled, and we have not attempted 
to solve for the bosonic propagator in this case, 
leaving it for future work. From the bosonic point of view the filling of a Landau level is presumably associated with the formation of a Bose condensate. Even in the absence of a magnetic field, the solution of the bosonic theory in the presence of a condensate required new tricks, and was successfully carried out only about a
a year ago \cite{Choudhury:2018iwf, Dey:2018ykx}. 
It would be very interesting to generalize these 
computations to include the effects of a background 
magnetic field in the presence of filled Landau 
levels. We hope to return to this problem in 
the near future. 

Although we have not, in this paper, directly 
used the bosonic description to obtain the 
spectrum of single particle states in the presence
of filled Landau levels, we can translate our 
fermionic results to bosonic language and so 
(assuming duality) obtain a detailed prediction 
for these results (see subsection \ref{brgc}). Provided that at least one Landau level is filled we find that the single  particle energy levels behave in a rather dramatic
 manner in the $|\lambda_B|\to 0$ (see subsection 
 \ref{bpc}). If we 
 take this limit with $m_{B}^{\rm cri}$ held fixed 
 as before, we find that the effective single particle 
 squared mass diverges like $\frac{1}{|\lambda_B|}$. 
 Moreover in the limit that $|\lambda_B|$ is small 
 we find that lowest chemical potential at which any 
 any given Landau level is completely filled is always 
 greater than the highest chemical potential at which it is 
 completely empty; i.e. the broadening of Landau levels 
 is always positive - and of order $\frac{1}{|\lambda_B|}$
 - in this limit. 
 
The divergence of effective masses and widths like $\frac{1}{|\lambda_B|}$ is a dramatic effect which suggests that the ungauged critical boson theory - like its free counterpart - is simply ill defined above a certain value of the chemical potential, and
that this run away behaviour is cured by  $|\lambda_B|$ coupling effects. A detailed exploration of this phenomenon (preferably from  both the bosonic and the fermion viewpoints) is likely to be a very interesting exercise; one that we leave to future work. 

This paper has been devoted to the study of particle propagators
and their poles - i.e. to the study of effective single particle energy levels - in the presence of a magnetic field. We have not carefully studied the thermodynamics (e.g. charge as a function of chemical potential) in the same background. We hope to return to this very interesting question (as well as to the equally interesting question of generalizing our results to finite temperature) in the future.

\section{Free particles in a magnetic field}

\subsection{Non-relativistic particles}

Consider the non relativistic Schrodinger equation 
for a spin-less particle of mass $|M|$ in a uniform
magnetic field $b$ described by Hamiltonian
\begin{equation}\label{Sch}
H= - \frac{(\nabla_i - i a_i)^2}{2|M|}
\end{equation}
We work in the in a rotationally invariant gauge 
\begin{equation} \label{rigc}
a_j = -b\frac{\epsilon_{ji} x^i} {2}
\end{equation}
All through this paper we restrict to the special case of $b>0$ (the case $b<0$ can be obtained from $b>0$ by a parity transformation).

The eigenstates and eigen energies of this problem 
are labelled by two integers $\nu$ and $l$. 
$$\nu= 0 \ldots \infty$$
is a `which Landau Level' label. 
$l= -\nu \ldots \infty$ labels the angular momentum of states. The eigen functions $\phi_{\nu,m}$ are given by (See \ref{bAppendix} for details) 
\begin{equation}\label{lagpol}
 \begin{aligned}
  \phi_{\nu,l}  =& i^{|l|}\left[\frac{b}{2\pi} \frac{(\nu-\frac{|l|}{2}+\frac{l}{2})!}{(\nu+\frac{|l|}{2}+\frac{l}{2})!} \right]^{1/2} e^{il\phi }e^{-u/2}u^{|l|/2}L_{\nu-\frac{|l|}{2}+\frac{l}{2}}^{|l|}(u),
  ~ E_{\nu,l}=\frac{b}{|M|} \left(\nu+\frac{1}{2} \right)
 \end{aligned}
\end{equation}
More explicitly, for $l> 0$\footnote{Here to obtain the second equality we are using following property$$ i^{l}\left[\frac{(\nu)!}{(\nu+l)!} \right]^{1/2} u^{l/2}L_{\nu}^{l}(u)=i^{-l}\left[\frac{(\nu+l)!}{(\nu)!} \right]^{1/2} u^{-l/2}L_{\nu+l}^{-l}(u)$$}
\begin{equation}
 \begin{aligned}
  \phi_{\nu,l}  =& i^{l}\left[\frac{b}{2\pi} \frac{(\nu)!}{(\nu+l)!} \right]^{1/2} e^{il\phi }e^{-u/2}u^{l/2}L_{\nu}^{l}(u)=\sqrt{\frac{2 \pi}{b}}e_{\nu,\nu+l}
 \end{aligned}
\end{equation}
and for $l \leq 0$
\begin{equation}
 \begin{aligned}
  \phi_{\nu,l}  =& i^{-l}\left[\frac{b}{2\pi} \frac{(\nu+l)!}{(\nu)!} \right]^{1/2} e^{il\phi }e^{-u/2}u^{-l/2}L_{\nu+l}^{-l}(u)=\sqrt{\frac{2 \pi}{b}}e_{\nu,\nu+l}
   \end{aligned}
\end{equation}
Where we have defined following wave functions
\begin{equation}\label{enmH}
e_{ m,n}(u, \phi)=  i^{m-n}\frac{b}{2 \pi} \left(\frac{n!}{m!} \right)^{1/2} e^{-i(m-n)\phi}
 u^{\frac{m-n}{2}}e^{- \frac{u}{2}} L_n^{m-n}(u)
\end{equation}
Above map between wave functions can be compactly rewritten as
\begin{equation} \label{map} 
\phi_{\nu,l}=\sqrt{\frac{2 \pi}{b}}e_{m,n}, ~~~ m=\nu, ~ n=\nu+l	
\end{equation} 
Note that, with this labelling, the variables 
$m, n$ range over the values
\begin{equation}\label{nmrange}
n, m = 0 , 1 \ldots 
\end{equation}

Note, of course, that the wave functions $\phi_{\nu, J}$ 
- and so $e_{n,m}$ - are orthonormal to each other 
and have been normalized to obey 
in the usual quantum mechanical sense 
\begin{equation}\label{orthog} 
\int d^2x \frac{2 \pi}{b} e^*_{n,m}({\vec x } )
e_{n',m'}({\vec x } ) = 
\delta_{n n'} \delta_{m m'}
\end{equation}

In the rest of this subsection we will determine the 
Greens function of the free non relativistic particle, 
i.e. we will find the function $\alpha$  that obeys the equation\footnote{Here $i$ runs over space variables $1,2$.} 
\begin{equation}\label{gfnr}
\left( \partial_t  -\frac{(\partial^x_i -ia_i(x))^2}{2|M|} \right) \alpha(x, x') 
= \delta^3 (x-x')
\end{equation}
where the derivative $\partial_t,\partial_i$ acts on the 
first argument of the Greens function.

\subsubsection{Symmetries of the Green's function} \label{ReducedVariables}

A uniform magnetic field 
\begin{equation}\label{FS}
a_3=0, ~~~\partial_i a_j-\partial_j a_j = \epsilon_{ij} b ~~~(i, j=1,2), ~~~\epsilon_{12}=1
\end{equation}
is invariant under both translations in the 
(12) plane as well as rotations in this plane.\footnote{Of course translations in the $3$ direction are trivially preserved 
	and are ignored through this discussion.}
These symmetries of the gauge invariant field strength are obscured by the fact that theres is no choice of background gauge that preserves all three symmetries.
To deal with this complication, we will find it useful to enumerate three gauges, chosen so that each gauge that preserves any one of three spacetime symmetries listed above but breaks the other two.

In this paper we always work in the rotational gauge $g_r$ 
\begin{equation} \label{rigc}
a_j = -b\frac{\epsilon_{ji} x^i} {2}. 
\end{equation}
which preserves the symmetry of rotations in the (12) plane but breaks the two translational 
symmetries. The $x_1$ translational gauge $g_1$ 
\begin{equation} \label{rigcy}
a_1 = -b x^2, ~~~a_2=0 
\end{equation}
preserves translations in the $x^1$ direction but 
breaks translations in $x^2$ and rotations. Finally the $x_2$ translational gauge $g_2$
\begin{equation} \label{rigcx}
a_2 = b x^1, ~~~a_1=0 
\end{equation}
preserves translations in the $2$ direction but 
breaks translations in $x^1$ and rotations. 
Note that 
\begin{equation}\label{gto}
\begin{split}
&a_i^{g_1}= a_i^{g_r} - \partial_i \left( \frac{b x^1 x^2}{2} \right) \\
&a_i^{g_2}= a_i^{g_r} +\partial_i \left( \frac{b x^1 x^2}{2} \right) \\
\end{split}
\end{equation}
Under a gauge transformation of the background 
field $a_\mu$, the propagator $\alpha$ and 
the self energy transforms as 
\begin{equation}\label{gaugetransf}
\begin{split} 
&a_\mu(x) \rightarrow a_\mu(x) + 
\partial_\mu \chi(x) \\
&\alpha(x,y) \rightarrow e^{i \chi(x)}
\alpha(x,y) e^{-i \chi(y)} \\
&\Sigma(x,y) \rightarrow e^{i \chi(x)}
\Sigma(x,y) e^{-i \chi(y)} \\
\end{split}
\end{equation}
In this paper we choose to work in the gauge $g_r$. In this gauge all Greens functions $\alpha(x,y)$ (and later in this paper self energies $\Sigma(x,y)$) are rotationally invariant. It also follows from \eqref{gaugetransf} and \eqref{gto} that 
$$ e^{i\frac{b x^1 x^2}{2}} \alpha(x,y) 
e^{-i\frac{b y^1 y^2}{2}}, ~~~e^{i\frac{b x^1 x^2}{2}} \Sigma(x,y) 
e^{-i\frac{b y^1 y^2}{2}}$$
are both translationally invariant in the 1 direction and so are functions only of $x^1-y^1$ while 
$$ e^{-i\frac{b x^1 x^2}{2}} \alpha(x,y) 
e^{i\frac{b y^1 y^2}{2}}, ~~~
e^{-i\frac{b x^1 x^2}{2}} \Sigma(x,y) 
e^{i\frac{b y^1 y^2}{2}}, ~~~$$
are both translationally invariant in the $2$ direction and 
so are function only of $x^2-y^2$.
It follows that both these conditions are met if and only if 
\begin{equation}\label{ussm}
\alpha(x, y)=e^{-i \frac{b}{2} \left(x^1 y^2 -x^2 y^1 
	\right) } \alpha_R(x-y), ~~~~
\Sigma(x,y) =e^{-i \frac{b}{2} \left(x^1 y^2 -x^2 y^1 
	\right) } \Sigma_R(x-y)
\end{equation}
where the functions $\alpha_R(x-y)$ and 
$\Sigma_R(x-y)$ are both simultaneously rotationally and translationally invariant.\footnote{Note that the `dressing factors' that relate $\alpha(x,y)$ to $\alpha_R(x,y)$ (and similarly for $\Sigma$) are 
	rotationally invariant.} 

The phase factor that occurs in \eqref{ussm} has a simple 
physical interpretation. Consider the integral 
\begin{equation}\label{intogf}
I(x,y)=\int_{x}^y a_i dx^i 
\end{equation}
where $a_i$ is the background listed in \eqref{rigc} and 
the integral on the RHS is take along the straight line from 
$x$ to $y$. It is easily verified that 
\begin{equation}\label{intogfn}
I(x,y)= b\frac{x^1y^2-x^2 y^1}{2}
\end{equation}
so that the Wilson line $e^{i I(x,y)}=e^{i b\frac{x^1y^2-x^2 y^1}{2}}$. 
It follows from \eqref{ussm} that 
\begin{equation}\label{alrwl}
\alpha_R({\vec x}-{\vec y})= \alpha(x, y)e^{iI(x,y)}
\end{equation}
Under a gauge transformation 
\begin{equation}\label{gtpsi}
\psi(x) \rightarrow e^{i \theta(x)}\psi(x), ~~~a_i(x) \rightarrow 
a_i(x) + \partial_i \theta(x), ~~~I(x,y) \rightarrow 
I(x,y) +\theta(y)-\theta(x)
\end{equation}
So that 
$$ \alpha(x,y) \rightarrow e^{i \theta(x)} \alpha(x,y) e^{-i\theta(y)}, ~~~e^{i I(x,y)} \rightarrow 
e^{-i \theta(x)} e^{i I(x,y)} e^{i\theta(y)}$$
It follows that $\alpha_R(x,y)= \alpha(x,y) e^{i I(x,y)}$ is background gauge invariant- and consequently both rotationally and translationally invariant.
\footnote{We thank A. Gadde for a discussion on this point.}  In words, $\alpha_R$ is simply the fermion propagator 
dressed by the straight line Wilson line that is needed 
to make it background gauge invariant, and is a
natural object to study in the presence of a nontrivial 
background gauge field.

\subsubsection{Differential equation 
for the Green's function}

Note that the first of \eqref{ussm} can be rewritten 
as 
\begin{equation} \label{rewa}
\alpha(x, y)=e^{i x^i a_i(y) } \alpha_R(x-y)
\end{equation}
where $a^j(y)$ given in \eqref{rigc}. It follows that 
\begin{equation}\label{partm}
\partial_i \alpha= e^{i \epsilon_{ij}x^i a^j(y) }
\left(\partial_ i + i a_i(y) \right) \alpha_R(x-y)
\end{equation}
It follows that the defining equation for the Greens function \eqref{gfnr} can be rewritten as 
\begin{equation}\label{gfnrn}
\left( \partial_{t}  -\frac{(\partial_{x_i} -ia_i(x-x'))^2}{2|M|} \right) \alpha_R(x-x') 
= \delta^3 (x-x')
\end{equation}
In \eqref{gfnrn} $x$ is the collective notation for $({\vec x}, t)$ and 
similarly for $x'$ and we have used the fact that 
$$e^{ix^i a_i(x') }\delta^3 (x-x')
=\delta^3 (x-x') .$$
Note that the differential operator that appears on the LHS of \eqref{gfnrn} 
depends only on $x-x'$, consistent with the result, derived in the previous subsection, 
that $\alpha_R$ is a function only of $x-x'$. 

\subsubsection{Solution for the Green's function}

It is useful to Fourier transform the invariant
Greens function in time. Let 
\begin{equation}\label{ftconv}
\alpha_R(x)= \int \frac{d \omega}{2 \pi} 
e^{i \omega x^3} \alpha_R(\omega, {\vec x})
\end{equation} 
It follows in the usual way that 
\begin{equation}\label{mit}
\begin{split}
&\alpha_R(\omega, {\vec x}-{\vec y})
= \sum_{\nu, l} \frac{\phi_{\nu, l}({\vec x})
\phi_{\nu, l}^*({\vec y})}{i\omega +E_\nu} \\
& E_\nu= \frac{b}{|M|}( \nu + \frac{1}{2}).\\
\end{split}
\end{equation} 
where the complete set of wave functions 
$\phi_{\nu,l}$ were defined in \eqref{lagpol}.
In the equation above, $\nu$ ranges from $0$ 
to infinity while $l$ ranges over the range $-\nu,-\nu+1,...$
As the denominator in \eqref{mit} is a function 
only of $\nu$, it follows that 
\eqref{mit} can be rewritten as 
\begin{equation}\label{mitn}
\begin{split}
&\alpha_R(\omega, {\vec x}-{\vec y})
= \sum_{\nu=0}^\infty \left( \frac{
	\sum_{l}\phi_{\nu, l}({\vec x})
	\phi_{\nu, l}^*({\vec y})}{i\omega +E_\nu} 
\right) \\
\end{split}
\end{equation}
As $ \alpha_R(\omega, {\vec x}-{\vec y})$ is a 
function only of the difference between its spatial positions for every value of $\omega$, it follows that the same must be true of residue of the $\alpha_R$ at $\omega=-iE_\nu$.  It follows, in other words, that the quantity 
$$\sum_{l}\phi_{\nu, l}({\vec x})
\phi_{\nu, l}^*({\vec y})$$
is translationally invariant and so  
\begin{equation}\label{mitnn}
\begin{split}
&\alpha_R(\omega, {\vec x})
= \sum_{\nu=0}^\infty \left( \frac{
	\sum_{l}\phi_{\nu, l}({\vec x})
	\phi_{\nu, l}^*({0})}{i\omega +E_\nu} 
\right) \\
\end{split}
\end{equation}
However the wave function $\phi_{\nu, l}(u,\phi)$, 
like any regular wave function at angular momentum $l$, behaves at small values of $u$
like $u^{|l|}$ (times, possibly, a positive power of $u$) and so vanishes at 
the origin unless $l=0$. So it follows that 
\begin{equation}\label{mitnnn}
\begin{split}
&\alpha_R(\omega, {\vec x})
= \sum_{\nu=0}^\infty \frac{
	\phi_{\nu, 0}({\vec x})
	\phi_{\nu,0}^*(0)}{i\omega +E_\nu}  \\
\end{split}
\end{equation}
or equivalently, using \eqref{enmH} 
\begin{equation}\label{mitnnn}
\begin{split}
&\alpha_R(\omega, {\vec x})
= \frac{2 \pi}{b} \sum_{n=0}^\infty \frac{
	e_{n,n}({\vec x})
	e_{n,n}^*(0)}{i\omega +E_n}  \\
\end{split}
\end{equation}
or, using $L_n(0)=1$ so that 
$e_{n,n}(0)= \frac{b}{2 \pi}$ (see \eqref{enm})
\begin{equation}\label{mitnnnn}
\begin{split}
&\alpha_R(\omega, {\vec x})
=  \sum_{n=0}^\infty \frac{
	e_{n,n}({\vec x})}{i\omega +E_n}  \\
\end{split}
\end{equation}

\subsubsection{Adding spin}

Upto this point we have studied the dynamics of 
a spin-less particle; however the discussion above is easily generalized to include the effects of spin. In 2+1 dimensions the little group of a massive particle is $SO(2)$; the irreducible representations of the covering group of $SO(2)$ are all one dimensional, and are labelled by a continuous spin parameter $s$ 
(defined so that the phase corresponding to a 
rotation of angle $\alpha$ is 
$e^{ \i \alpha s}.)$. The Schrodinger equation for a spin $s$ particle on an arbitrary manifold differs from the corresponding equation for a 
spin-less particle even in the absence of a magnetic field (the derivative is generalized to a covariant derivative including the spin connection). In the special case of flat space
studied in this paper this difference goes away, 
and the spin of the particle enters its Schrodinger equation only through a $B. \mu$ coupling in the Hamiltonian. This coupling modifies the Hamiltonian \eqref{Sch} to 
\begin{equation}\label{Sch}
H= - \frac{(\nabla_i - i a_i)^2}{2|M|}-\frac{bg s}{2|M|}
\end{equation}
where $s$ is the particle's spin and $g$ is the 
`g' factor - the anomalous magnetic moment of the particle (the quantity that is 2 for a 
relativistic fermion). The wave functions 
\eqref{lagpol} remain unchanged, but their eigen energies are now given by 
\begin{equation} \label{lagpoln}
 E_{\nu}= 
\frac{b}{|M|} \left( \nu + \frac{1}{2} -\frac{g s}{2} \right) 
\end{equation}
The Greens function continues to take the form  
\eqref{mitnnnn} with $E_\nu$ is given by \eqref{lagpoln}. 

\subsection{Free relativistic bosons in a magnetic field}

The free Klein Gordon equation in the presence 
of a background field is given by 
\begin{equation}\label{gfnrr}
-\left( \partial_t^2  +\left(\partial^x_i -ia_i(x) \right)^2\right) \phi  + M^2 \phi =0, 
\end{equation}
and the corresponding propagator obeys the equation 
\begin{equation}\label{zz}
 \left(-( \partial_t^2  +\left(\partial^x_i -ia_i(x) \right)^2) +M^2\right) \alpha_R(x)
= \delta^3 (x)
\end{equation}
The eigen solutions to \eqref{gfnrr} are given by 
\begin{equation}\label{solns}
\phi= \phi_{\nu, l} e^{\pm i t \sqrt{M^2+
2b\left(\nu+ \frac{1}{2} \right)} } 
\end{equation}
(recall we are in Euclidean space). The  Greens function $\alpha_R$ is given by 
\begin{equation}\label{mitrb}
\begin{split}
\alpha_R(\omega, {\vec x})
&=  \sum_{n=0}^\infty \frac{
	e_{n,n}({\vec x})}{\omega^2 +M^2 + 2b\left(n+ \frac{1}{2} \right)}  \\
&=\sum_{n=0}^\infty \frac{e_{n,n}({\vec x})}{2\sqrt{ M^2 + 2b\left(n+ \frac{1}{2}\right)}}
\left( \frac{1}{i \omega+\sqrt{ M^2 + 2b\left(n+ \frac{1}{2}\right)}} + \frac{1}{-i\omega+\sqrt{ M^2 + 2b\left(n+ \frac{1}{2}\right)}} \right)
\end{split}
\end{equation}
Upon setting $\omega=-i|M|+ \omega_{NR}$ and taking the non relativistic limit $M \to \infty$ at fixed $\omega_{NR}$ we find that the propagator \eqref{mitrb} reduces to the propagator \eqref{mitnnnn} (upto proportionality constants), as expected on general grounds. 

\subsection{Free relativistic fermion in a magnetic field}

Consider a free Dirac particle governed by the 
action
\begin{equation}\label{ffs}
S= \int d^3x {\bar \psi} \left( \gamma^\mu D^{x}_\mu  +m_F\right) \psi 
\end{equation}
placed in a uniform magnetic field. The equation of motion for this particle  
\begin{equation}\label{freeop}
\left(\gamma^\mu D_\mu  +m_F \right) \psi=0
\end{equation}
can be cast into the Schrodinger form\footnote{Note that the operator $\gamma^\mu D_\mu$ 
has no good Hermiticity properties. For this 
reason it is not useful to solve 
\eqref{sdenof} by expanding in eigen solutions of the operator $\gamma^\mu D_\mu +m_F$. }
\begin{equation}\label{sdenof}
\left( \partial_3 + H \right)\psi = 0
\end{equation}
with
\begin{equation}\label{hdef} 
H=\gamma^3 \gamma^i D_i+ \gamma^3 m_F
\end{equation}
The operator $H$ is simply the Hamiltonian 
of the Minkowskian Dirac wave equation. 
\footnote{In other words the 
i.e. the Minkowskian Dirac equation is 
\begin{equation}\label{Dirsch}
i \partial_t \psi = H \psi 
\end{equation} }
As usual, the Hamiltonian $H$-  viewed as 
a differential operator that acts on 
spinors that are functions of $x^1$ and 
$x^2$ - is Hermitian under the standard 
inner product 
\begin{equation}\label{inner product}
(\chi, \psi )= \int d^2x \chi^\dagger(\vec{x}) \psi(\vec{x})
\end{equation}

Let $\psi_n$ denote the normalized eigenfunctions of the operator $H$ with eigenvalue $E_n$, i.e.
\begin{equation}\label{hev}
H \psi_n(\vec{x}) = E_n \psi_n(\vec{x})
\end{equation}
On general eigenfunctions provide an orthonormal basis for (two dimensional) spinor space and so obey the completeness relation 
\begin{equation} \label{completeness}
\sum_n \psi_n(\vec{x}) \psi^\dagger_n(\vec{x'})
=   I \ \delta^2(\vec{x}-\vec{x}')
\end{equation}
where $I$ is the identity matrix in $2 \times 2$ spinor space. 

We now describe the explicit form of the 
eigenfunctions $\psi_n$ and their energy eigen 
spectrum. Let us start with the energy spectrum. 
The spectrum has positive and negative energy 
states. The energies of positive energy states 
are given by  $E=\xi^+_\nu$ where 
\begin{equation}\label{epesp}
(\xi^+_\nu)^2= m_F^2 + 2b \left( \nu+ \frac{1}{2} -
\frac{{\rm sgn}(m_F)}{2} \right)
\end{equation}
The spectrum of negative energy states is given by $E=\xi^-_\nu$ where 
\begin{equation}\label{epesn}
(\xi^-_\nu)^2= m_F^2 + 2b \left( \nu+ \frac{1}{2} +\frac{{\rm sgn}(m_F)}{2} \right)
\end{equation}
These can be combined into the compact expression
\begin{equation}\label{Efer}
E^{\eta \nu J}= \eta \sqrt{m_F^2 + 2b \left( \nu+ \frac{1}{2} -\eta \frac{{\rm sgn}(m_F)}{2} \right) }
\end{equation}
Where $\eta=\pm 1$ for positive/negative energy states.
In both \eqref{epesp} and \eqref{epesn}, the
parameters $\nu$ range over the values 
$$0, 1 , 2 \ \ldots \infty$$
As in previous subsections, in addition to the 
`which Landau Level' label, our eigen energies 
are labelled by their angular momentum $J$. For the positive/negative energy states it is convenient to set 
\begin{equation}\label{Jorb}
J=l + \eta \frac{{\rm sgn}(m_F)}{2}
\end{equation}
For both positive and negative energies,  $l$ (physically the orbital angular momentum) ranges over the values 
$$ -\nu, -\nu+1, -\nu +2 \ldots$$

Let the eigen functions be denoted by 
$\psi^{\eta \nu J}$. Recall that 
$\psi^{\eta \nu J}$ is a two component spinor, 
whose components we denote by 
\begin{equation}\label{psisg}
\psi^{\eta \nu J}=\begin{bmatrix}
\psi_+^{\eta \nu J} \\
\psi_-^{\eta \nu J}
\end{bmatrix}
\end{equation}
After solving for the eigenfunctions we find following explicit results\footnote{$L_n^l$ is associated Laguerre polynomial. For non-negative integer $n$, $L_n^l(x)$ is 
	a polynomial in $x$ of degree $n$  normalised as $ L_n^l(0)=\frac{\Gamma(n+l+1)}{\Gamma(n+1) \Gamma(l+1)}$. For negative integer $n$,  $ L_n^l(x)$ is taken to be zero by definition. In the 
	exceptional range of parameters, the 
	expressions \eqref{psisg} involve the 
	quantity $L_{-1}^{l_\pm}$.
	Consequently the final expressions for $\psi_\pm$ listed in \eqref{psexp} only involve the functions $L_n^l(x)$ for $n \geq 0$. }

\begin{equation}\label{psexp}
\begin{split}
\psi_{\pm}^{\eta \nu J}(u,\phi)= c_{\pm}\frac{ e^{i  \rm{sgn}(J) l_{\pm}  \phi}}{\sqrt{2 \pi}} u^{\frac{l_{\pm}}{2}} e^{-\frac{u}{2}} L_{n_{\pm}}^{l_{\pm}}(u)
\end{split} 
\end{equation} 
where $(u = \frac{b}{2}x_s^2$, $\phi)$ are coordinates on space and \footnote{Here $|x|$ stands for absolute value of $x$. }
\begin{equation}\label{coefficientsT}
\begin{split}
& \ l_+=|J-\frac{1}{2}|, \ l_-=|J-\frac{1}{2}|+\rm{sgn}(J), \  n_+=n, \ n_-=n-\frac{1}{2}\rm{sgn}(J)(1+\rm{sgn}(bJ))\\
&  l=J - \eta \frac{{\rm \rm{sgn}}(m_F)}{2}, ~ n=\nu+l ~~~(l \leq 0), ~~~n= \nu - \eta \frac{{\rm \rm{sgn}}(m_F)}{2} + \frac{1}{2}~~~(l > 0)\\
 &  c_-=c_+ \left(\frac{\sqrt{2|b|}}{m+\rm{sgn}(J)E^{\eta \nu J}}\right)^{\rm{sgn}(J)}
\\
& c_+=\left(\frac{1}{|b|} \left(\frac{ \Gamma(1+n_++l_+)}{\Gamma(1+n_+)}+
\left(\frac{c_-}{c_+}\right)^2\frac{ \Gamma(1+n_-+l_-)}{\Gamma(1+n_-)}
 \right)   \right)^{-1/2}
\end{split}
\end{equation}

In the non relativistic limit the positive energy 
part (i.e. $\eta >0$) of the  Dirac equation 
reduces to a Schrodinger equation for a particle 
of spin $\frac{m_F}{2}$. The effective Schrodinger wave 
function is proportional to $\psi_+$ when $m_F>0$, but 
$\psi_-$ when $m_F<0$. Let us consider the two cases 
in turn. When $m_F>0$ it follows from \eqref{coefficientsT} that $l_+=|l|$ and $n_+=\nu$ if 
$l > 0$ or $n_+=\nu+l$ if $l \leq 0$. 
Similarly, 
it follows from \eqref{coefficientsT}  that if 
$m_F<0$ then $l_-=|l|$ and $n_-= \nu$ if $l>0$ 
and $n_-= \nu+l$ if $l\leq0$. In perfect agreement 
with the expectations \eqref{lagpol}. The non 
relativistic limit of \eqref{epesp} also agrees 
perfectly with \eqref{lagpoln} when $g= 2$ and 
$s=\frac{{\rm sgn}(m_F)}{2}$.

The Free Dirac Green's function in a background 
magnetic field satisfies
\begin{equation}\label{defor}
	\begin{aligned}
		 (\gamma^\mu D^{x}_\mu  +m_F)\alpha_R = -  \delta^3(x)
	\end{aligned}
\end{equation}
In order to solve this equation \eqref{defor} 
we expand 
\begin{equation}
\label{alphaexp}
\alpha_R(x-x')= 
\int \frac{dw}{2\pi} e^{i\omega(t-t')}
\sum_{n,m} \psi_n(\vec{x})
c_{nm}(\omega) \psi^\dagger_m(\vec{x'})
\end{equation}
Plugging this expansion into \eqref{defor}
and using \eqref{completeness}
\begin{equation}
\label{alphasol} 
\int \frac{dw}{2\pi} \sum_{n,m}\left( i\omega + E_n \right)
e^{i\omega(t-t')}
  \psi_n(\vec{x}) c_{nm}(\omega)\psi^\dagger_m(\vec{x'})= 
-\int \frac{d \omega}{2 \pi} e^{ i \omega(t-t')}  \sum_{n} \gamma^3 \psi_n(\vec{x})\psi^\dagger_n(\vec{x'})
\end{equation} 
Multiplying both sides of this equation 
with $\psi^\dagger_a(\vec{x}), \psi_b(\vec{x'})$ (respectively from left and right), 
integrating over $\vec{x}$ and $\vec{x'}$ and using 
orthonormality we find 
\begin{equation} \label{cnmr}
c_{nm}(\omega) = \frac{1}{-i \omega-E_n} \int d^2 x \psi^\dagger_n(\vec{x}) 
	\gamma^3 \psi_m(\vec{x}) 
\end{equation}
Plugging \eqref{cnmr} into \eqref{alphaexp}, 
and using \eqref{completeness} to perform the 
sum over $m$ we find  
\begin{equation}
\label{alphaexp}
\alpha_R(x-x')= 
i\int_{-\infty}^\infty \frac{dw}{2\pi} e^{i\omega(t-t')}
\sum_{n} 
\frac{\psi_n(\vec{x}) \psi^\dagger_n(\vec{x'}) \gamma^3}
{\omega -i E_n}
\end{equation}
With explicit form of these eigen functions at hand this equation takes the form 
\begin{equation}
\label{alphaexpm}
\alpha_R(x-x')= -
\int_{-\infty}^\infty \frac{dw}{2\pi} e^{i\omega(t-t')}
\sum_{\nu=0}^\infty  \left(
\sum_{l=-\nu}^\infty\frac{\psi^{+\nu l}(\vec{x}) (\psi^{+ \nu l})^\dagger (\vec{x'}) \gamma^3}
{i \omega + \xi^+(\nu)} + \sum_{l=-\nu}^\infty\frac{\psi^{-\nu l}(\vec{x}) (\psi^{- \nu l})^\dagger(\vec{x'}) \gamma^3}
	{i \omega  -\xi^-(\nu)} \right)
\end{equation}
As in the discussion around \eqref{mit} the formula 
\eqref{alphaexp} can be simplified by using the 
translational invariance of the Greens function 
$\alpha_R$ to set the argument of every occurrence of  $\psi^\dagger$ to zero. We then use the fact that 
$\psi^{\pm \nu l}$ vanish unless $l=0$. It follows that 
\eqref{alphaexp} can be replaced by 
\begin{equation}
\label{alphaexpm}
\alpha_R(x-x')= -
\int_{-\infty}^\infty \frac{dw}{2\pi} e^{i\omega(t-t')}
\sum_{\nu=0}^\infty  \left(
\frac{\psi^{+\nu 0}(\vec{x}) (\psi^{+ \nu 0})^\dagger (0) \gamma^3}
{i \omega + \xi^+(\nu)} + \frac{\psi^{-\nu 0}(\vec{x}) (\psi^{- \nu 0})^\dagger(\vec{x'}) \gamma^3}
{i \omega  -\xi^-(\nu)} \right)
\end{equation}
Note that the denominators $\xi^\pm (\nu)$ obey the 
relationship 
\begin{equation} \label{rela}
\xi^+_\nu= \xi^{-}_{\nu-{\rm sgn}(m_F)}
\end{equation}
In other words if we choose
\begin{equation}\label{nundic}
\begin{split}
\nu=
\begin{cases}
&\tilde{n}+ \frac{1}{2} +\frac{{\rm sgn}(m_F)}{2}~~~~~~{\rm +ve} ~~{\rm energy}\\
	&\tilde{n}+ \frac{1}{2} -\frac{{\rm sgn}(m_F)}{2}
		~~~~~~{\rm -ve} ~~{\rm energy}\\
	\end{cases}
	\end{split}
	\end{equation}
Then we have 
\begin{equation}\label{bothchi}
(\xi^+_\nu)^2=(\xi^-_\nu)^2=m_F^2+ 2 |b|(\tilde{n}+1) \equiv \xi_{\tilde{n}}
\end{equation}
For this reason it is sometimes convenient to combine 
together the positive and negative energy Landau Levels with 
the same value of ${\tilde n}$ \footnote{Note that the levels we 
combine do not have the same value of $\nu$; their $\nu$ levels differs 
by unity, as follows from \eqref{nundic}.} upon doing 
this we find that the propagator is given by 
\begin{equation}\label{alphafreeexpansion}
\begin{split}
&\alpha_R^F(x-x')= 
\int_{-\infty}^\infty \frac{dw}{2\pi} e^{i\omega(t-t')}
\alpha_R^F(\omega,\vec{x}-\vec{x'})\\
&\alpha_R^F(\omega,\vec{x})=\gamma^+\alpha_{R,+}^F(\omega,\vec{x})+\gamma^-\alpha_{R,-}^F(\omega,\vec{x})+\gamma^3\alpha_{R,3}^F(\omega,\vec{x})+I\alpha_{R,I}^F(\omega,\vec{x}) \\
&\alpha_{R,+}^F(u,\phi)= -\sum_{n=0}^{\infty}\frac{1}{\Pi(\omega,n)} ( i \sqrt{b(n+1)}) e_{n+1,n}(u,\phi)\\
&\alpha_{R,-}^F(u,\phi)= -\sum_{n=0}^{\infty}\frac{1}{\Pi(\omega,n)} (- i \sqrt{b(n+1)}) e_{n+1,n}(u,\phi)^*\\
&\alpha_{R,3}^F(u,\phi)= -\frac{1}{2}\sum_{n=0}^{\infty} \left(  \frac{(i\omega-m_F)}{\Pi(\omega,n-1)} + \frac{(i\omega+m_F)}{\Pi(\omega,n)}\right)e_{n,n}(u,\phi)\\
&\alpha_{R,I}^F(u,\phi)=  -\frac{1}{2}\sum_{n=0}^{\infty} \left(  \frac{(i\omega-m_F)}{\Pi(\omega,n-1)} - \frac{(i\omega+m_F)}{\Pi(\omega,n)}\right)e_{n,n}(u,\phi)\\
&-\Pi(\omega,n)=-(i \omega + \xi_n) (i \omega-\xi_n)=  \omega^2+ m_F^2 + 2 |b|(n+1) 
\end{split}
\end{equation}

\subsection{Free relativistic particle of spin s in a magnetic field }

We have already seen that a non relativistic particle of spin 
$s$ and $g$ factor 2 has a Greens Function with poles located at $\omega=i E_\nu$ 
where
\begin{equation} \label{lagpoln}
E_{\nu}= 
\frac{b}{|M|} \left( \nu + \frac{1}{2} - s \right) 
\end{equation}

On the other hand we have also seen that the poles of a 
relativistic particle of spin $s=0$ or $s \pm \frac{1}{2}$ occur at $i \chi^+_\nu$ or $-i\chi^-_\nu$ 
where 
\begin{equation}\label{epespms} 
\begin{split} 
&(\chi^+_\nu)^2= M^2 + 2b \left( \nu+ \frac{1}{2} -s \right)\\
&(\chi^-_\nu)^2= M^2 + 2b \left( \nu+ \frac{1}{2} +s \right)\\
\end{split}
\end{equation}

It is natural to suggest that the propagator for a `free particle of arbitrary spin $s$'- to the extent that one can make sense of such a notion - should also have poles located at 
the values listed in \eqref{epespms}. Moreover at these values of $\nu$ we should expect to find an infinite degeneracy of 
states, with angular momenta given - respectively for positive and negative energy states - by  
\begin{equation}\label{Jorb}
J=l \pm s
\end{equation}
with the `orbital angular momentum $l$' ranging over the values 
$$ -\nu, -\nu+1, -\nu +2 \ldots$$

\section{A twisted convolution (Moyal star product)}

When we turn on interactions in the next section, it will turn out that our matter 
propagators obey a Schwinger Dyson equation that 
is given in terms of a twisted convolution $*_b$ defined as follows. Given any two functions $A$ and 
$B$ on $R^2$ we 
define
\begin{equation}\label{stp}
\left( A *_{b} B \right) (x)= \int d^2 w
A(x-w) e^{-i \frac{b}{2} \epsilon_{ij}(x^i-w^i) (w^j) } B(w)
\end{equation}
In Fourier space \eqref{stp} is simply the famous Moyal star product. 
It follows that the $*_b$ product is associative but non commutative. 

Quite remarkably, the eigen functions $e_{n,m}$ defined above are the basis functions used 
in the discussions of non-commutative solitons \cite{Gopakumar:2000zd} (after flip of position and momentum) and so obey the following identity
\begin{equation} \label{mrules} 
\begin{split}
& (e_{n,p} ~*_b ~e_{p', m})(\vec{x}) = \delta_{p p'} e_{n, m}(\vec{x}) \\
\end{split}
\end{equation}
Now we turn to an algebraic exercise that will be useful in next section. The property \eqref{mrules} can be used to define inverse of twisted convolution as we demonstrate below.
Consider $2 \times 2$ matrix 
valued function given by
\begin{equation}\label{simpalg}
\alpha= \sum_{m=0}^{\infty} \left(\alpha_{+,m+1} e_{m+1, m}\gamma^+ +\alpha_{-,m} e_{m,m+1}\gamma^-+ 
\alpha_{3,m} e_{m,m} \gamma^3  
+ \alpha_{I m} e_{m.m} I\right)
\end{equation} 
and another similar function given by 
\begin{equation}\label{simpalgb}
\beta= \sum_{m=0}^{\infty} \left(\beta_{+,m+1} e_{m+1, m}\gamma^+ +\beta_{-,m} e_{m,m+1}\gamma^-+ 
\beta_{3,m} e_{m,m} \gamma^3  
+ \beta_{I m} e_{m.m} I\right)
\end{equation}
Then, using \eqref{mrules}, it is easily verified that 
\begin{equation}\label{albe}
(\alpha*_b \beta)({\vec x} - {\vec y}) = (\beta *_b \alpha ({\vec x} - {\vec y}))
= I \delta^{2}({\vec x} - {\vec y})
\end{equation} 
if and only if the coefficients $\alpha_n$ 
are determined in terms of $\beta_n$ as follows. For  $n>0$
\begin{equation}\label{starinverse}
\begin{split}
&\alpha_{+,n}= \frac{1}{\Pi(n-1)} \beta_{+,n}\\
&\alpha_{-,n}= \frac{1}{\Pi(n)} \beta_{-,n}\\
&\alpha_{3,n}= \frac{1}{2} \left(  \frac{\beta_{3,n-1}-\beta_{I,n-1}}{\Pi(n-1)} + \frac{\beta_{3,n+1}+\beta_{I,n+1}}{\Pi(n)}\right)\\
&\alpha_{I,n}=  \frac{1}{2} \left(  \frac{\beta_{3,n-1}-\beta_{I,n-1}}{\Pi(n-1)} - \frac{\beta_{3,n+1}+\beta_{I,n+1}}{\Pi(n)}\right)\\
\end{split}
\end{equation}
and for $n=0$\footnote{Roughly speaking \eqref{stin2} is analytical continuation of \eqref{starinverse} once we use the definition that
	\begin{equation}
	\beta_{+,0}= 0
	\end{equation}.}
\begin{equation}\label{stin2}
\begin{split}
&\alpha_{+,0}= 0\\
&\alpha_{-,0}= \frac{1}{\Pi(0)} \beta_{-,0}\\
&\alpha_{3,0}= \frac{1}{2} \left(  \frac{1}{\beta_{3,0}+\beta_{I,0}} + \frac{\beta_{3,1}+\beta_{I,1}}{\Pi(0)}\right)\\
&\alpha_{I,n}=  \frac{1}{2} \left(  \frac{1}{\beta_{3,0}+\beta_{I,0}}- \frac{\beta_{3,1}+\beta_{I,1}}{\Pi(0)}\right)\\
\end{split}
\end{equation}
where we have defined
\begin{equation}\label{npdef}
\Pi(n)=(\beta_{3,n}-\beta_{I,n})(\beta_{3,n+1}+\beta_{I,n+1})+2 \beta_{+,n+1}\beta_{-,n}
\end{equation}

\subsection{Twisted convolution and free propagators}

\subsubsection{Free non-relativistic particles of 
	spin $s$}

Consider the function 
\begin{equation}\label{do} 
K^{NR}_{R}(\omega, {\vec x}) 
=  \sum_{n=0}^\infty \left( i\omega +E_n \right)
	e_{n,n}({\vec x}) 
\end{equation}
with $E_n$ given by \eqref{lagpoln}.
It follows from \eqref{mrules} and \eqref{mitnnnn}that 
\begin{equation} \label{dersc}
(K^{NR}_{R}(\omega)*_b \alpha_R(\omega))({\vec x})= \delta^2({\vec x})
\end{equation}
(where $\alpha_R$ is given by \eqref{mitnnnn} ).
More generally, given any function 
$A(\omega, {\vec x})$, we have  
\begin{equation}\label{dof}
\left( i\omega  -\frac{(\partial_{x_i} -ia_i(x-x'))^2}{2|M|} -\frac{bg s}{2|M|}
\right) A(\omega, {\vec x})
= \left( K^{NR}_R(\omega) *_b A(\omega) \right)({\vec x})
\end{equation}
In other words, the action of the Schrodinger 
operator on any wave function can be obtained 
by star convoluting $K^{NR}_R$ with that function. 

\subsubsection{Free relativistic boson}

As in the previous subsection if we define 
\begin{equation}\label{rsp}
K^{B}_{R}(\omega, {\vec x})= 
\sum_{n=0}^\infty \left( \omega^2 +M^2 + 2b\left(n+ \frac{1}{2} \right) \right)
e_{n,n}({\vec x}) 
\end{equation}
it follows that 
\begin{equation} \label{dersc}
(K^{B}_{R}(\omega) *_b \alpha_R(\omega))({\vec x})= \delta^2({\vec x})
\end{equation}
(where $\alpha_R$ is now the reduced propagator 
for the relativistic scalar theory given in \eqref{mitrb}) and, more generally, 
\begin{equation}\label{dof}
\left(\omega^2-  \left(\partial^x_i -ia_i(x) \right)^2 +M^2\right) A(\omega, {\vec x})
= \left(K^{B}_{R}(\omega) *_b A(\omega) \right)({\vec x})
\end{equation}

\subsubsection{Free relativistic fermion}

The matrix valued function $K^F_R$ given by\footnote{That is \begin{equation}\label{cdast}
K_{R,+,n}^F= i \sqrt{b(n)},~ K_{R,-,n}^F= i \sqrt{b(n+1)},~ K_{R,3,n}^F= i \omega,~ K_{R,I,n}^F=m_F
\end{equation}} 
\begin{equation}\label{kfrd}
K^F_R = \sum_{m=0}^{\infty} \left(i \sqrt{b(m+1)} e_{m+1, m}\gamma^+ +i \sqrt{b(m+1)} e_{m,m+1}\gamma^-+ 
i \omega e_{m,m} \gamma^3  
+ m_F e_{m.m} I\right)
\end{equation}
obeys the identity
\begin{equation}\label{anfun}
(K^F_R(\omega) *_b \alpha_R(\omega))(\vec{x})= -  \delta^2(\vec{x})
\end{equation}
More generally $K^F_R$ obeys the identity
\begin{equation}\label{anfun}
(\gamma^3 i \omega+\gamma^i D_i  +m_F)A(\omega, {\vec x})=(K^F_R(\omega) *_b A(\omega))({\vec x})
\end{equation}
where $A$ is any normalizable $2 \times 2$ matrix valued function on $R^2$.

\section{Regular fermions coupled to Chern Simons gauge fields} \label{regfer}

\subsection{Gap equation in a uniform background gauge field}

Consider $SU(N_F)$ regular fermions at level $k_F - \frac{1}{2} {\rm sgn}(k_F)$\footnote{Note that integrating out the fermion shifts the level to $k_F - \frac{1}{2}{\rm sgn}(k_F)+\frac{1}{2}{\rm sgn}(m_F)$. For $\rm{sgn}(m_Fk_F)=1$ low energy Chern Simons level becomes $k_F$, whereas for ${\rm sgn}(m_Fk_F)=-1$ low energy Chern Simons level becomes ${\rm sgn}(k_F)(|k_F|-1)$. At least in absence of a magnetic field, this is consistent with duel bosonic theory being in unhiggsed/higgsed phase (see \cite{Choudhury:2018iwf},\cite{Dey:2018ykx} for more details).  }. As usual we use the symbol $\kappa_F$ to denote 
\begin{equation} \label{kf}
\kappa_F={\rm sgn}(k_F) \left( |k_F|+ N_F \right)
\end{equation} 
Our fermionic theory has a  $U(1)$ global symmetry - the symmetry under which the fundamental field $\psi$ has charge 
$+1$ and the anti-fundamental field ${\bar \psi}$ has charge $-1$. In the presence of a background gauge field $a_\mu$ 
for this global $U(1)$ symmetry, the Euclidean action for our theory takes the form
\begin{equation}\label{eucact}
S_F= \frac{i \kappa_F}{4 \pi} 
\int {\rm Tr} \left( A dA + \frac{2}{3} A^3 
\right) + \int {\bar \psi}  \gamma^\mu \tilde{D}_\mu \psi + m_F {\bar \psi} \psi 
\end{equation}  
where 
\begin{equation} \label{Ddef}
\tilde{D}_\mu \psi  = \left( \partial_\mu -i A_\mu  -i a_\mu \right) 
\psi
\end{equation}
Let 
\begin{equation}\label{fermprop}
\langle \psi(x)_{A\alpha } 
{\bar \psi}^{B\beta}(y)\rangle = 
\delta_A^B \alpha(x, y)_\alpha^\beta
\end{equation} 
($A$ is an $SU(N_F)$ fundamental gauge index, 
- note the subscript $F$ stands for Fermion not flavour - $B$ is an $SU(N_F)$ gauge anti-fundamental index and 
$\alpha$ and $\beta$ are spinor indices).

In the large $N$ limit and in the lightcone gauge used in this paper (or more generally in any gauge in which the cubic term drops out of the Chern Simons action) the propagator $\alpha(x,y)$ 
obeys the `Schwinger Dyson' or gap equation 
\begin{equation}\label{sde} \begin{split} 
&\left( \gamma^\mu D^{x}_\mu  +m_F\right) 
\alpha(x,y) + \int d^3 w~ 
\Sigma(x, w) \alpha(w, y) = - \delta^{3}(x-y)\\
& \Sigma(x,y)=-\frac{N}{2}   \gamma^\mu \alpha(x,y) \gamma^\nu G_{\mu\nu}(x-y)\\
\end{split} 
\end{equation}
where spinor indices (which contract by the rules of matrix multiplication) have been omitted for 
simplicity and where 
\begin{equation}\label{dmx}
 D^{x}_\mu= \partial^x_\mu -i a_\mu(x)
 \end{equation}
and 
\begin{equation}\label{Aprop}
\langle A_\mu(x) A_\nu(y) \rangle 
= G_{\mu\nu}(x-y)
\end{equation}

As we have explained above, in this 
paper we turn on a uniform magnetic field of 
magnitude $b$ in the $12$ plane and work in the 
rotationally symmetric gauge \eqref{rigc}.
As explained in Subsection \ref{ReducedVariables} above, the 
Greens function $\alpha$ and the self energy $\Sigma$ in this 
gauge are given in terms of translationally and rotationally 
invariant functions by the formulae 
\eqref{ussm} (which we reproduce here 
for convenience)
\begin{equation}\label{ussmr}
\alpha(x, y)=e^{-i \frac{b}{2} \left(x^1 y^2 -x^2 y^1 
	\right) } \alpha_R(x-y), ~~~~
\Sigma(x,y) =e^{-i \frac{b}{2} \left(x^1 y^2 -x^2 y^1 
	\right) } \Sigma_R(x-y)
\end{equation}
where $\alpha_R(x-y)$ and 
 $\Sigma_R(x-y)$ are both simultaneously rotationally and translationally invariant. 

Plugging \eqref{ussmr} into \eqref{sde} yields the following gap 
equations for $\alpha_R(x-y)$ and 
$\Sigma_R(x-y)$. 
\begin{equation} \label{sdentt}
\begin{split}
&\Sigma_R(x-y)=-\frac{N}{2}   \gamma^\mu \alpha_R(x-y) \gamma^\nu G_{\mu\nu}(x-y)\\
&\left( \gamma^\mu D^{(x-y)}_\mu  +m_F\right) 
\alpha_R(x-y) + \int d^3 w~ 
\Sigma_R(x-w) e^{-i \frac{b}{2} \epsilon_{ij}(x^i-w^i) (w^j-y^j) } \alpha_R(w-y) = - \delta^{3}(x-y)\\
\end{split}
\end{equation}
It is convenient to view our equations in Fourier space in time
\begin{equation}\label{ftconva} \begin{split} 
&\alpha_R(x)= \int \frac{d \omega}{2 \pi} 
e^{i \omega x^3} \alpha_R(\omega, {\vec x})\\
&\Sigma_R(x)= \int \frac{d \omega}{2 \pi} 
e^{i \omega x^3} \Sigma_R(\omega, {\vec x})\\
\end{split}
\end{equation} 

For definiteness in this paper we work in $A_-(x)=0$ gauge, in which gauge field propagator takes following form
\begin{equation}\label{GaugeProp}
G_{\mu\nu}(x)=-\frac{2i}{\kappa_F}\epsilon_{\mu\nu-}\frac{\delta(x^3)}{x^+} \implies G_{\mu\nu}(\omega,{\vec x})=-\frac{2i}{\kappa_F}\epsilon_{\mu\nu-}\frac{1}{x^+} 
\end{equation}
It follows 
from the first of \eqref{sdentt} that 
\begin{equation}\label{sigom}
 \Sigma_R(x-y) \propto \delta(x^3-y^3)
 \end{equation}
It follows that the factor of $\alpha_R(x-y)$ on the RHS of the first 
of \eqref{sdentt} is always evaluated at $x^3=y^3$. Note that the equal time propagator is given in Fourier space by  
\begin{equation}\label{halphahat}
\alpha_R(x-y)|_{x^3=y^3}=\hat{\alpha}_{R}({\vec x}-{\vec y}) \equiv \int_{-\infty}^{+\infty}\frac{d\omega}{2 \pi} \alpha_{R}(\omega,{\vec x}-{\vec y})
\end{equation}
The gap equations can be recast as 
\begin{equation} \label{sdento}
\begin{split}
&\Sigma_R(\omega,{\vec x}-{\vec y})=-\frac{N}{2}   \gamma^\mu {\hat \alpha_R({\vec x}-{\vec y})} \gamma^\nu G_{\mu\nu}(\omega,{\vec x}-{\vec y})\\
&\left( i \omega \gamma^3 
+ {\vec D}^{(x-y)} \gamma^i + m_F
\right) \alpha_R(\omega, {\vec x}-{\vec y}) + \left( \Sigma_R(\omega)
*_b \alpha_R(\omega) \right)({\vec x}-{\vec y})= - \delta^{2}({\vec x}-{\vec y})
\end{split}
\end{equation}
where $*_b$ is the twisted convolution defined in \eqref{stp}. 
It follows immediately from \eqref{GaugeProp} (which in particular asserts that
$G_{\mu\nu}(\omega,{\vec x})$ is independent of $\omega$) that  
$\Sigma_R(\omega, {\vec x})$ is actually independent of $\omega$.

\subsection{Solving for $\alpha_R$ in terms of $\Sigma_R$}

In this subsection we use the second of \eqref{sdento} \begin{equation}\label{sdenofsspr}
\left( i \omega \gamma^3 
+ {\vec D}^{(x-y)} \gamma^i + m_F
\right) \alpha_R(\omega, {\vec x}-{\vec y}) + \left( \Sigma_R(\omega) 
*_b \alpha_R(\omega) \right)({\vec x}-{\vec y})= - \delta^{2}({\vec x}-{\vec y})
\end{equation}
to evaluate the propagator $\alpha_R$ in terms of $\Sigma_R$. Using 
\eqref{kfrd} and \eqref{anfun} (recall these equations allow us to 
replace the Dirac differential operator by a star product with the function $K^F_R)$, \eqref{sdenofsspr} may be rewritten 
as 
\begin{equation}\label{gapeqstarform}
\begin{split}
&(K_R(\omega) 
*_b \alpha_R(\omega))({\vec x}-{\vec y})= - \delta^{2}({\vec x}-{\vec y})\\
\end{split}\end{equation}
where, \footnote{Recall $\Sigma_R(\omega,\vec{x})$ is independent of $\omega$ as argued in previous section.}
\begin{equation}\label{wheyu}
\begin{split}
&K_R(\omega,\vec{x})= K_R^F(\omega,\vec{x})+ \Sigma_R(\omega,\vec{x}) \\
& K_R^F(\omega,\vec{x})=\sum_{m=0}^{\infty} \left(i \sqrt{b(m+1)} e_{m+1, m}\gamma^+ +i \sqrt{b(m+1)} e_{m,m+1}\gamma^-+ 
i \omega e_{m,m} \gamma^3  
+ m_F e_{m.m} I\right)\\
&\Sigma_R(\omega,\vec{x})=\sum_{m=0}^{\infty} \left(\Sigma_{+,m+1} e_{m+1, m}\gamma^+ +\Sigma_{-,m} e_{m,m+1}\gamma^-+ 
\Sigma_{3,m} e_{m,m} \gamma^3  
+ \Sigma_{I m} e_{m.m} I\right)
\end{split}
\end{equation}
We will now use \eqref{gapeqstarform} to solve for $\alpha_R$ in 
terms of the function $K_R$ (and so, effectively, in terms of $\Sigma$).

We now note that \eqref{gapeqstarform} is of the form \eqref{albe} 
with the role of $\beta$ played by $K_R$. It follows from 
\eqref{starinverse} and \eqref{stin2} that  for $n>0$
\begin{equation}\label{alphasol}
\begin{split}
&\alpha_{R,+,n}= -\frac{1}{\Pi(n-1)} K_{R,+,n}\\
&\alpha_{R,-,n}= -\frac{1}{\Pi(n)} K_{R,-,n}\\
&\alpha_{R,3,n}= -\frac{1}{2} \left(  \frac{K_{R,3,n-1}-K_{R,I,n-1}}{\Pi(n-1)} + \frac{K_{R,3,n+1}+K_{R,I,n+1}}{\Pi(n)}\right)\\
&\alpha_{R,I,n}= - \frac{1}{2} \left(  \frac{K_{R,3,n-1}-K_{R,I,n-1}}{\Pi(n-1)} - \frac{K_{R,3,n+1}+K_{R,I,n+1}}{\Pi(n)}\right)\\
\end{split}
\end{equation}
and for $n=0$
\begin{equation}\label{alphasol2}
\begin{split}
&\alpha_{R,+,0}= 0\\
&\alpha_{R,-,0}= -\frac{1}{\Pi(0)} K_{R,-,0}\\
&\alpha_{R,3,0}= -\frac{1}{2} \left(  \frac{1}{K_{R,3,0}+K_{R,I,0}} + \frac{K_{R,3,1}+K_{R,I,1}}{\Pi(0)}\right)\\
&\alpha_{R,I,0}= - \frac{1}{2} \left(  \frac{1}{K_{R,3,0}+K_{R,I,0}}- \frac{K_{R,3,1}+K_{K,I,1}}{\Pi(0)}\right)\\
\end{split}
\end{equation}
where
\begin{equation}
\Pi(n)=(K_{R,3,n}-K_{R,I,n})(K_{R,3,n+1}+K_{R,I,n+1})+2 K_{R,+,n+1}K_{R,-,n}
\end{equation}

The RHS of \eqref{alphasol} and \eqref{alphasol2} are complicated
functions of the unknown coefficients in the expansion of $\Sigma$. We have already seen, however, that these otherwise unknown coefficients are independent of $\omega$. For this reason it is possible to make the $\omega$ dependence in \eqref{seom} \eqref{alphasol} and \eqref{alphasol2} completely 
explicit. To proceed we borrow the results \eqref{sgp3sol}
(which implies the expansion of $\Sigma$ \eqref{seom}) from the next subsection. Plugging that expansion into \eqref{alphasol} and \eqref{alphasol2}, using the fact that the coefficient 
functions in \eqref{seom} are independent of $\omega$, and 
doing a little algebra we find that the quantity $\Pi(n)$ is 
a quadratic expression in $\omega$; more precisely  
\begin{equation}\label{factorizedprop}
\begin{split}
\Pi(n)&=-(\omega-i \zeta_+^F(n))(\omega+i \zeta_-^F(n))\\
\zeta_\pm^F(n)&=\pm \frac{1}{2}\left( (K_{R,I,n+1} -K_{R,I,n})\pm \left( (K_{R,I,n+1} -K_{R,I,n})^2 \right. \right. \\ 
& ~~~~~~~~~~~~~~~~~~~~~~~~~~~~~~ \left.  \left. +4  K_{R,I,n+1}K_{R,I,n}-8K_{R,+,n+1}K_{R,-,n}\right)^{1/2} \right)\\
&=\pm \frac{1}{2}( (\Sigma_{R,I,n+1} -\Sigma_{R,I,n})\pm ( (\Sigma_{R,I,n+1} -\Sigma_{R,I,n})^2 \\ 
& ~~ ~ +4  (m_F+\Sigma_{R,I,n+1})(m_F+\Sigma_{R,I,n})-8i\sqrt{b(n+1)}(i\sqrt{b(n+1)} +\Sigma_{R,+,n+1}))^{1/2} )
\end{split}
\end{equation}
We also define the quantity 
\begin{equation} \label{seex}
\zeta_{ex}= K_{R, I, 0}= m_F+\Sigma_{R, I, 0}
\end{equation}

The quantities $\zeta_\pm^F(n)$ are the positive/negative energy poles of the exact fermion propagator which reduce, in the 
limit $\lambda_F \to 0$ to the poles $\zeta^\pm_\nu$ (see 
\eqref{epesp} and \eqref{epesn} with the relationship 
between $\nu$ and $n$ \eqref{nundic}) for $n \geq 0$. 
On the other hand the quantity $\zeta_{ex}$ reduces to $m_F$, i.e. to the 
energy of the exceptional state at zero $\lambda_F$ (see under \eqref{bothchi}).

The equations \eqref{alphasol} and \eqref{alphasol2} 
respectively may now be rewritten as 
\begin{equation}
\begin{split} \label{alphaform}
&\alpha_{R,+,n}= \frac{( i\sqrt{b(n)}+ \Sigma_{R,+,n})}{(\omega-i \zeta_+^F(n-1))(\omega+i \zeta_-^F(n-1))}\\
&\alpha_{R,-,n}= \frac{i\sqrt{b(n+1)}}{(\omega-i \zeta_+^F(n))(\omega+i \zeta_-^F(n))}\\
&\alpha_{R,3,n}= \frac{1}{2} \left(  \frac{i \omega-m_F- \Sigma_{R,I,n-1}}{(\omega-i \zeta_+^F(n-1))(\omega+i \zeta_-^F(n-1))} + \frac{i \omega+m_F+ \Sigma_{R,R,I,n+1}}{(\omega-i \zeta_+^F(n))(\omega+i \zeta_-^F(n))}\right)\\
&\alpha_{R,I,n}= \frac{1}{2} \left(  \frac{i \omega-m_F- \Sigma_{R,I,n-1}}{(\omega-i \zeta_+^F(n-1))(\omega+i \zeta_-^F(n-1))} - \frac{i \omega+m_F+ \Sigma_{R,I,n+1}}{(\omega-i \zeta_+^F(n))(\omega+i \zeta_-^F(n))}\right)\\
\end{split}
\end{equation}
and 
\begin{equation}\label{alphasol2}
\begin{split}
&\alpha_{R,+,0}= 0\\
&\alpha_{R,-,0}=  \frac{i\sqrt{b}}{(\omega-i \zeta_+^F(0))(\omega+i \zeta_-^F(0))}\\
&\alpha_{R,3,0}= \frac{1}{2} \left( \frac{i}{\omega -i\zeta_{ex}} +\frac{i \omega+m_F+ \Sigma_{R,R,I,1}}{(\omega-i \zeta_+^F(0))(\omega+i \zeta_-^F(0))}\right)\\
&\alpha_{R,I,0}=  \frac{1}{2} \left(  \frac{i}{ \omega - i \zeta_{ex}}-\frac{i \omega+m_F+ \Sigma_{R,R,I,1}}{(\omega-i \zeta_+^F(0))(\omega+i \zeta_-^F(0))}\right)\\
\end{split}
\end{equation}

\subsection{Recursion relations for the coefficients of $\Sigma$}

In this subsection we will attempt to complete the process of solving the gap equations by plugging \eqref{alphasol} and \eqref{alphasol} into 
the first of \eqref{sdento} 
\begin{equation}\label{gapeq}
\Sigma_R({\vec x}-{\vec y})=-\frac{N}{2}   \gamma^\mu {\hat \alpha_R({\vec x}-{\vec y})} \gamma^\nu G_{\mu\nu}({\vec x}-{\vec y})
\end{equation}
to obtain a closed set of equations 
for the coefficient functions of the quantity $\Sigma$. 

As we have noted above, in this paper we follow earlier studies of matter 
Chern Simons theories (starting with  \cite{Giombi:2009wh} ) to work in lightcone 
gauge $A_-=0$. In this gauge  we have 
\begin{equation}\label{gcgp}
G_{\mu\nu}({\vec x})=-\frac{2i}{k}\epsilon_{\mu\nu-}\frac{1}{x^+} 
\end{equation}
Note that the only non-zero components of Chern-Simons field propagator are  $G_{+3},G_{3+}$. This immediately implies\footnote{Here we are using the linear algebra identity 
that for any matrix $\alpha_R $ we have
\begin{equation}
 \gamma^3 \alpha_R \gamma^+-\gamma^+ \alpha_R \gamma^3=2(\alpha_{R,I} \gamma^+-\alpha_{R,I}I)
\end{equation}}
\begin{equation}\label{sgp3sol}
 \Sigma_{R,-}=0, ~ \Sigma_{R,3}=0
\end{equation}
so that the last of \eqref{wheyu} (the expansion of $\Sigma$ in 
terms of its components) simplifies to 
\begin{equation}\label{seom}
\Sigma= \sum_{n=0}^\infty 
(\Sigma_{R, +, n}  e_{n, n-1} \gamma^++ 
\Sigma_{R, I, n} e_{n, n}I)
\end{equation} 
where $\Sigma_{R, + n}$ and $\Sigma_{R, + n}$
are $\omega$ independent numbers.

We now use \eqref{enmz} to expand the RHS of \eqref{gapeq} in a linear  sum of the basis elements $e_{n,m}$ using \eqref{decomp} and equate coefficients on both 
sides of \eqref{gapeq}. We obtain the equations
\begin{equation}\label{sigmge}
\begin{split}
&\Sigma_{R,I,n'}=2i\sqrt{b}\lambda_F \sum_{n=n'+1}^{\infty} \frac{1}{\sqrt{n}}\hat{\alpha}_{R,-,n-1} \\
&\Sigma_{R,+, n'+1}=2i\sqrt{b}\lambda_F \sum_{n=0}^{n'} \frac{1}{\sqrt{n'+1}}\hat{\alpha}_{R,I,n} \\ 
\end{split}
\end{equation} 
where  ${\hat \alpha_R}$ was defined in \eqref{halphahat} and
as usual we have expanded  
\begin{equation}\label{simpalg}
{\hat \alpha}= \sum_{m=0}^{\infty} \left({\hat \alpha_{+,m+1}} e_{m+1, m}\gamma^+ +{\hat \alpha_{-,m}} e_{m,m+1}\gamma^-+ 
{\hat \alpha_{3,m}} e_{m,m} \gamma^3  
+ {\hat \alpha_{I m}} e_{m.m} I\right)
\end{equation} 
Since we have already solved for the coefficients of 
$\alpha$ - and so the coefficients of ${\hat \alpha}$ - in terms 
of the coefficients in \eqref{seom}, \eqref{sigmge} are closed 
equations for the unknown coefficients in the expansion of 
$\Sigma$ (i.e. the equation \eqref{seom}). We will make these 
equations more explicit below. 

To end this subsection we note that \eqref{sigmge} is the 
analogue of the integral equation 2.67 in \cite{Giombi:2009wh}; 
as we will see later \eqref{sigmge} actually reduce to 2.67 of 
\cite{Giombi:2009wh} in the zero $b$ limit. 

The authors of \cite{Giombi:2009wh} were able to solve their
integral equation by differentiating it 
(see 2.68 of that paper). In order to solve 
the equation \eqref{sigmge} we adopt a 
similar manoeuver; we simply subtract the 
equations \eqref{sigmge} at $n$ and $n-1$
to obtain 
\begin{equation}\label{gapeqfinal}
 \begin{split}
  & \sqrt{n+1}(\Sigma_{R,I,n+1}-\Sigma_{R,I,n})=-2i\sqrt{b}\lambda_F \hat{\alpha}_{R,-,n}\\
  & \sqrt{n+1}\Sigma_{R,+,n+1}-\sqrt{n}\Sigma_{R,+,n}=+2i\sqrt{b}\lambda_F \hat{\alpha}_{R,I,n}
 \end{split}
\end{equation}

\subsection{Reality properties} \label{intfe}

In this brief subsection we momentarily break the flow of our 
presentation in order to discuss the reality properties of 
various quantities. It is not 
difficult to argue that the quantities
\begin{equation}\label{realprop}
 \Sigma_{R, I, n}, ~~~K_{R, I, n}, ~~~
{\hat \alpha}_{R,I,n}, ~~~\Pi(n)
\end{equation}
are all real, while 
\begin{equation}\label{improp}
\Sigma_{R, -, n}, ~~~K_{R, \pm, n}, ~~~
{\hat \alpha}_{R, \pm ,n}
\end{equation}
are purely imaginary. In order to see that this is the case we first note that these  reality properties are manifest at $\lambda_F=0$ \footnote{It is obvious that \eqref{realprop} \eqref{improp} are true of $K_R^F$ . In order to verify 
the corresponding claims for ${\hat \alpha}_R^F$ we use \eqref{alphafreeexpansion} 
 and perform the integral 
over $\omega$ needed to obtain 
${\hat \alpha}_R^F$: note that imaginary 
terms in the third and fourth lines of 
\eqref{alphafreeexpansion} integrate to 
zero.}. We next verify that the reality 
properties \eqref{improp} and 
\eqref{realprop} are consistent with our
gap equations. The consistency of these 
reality properties with \eqref{sigmge} 
is manifest. In order to verify that 
\eqref{improp} and \eqref{realprop} are 
consistent with \eqref{alphasol} we 
integrate both sides of \eqref{alphasol} 
over $\omega$ from $-\infty$ to $\infty$; 
it is not difficult to convince oneself that 
the result of such an integral is consistent 
with the reality properties assumed assigned 
to the LHS of \eqref{alphasol}
\footnote{The expressions in the  RHS of the first two lines of \eqref{alphasol} are of the schematic form $i A(i\omega)$ while the 
expressions on the RHS of the third and 
fourth lines of the same equation take the 
schematic form $B(i\omega)$ where $A(z)$ and 
$B(z)$ complex functions that are real valued
when $z$ is restricted to the real line. It 
follows that the integral of the RHS of these
equations - performed using any regulation 
scheme that preserves $\omega \leftrightarrow -\omega$ symmetry (such as the regulation scheme employed in this paper) - is imaginary for the first two lines of \eqref{alphasol}
but real for the next two lines of \eqref{alphasol}, establishing the 
consistency of \eqref{alphasol} with the 
reality properties listed above. }

In this paper we  solve our gap equations recursively, order by order in the coupling 
constant. It follows immediately from the 
discussion of the last paragraph that 
each order in the recursive expansion
of $\Sigma_R$ and ${\hat \alpha}_R$ - 
and hence the full solution - also obey 
the reality properties \eqref{improp} and 
\eqref{realprop}, establishing our claim. 

In the rest of this subsection we will 
solve the recursion relations for the self 
energy unknowns. We will treat the self 
energies at generic values of $n$ (i.e., those relevant to the propagator terms  \eqref{alphasol}) 
and exceptional values of $n$ (i.e. those 
relevant to the propagator terms \eqref{alphasol2})  separately.

\subsection{The Integral over $\omega$}

We now return to the main flow of this section. We 
will now evaluate the quantities  ${\hat \alpha}_{R, -, n}$ and
${\hat \alpha}_{R, I, n}$ by integrating the RHS of  \eqref{alphaform} and \eqref{alphasol2} over $\omega$.
We will then insert the resultant expressions into \eqref{sigmge}, turning that equation into a 
closed equation for the coefficients of $\Sigma$. 

The integrals over $\omega$ that we need to evaluate are 
\begin{equation}\label{firstint}
\begin{aligned}
	I_1=\int_{-\infty}^{+\infty}\frac{d\omega}{2 \pi} \alpha_{R,-,n}
= \int_{-\infty}^{+\infty}\frac{d\omega}{2 \pi}~\frac{i\sqrt{b(n+1)}}{(\omega-i \zeta_+^F(n))(\omega+i \zeta_-^F(n))}
\end{aligned}
\end{equation}
and for $n \neq 0$
\begin{equation}\label{secondint}
\begin{aligned}
	I_2=\int_{-\infty}^{+\infty}\frac{d\omega}{2 \pi} 
\alpha_{R,I,n}  = \int_{-\infty}^{+\infty}\frac{d\omega}{4 \pi} \left(  \frac{i \omega-m_F- \Sigma_{R,I,n-1}}{(\omega-i \zeta_+^F(n-1))(\omega+i \zeta_-^F(n-1))} - \frac{i \omega+m_F+ \Sigma_{R,I,n+1}}{(\omega-i \zeta_+^F(n))(\omega+i \zeta_-^F(n))}\right)
\end{aligned}
\end{equation}
and also the $n=0$ version of this integral 
\begin{equation}\label{secondinth}
\begin{aligned}
	\tilde{I}_2=\int_{-\infty}^{+\infty}\frac{d\omega}{2 \pi} 
\alpha_{R,I,0}  = \int_{-\infty}^{+\infty}
\frac{d \omega }{4 \pi} \left( \frac{i}{\omega - i \zeta_{ex}}-\frac{i \omega+m_F+ \Sigma_{R,R,I,1}}{(\omega-i \zeta_+^F(0))(\omega+i \zeta_-^F(0))}\right)
\end{aligned}
\end{equation}

The contour for the integral over $\omega$ has so far been taken to the  real axis in the complex $\omega$ plane. However there is 
a natural physical context - one that will be of interest to us in this 
paper - in which this contour is deformed. This happens when we turn
on a chemical potential $\mu$ for the same global symmetry for which 
we have turned on a magnetic field (recall this is the symmetry under 
which all components of $\psi$ carry charge unity while all components 
of ${\bar \psi}$ carry charge $-1$.) Turning on such a chemical
potential is equivalent to making the replacement 
$$\partial_t \psi  \rightarrow (\partial_t - \mu), ~~~
\partial_t {\bar \psi}  \rightarrow (\partial_t + \mu) {\bar \psi}
$$
which, in turn, is equivalent to the replacement 
$$\omega \rightarrow \omega + i\mu$$
in all the formulae above. If we change variables from $\omega$ to 
$\omega + i \mu$, all the formulae above retain their old form (i.e. 
the form they had in the absence of the chemical potential) with one 
single change; the contour of $\omega$ integration, which earlier had
been the real axis, changes to the line 
\begin{equation}\label{contour}
{\rm Im} (\omega)= i \mu
\end{equation}
 
It follows that in order to solve our problem at 
arbitrary values of the chemical potential, we need to  evaluate 
the integrals \eqref{firstint}, \eqref{secondint} and \eqref{secondinth}
contours of the form \eqref{contour} at arbitrary values of 
$Im (\omega)$. We now turn to this problem.

\subsubsection{The first integral}

Let us consider the integral \eqref{firstint}. If the contour of 
integration lies either above or below both poles in the integrand 
on the RHS of \eqref{firstint} then the integral clearly vanishes.\footnote{ In this case we can deform the contour to an arc at infinity. As the integrand decays like $\frac{1}{\omega^2}$, the 
contribution from this arc vanishes.}
In general the integral evaluates to
\begin{equation}
	\begin{aligned}
		I_1=&i\sqrt{b(n+1)}\frac{\Theta(\tilde{\zeta}_+^F(n))\Theta(\tilde{\zeta}_-^F(n))-\Theta(-\tilde{\zeta}_+^F(n))\Theta(-\tilde{\zeta}_-^F(n))}{ \tilde{\zeta}_+^F(n)+ \tilde{\zeta}_-^F(n)}  \\
	\end{aligned}
\end{equation}
For convenience we defined   
\begin{equation}
	\begin{aligned}
		\tilde{\zeta}_\pm^F=\zeta_\pm^F\mp \mu
	\end{aligned}
\end{equation}
It follows that if $\zeta_+^F(n)>-\zeta_-^F(n)$ (as we expect to be the case on physical grounds) and the contour passes between the two poles 
then 
\begin{equation}\label{iors}
\begin{split}
I_1=i\sqrt{b(n+1)}\frac{1}{\zeta_+^F(n) +\zeta_-^F(n)} 
\end{split}
\end{equation}

\subsubsection{The second integral}  

When the integration contour lies either above or below all 
four poles in the integrand of \eqref{secondint}, once again 
the integral $I_2$ vanishes. In general the integral evaluates to
\begin{equation}
	\begin{aligned}
		I_2=&\frac{1}{2}\left( \frac{1}{2}\frac{ \tilde{\zeta}_+^F(n)- \tilde{\zeta}_-^F(n)}{ |\tilde{\zeta}_+^F(n)|+ |\tilde{\zeta}_-^F(n)|}-\frac{1}{2}\frac{ \tilde{\zeta}_+^F(n-1)- \tilde{\zeta}_-^F(n-1)}{ |\tilde{\zeta}_+^F(n-1)|+ |\tilde{\zeta}_-^F(n-1)|} \right)\\
		&-\frac{1}{2}\left((K_{R,I,n+1}-\mu)\frac{\Theta(\tilde{\zeta}_+^F(n))\Theta(\tilde{\zeta}_-^F(n))-\Theta(-\tilde{\zeta}_+^F(n))\Theta(-\tilde{\zeta}_-^F(n))}{ \tilde{\zeta}_+^F(n)+ \tilde{\zeta}_-^F(n)} \right) \\
		&-\frac{1}{2}\left((K_{R,I,n-1}+\mu)\frac{\Theta(\tilde{\zeta}_+^F(n-1))\Theta(\tilde{\zeta}_-^F(n-1))-\Theta(-\tilde{\zeta}_+^F(n-1))\Theta(-\tilde{\zeta}_-^F(n-1))}{ \tilde{\zeta}_+^F(n-1)+ \tilde{\zeta}_-^F(n-1)} \right)
	\end{aligned}
\end{equation}

 In order to make 
the discussion of this subsection concrete, let us make the 
reasonable assumption  that 
\begin{equation}\label{reas}
 -\zeta_-^F(n) < -\zeta_-^F(n-1) <\zeta_+^F(n-1)<\zeta_+^F(n)
 \end{equation} 
Under the assumption \eqref{reas} the conclusion of this 
paragraph can be restated as follows: the integral on the 
RHS of \eqref{secondint} vanishes provided when, on the contour of integration ${\rm Im}(\omega)> \zeta_+(n)$ or 
${\rm Im}(\omega)< -\zeta_-(n)$.

More generally, always assuming \eqref{reas} we have 
\begin{equation}\label{valit} \begin{split}
I_2= \begin{cases}
& 0 ~~~ \mu> \zeta^F_+(n)\\
&A   ~~~\zeta_+^F(n-1)< \mu <\zeta_+^F(n)\\
& B  ~~~ -\zeta_-^F(n-1)< \mu < \zeta_+^F(n-1) \\
&C ~~~ -\zeta_-^F(n)< \mu < -\zeta_-^F(n-1)\\
& 0~~~ \mu< -\zeta^F_-(n)\\
\end{cases}
\end{split}
\end{equation}
where $A,B,C$ are given by 
\begin{equation}
	\begin{aligned}
		A=&\frac{1}{2}\left( \frac{1}{2}\frac{ \zeta_+^F(n)- \zeta_-^F(n)}{ \zeta_+^F(n)+ \zeta_-^F(n)}+\frac{1}{2} \right)-\frac{1}{2}\left(\frac{K_{R,I,n+1}}{ \zeta_+^F(n)+ \zeta_-^F(n)} 
		\right)\\
		B=&\frac{1}{2}\left( \frac{1}{2}\frac{ \zeta_+^F(n)- \zeta_-^F(n)}{ \zeta_+^F(n)+ \zeta_-^F(n)}-\frac{1}{2}\frac{ \zeta_+^F(n-1)- \zeta_-^F(n-1)}{ \zeta_+^F(n-1)+ \zeta_-^F(n-1)} \right)\\
		&-\frac{1}{2}\left(\frac{K_{R,I,n+1}}{ \zeta_+^F(n)+ \zeta_-^F(n)} \right) 
		-\frac{1}{2}\left(\frac{K_{R,I,n-1}}{ \zeta_+^F(n-1)+ \zeta_-^F(n-1)} \right)\\
		C=&\frac{1}{2}\left( \frac{1}{2}\frac{ \zeta_+^F(n)- \zeta_-^F(n)}{ \zeta_+^F(n)+ \zeta_-^F(n)}-\frac{1}{2} \right)-\frac{1}{2}\left(\frac{K_{R,I,n+1}}{ \zeta_+^F(n)+ \zeta_-^F(n)} 
		\right)&
	\end{aligned}
\end{equation}

\subsubsection{The third integral} 

In general the integral evaluates to
\begin{equation}
	\begin{aligned}
		\tilde{I}_2=&\frac{1}{2}\left( \frac{1}{2}\frac{ \tilde{\zeta}_+^F(0)- \tilde{\zeta}_-^F(0)}{ |\tilde{\zeta}_+^F(0)|+ |\tilde{\zeta}_-^F(0)|}-\frac{1}{2}\rm{sgn}(\zeta_{ex}-\mu) \right)\\
		&-\frac{1}{2}\left((K_{R,I,1}-\mu)\frac{\Theta(\tilde{\zeta}_+^F(0))\Theta(\tilde{\zeta}_-^F(0))-\Theta(-\tilde{\zeta}_+^F(0))\Theta(-\tilde{\zeta}_-^F(0))}{ \tilde{\zeta}_+^F(0)+ \tilde{\zeta}_-^F(0)} \right) \\
	\end{aligned}
\end{equation}
Working under the reasonable assumption 
\begin{equation}\label{reasex}
-\zeta_-^F(0) < \zeta_{ex} <\zeta_+^F(0)
\end{equation}
we find 
\begin{equation}\label{valitne} \begin{split}
\tilde{I}_2= \begin{cases}
& 0, ~~~ \mu> \zeta^F_+(0)\\
&\frac{1}{2}\left( \frac{1}{2}\frac{ \zeta_+^F(0)- \zeta_-^F(0)}{ \zeta_+^F(0)+ \zeta_-^F(0)}+\frac{1}{2}\right)
		-\frac{1}{2}\left(K_{R,I,1}\frac{1}{ \zeta_+^F(0)+ \zeta_-^F(0)} \right)   ~~~\zeta_{ex}< \mu <\zeta_+^F(0)\\
&\frac{1}{2}\left( \frac{1}{2}\frac{ \zeta_+^F(0)- \zeta_-^F(0)}{ \zeta_+^F(0)+ \zeta_-^F(0)}-\frac{1}{2}\right)
		-\frac{1}{2}\left(K_{R,I,1}\frac{1}{ \zeta_+^F(0)+ \zeta_-^F(0)} \right)  ~~~ -\zeta_-^F(0)< \mu < \zeta_{ex} \\
&0 ~~~ \mu < -\zeta_-^F(0)\\
\end{cases}
\end{split}
\end{equation}

\subsection{Explicit Gap Equations}

In this subsection we will input the results of the previous 
subsection into \eqref{sigmge} in order to turn this equation 
into a set of closed equations for the components of $\Sigma$. 
Our final results depend on the value of $\mu$. Through the rest 
of this paper we assume \eqref{reas}. It is useful to introduce some 
terminology. 

If $\zeta^F_+(m-1)<\mu <\zeta^F_+(m)$ we say that $\mu$ 
lies in the $(m-1)^{th}$ positive band. If $\zeta_{ex}<\mu< \zeta^F_+(0)$ we say that $\mu$ lies in the exceptional positive 
band. If $-\zeta^F_-(0) <\mu<\zeta_{ex}$ we say that that 
$\mu$ lies in the exceptional negative band. Finally if 
$-\zeta^F_-(m)<\mu <-\zeta^F_+(m-1)$ we say that $\mu$ lies in 
the $(m-1)^{th}$ negative band.

Some notation: we define
\begin{equation} \label{reschi}
\begin{split}
\chi_n=\frac{1}{\zeta_+^F(n)+\zeta_-^F(n)  }
\end{split}
\end{equation}
The quantity $\chi_n$ determines the sum of $\zeta_\pm^F(n)$ and 
so carries less information than the individual values of 
$\zeta_+^F(n)$ and $\zeta_-^F(n)$. However the additional information 
needed to reconstruct these two quantities individually is obtained 
quite easily directly from the gap equations. Using \eqref{factorizedprop} it follows 
that
\begin{equation} \label{difzeta}
\zeta^+(n)- \zeta^-(n) = \Sigma_{R,I,n+1} -\Sigma_{R,I,n}.
\end{equation} 
Now the RHS of \eqref{difzeta} can be evaluated using \eqref{gapeqfinal} (and the results for the integral over $\omega$ 
presented in subsection \ref{intfe}. The specific results for 
$\zeta^\pm(n)$ in terms of $\chi_n$ depend on details and will be 
presented below.

\subsubsection{Positive Band M}
To start with let us assume that $\mu$ lies in the $M^{th}$ 
positive band with $M\geq 0$ i.e. that $\zeta_+(M) < \mu < \zeta_+(M+1)$. 
In this case the gap equations become 
\begin{equation}\label{gapeqmupm}
	\begin{aligned}
		& \Sigma_{R,I,M+1}=\Sigma_{R,I,M}= \dots =\Sigma_{R,I,1}=\Sigma_{R,I,0} =-2 \lambda_F \sum_{k=M+1}^{\infty}b \chi_k\\
		& \Sigma_{R,I, n'+1}= -2 \lambda_F \sum_{n=n'+1}^{\infty} b \chi_n, ~~~(n' \geq M+1)\\
		& \sqrt{M+1}\Sigma_{R,+,M+1}=\sqrt{M}\Sigma_{R,+,M}=\dots=\Sigma_{R,+,1}=0\\
	&	\sqrt{M+2}\Sigma_{R,+,M+2}= \sqrt{b}(i\lambda_F^2 (\sqrt{b} \chi_{M+1})^2-i \lambda_F K_{R,I,M+2}\chi_{M+1}+\frac{1}{2}i \lambda_F)\\
	 &  \sqrt{n'+1}\Sigma_{R,+,n'+1}=\sqrt{M+2}\Sigma_{R,+,M+2}- \sum_{n=M+2}^{n'} \sqrt{b}i\lambda_F (K_{R,I,n-1}\chi_{n-1}+K_{R,I,n+1}\chi_{n})\\
	 &~~~~~~~~~~~~~~~~~~~~~~~~~ +\sum_{n=M+2}^{n'} i \sqrt{b}\lambda_F^2( (\sqrt{b}\chi_n)^2-(\sqrt{b}\chi_{n-1})^2)
	 ~~~(n'\geq M+2) \\
	\end{aligned}
\end{equation}

\subsubsection{Positive Exceptional Band}
In the positive exceptional band (i.e. for $\zeta_{ex}<\mu< \zeta_+(0)$) on the other hand
we have 
\begin{equation}\label{gapeqmupe}
\begin{aligned}
& \Sigma_{R,I, n'}= -2 \lambda_F \sum_{n=n'}^{\infty} b \chi_n, ~~~(n \geq 0, ~{\rm i.e} ~{\rm all} ~n)\\
&\Sigma_{R,+,1}= \sqrt{b}(i\lambda_F^2 (\sqrt{b} \chi_{0})^2-i \lambda_F K_{R,I,1}\chi_{0}+\frac{1}{2}i \lambda_F)\\
&  \sqrt{n'+1}\Sigma_{R,+,n'+1}=\Sigma_{R,+,1}- \sum_{n=1}^{n'} \sqrt{b}i\lambda_F (K_{R,I,n-1}\chi_{n-1}+K_{R,I,n+1}\chi_{n})\\
&+\sum_{n=1}^{n'} i \sqrt{b}\lambda_F^2( (\sqrt{b}\chi_n)^2-(\sqrt{b}\chi_{n-1})^2)
~~~(n'\geq 1) \\
\end{aligned}
\end{equation}

\subsubsection{Negative Exceptional Band}
In the negative exceptional band (i.e. 
$-\zeta_-(0) < \mu <\zeta_{ex}$) the equations are
\begin{equation}\label{gapeqmune}
\begin{aligned}
& \Sigma_{R,I, n}= -2 \lambda_F \sum_{n}^{\infty} b \chi_k, ~~~(n \geq 0, ~{\rm i.e} ~{\rm all} ~n)\\
&\Sigma_{R,+,1}= \sqrt{b}(i\lambda_F^2 (\sqrt{b} \chi_{0})^2-i \lambda_F K_{R,I,1}\chi_{0}-\frac{1}{2}i \lambda_F)\\
&  \sqrt{n'+1}\Sigma_{R,+,n'+1}=\Sigma_{R,+,1}- \sum_{n=1}^{n'} \sqrt{b}i\lambda_F (K_{R,I,n-1}\chi_{n-1}+K_{R,I,n+1}\chi_{n})\\
&+\sum_{n=1}^{n'} i \sqrt{b}\lambda_F^2( (\sqrt{b}\chi_n)^2-(\sqrt{b}\chi_{n-1})^2)
~~~(n'\geq 1) \\
\end{aligned}
\end{equation}

\subsubsection{Negative band M }
In the $M^{th}$ negative band for $M \geq 0$ 
(i.e. $-\zeta_-(M+1)< \mu < -\zeta_-(M)$) we have 
\begin{equation}\label{gapeqmunm}
\begin{aligned}
& \Sigma_{R,I,M+1}=\Sigma_{R,I,M}= \dots =\Sigma_{R,I,1}=\Sigma_{R,I,0} =-2 \lambda_F \sum_{k=M+1}^{\infty}b \chi_k\\
& \Sigma_{R,I, n'+1}= -2 \lambda_F \sum_{n=n'+1}^{\infty} b \chi_n, ~~~(n' \geq M+1)\\
& \sqrt{M+1}\Sigma_{R,+,M+1}=\sqrt{M}\Sigma_{R,+,M}=\dots=\Sigma_{R,+,1}=0\\
&	\sqrt{M+2}\Sigma_{R,+,M+2}= \sqrt{b}(i\lambda_F^2 (\sqrt{b} \chi_{M+1})^2-i \lambda_F K_{R,I,M+2}\chi_{M+1}-\frac{1}{2}i \lambda_F)\\
&  \sqrt{n'+1}\Sigma_{R,+,n'+1}=\sqrt{M+2}\Sigma_{R,+,M+2}- \sum_{n=M+2}^{n'} \sqrt{b}i\lambda_F (K_{R,I,n-1}\chi_{n-1}+K_{R,I,n+1}\chi_{n})\\
&+\sum_{n=M+2}^{n'} i \sqrt{b}\lambda_F^2( (\sqrt{b}\chi_n)^2-(\sqrt{b}\chi_{n-1})^2)
~~~(n'\geq M+2) \\
\end{aligned}
\end{equation}

\subsection{Solution of the gap equations}

\subsubsection{$\mu$ in the M positive/negative band}

To start with let us focus on the gap equations when $\mu$ lies in the 
$M^{th}$ positive/negative band with $M\geq0$. 

{\bf Solution for $\zeta_\pm^F(n)$ for $n \geq M+1$ }

When $\mu$ lies in either the positive or negative $M^{th}$ 
band for $n'\geq M+2$ we have 
\begin{equation}\label{gapeqmu}
	\begin{aligned}
		& \Sigma_{R,I,n'+1}=-2 \lambda_F \sum_{n=n'+1}^{\infty}b \chi_n~~~\\
		 &  \sqrt{n'+1}\Sigma_{R,+,n'+1}=\sqrt{M+2}\Sigma_{R,+,M+2}- \sum_{n=M+2}^{n'} \sqrt{b}i\lambda_F (K_{R,I,n-1}\chi_{n-1}+K_{R,I,n+1}\chi_{n})\\
  & ~~~~~~~~~~~~~~~~~~~~~~~~~~~~~~~~~~~~~~~~~~~~~~~~ +\sum_{n=M+2}^{n'} i \sqrt{b}\lambda_F^2( (\sqrt{b}\chi_n)^2-(\sqrt{b}\chi_{n-1})^2) ~~~\\
	\end{aligned}
\end{equation}

For $n\geq M+2$ it follows that
\begin{equation}
\begin{split}\label{gapeqsimplified}
& \frac{1}{\sqrt{b}}\left(\Sigma_{R,I,n+1}-\Sigma_{R,I,n}\right)=2  \lambda_F  \ \sqrt{b}\chi_n\\
& \frac{1}{\sqrt{b}} \left( \sqrt{n+1}\Sigma_{R,+,n+1}-\sqrt{n}\Sigma_{R,+,n} \right)=- i\lambda_F (K_{R,I,n-1}\chi_{n-1}+K_{R,I,n+1}\chi_{n})\\
&~~~~~~~~~~~~~~~~~~~~~~~~~~~~~~~~~~~~~~~~~~~~~~~~~~~~~~~~~~~~~+i \lambda_F^2( (\sqrt{b}\chi_n)^2-(\sqrt{b}\chi_{n-1})^2)\\
\end{split}
\end{equation}

Our task is to solve \eqref{gapeqsimplified} 
for the variables $\Sigma_{R,I,n}$ and 
$\Sigma_{R,+,n+1}$. The RHS
of \eqref{gapeqsimplified} is a rather complicated function of these variables 
(the complication arrises because we are required to substitute \eqref{factorizedprop} and \eqref{reschi} into this RHS in order to 
express it in terms of the variables of the problem). Quite remarkably, however, the 
difference equations \eqref{gapeqsimplified} in fact can be solved exactly. The key to finding the solution of these equations is the observation that \eqref{gapeqsimplified} imply that the quantities 
\begin{equation}\label{definition}
\begin{split}
I_n= \lambda_F^2  (\sqrt{2b}\chi_n)^2+\frac{1}{(\sqrt{2b}\chi_n)^2}
\end{split}
\end{equation}
obey the remarkably simple 
recursion relation  (see Appendix \ref{recrel} for a derivation)
\begin{equation}\label{recursion}
\begin{split}
I_{n+1}-I_{n}=4 ~~~(n \geq M+2)
\end{split}
\end{equation}
It follows that 
\begin{equation}\label{sol}
\begin{split}
I_n=I_{M+2}+4(n-(M-2)) ~~~(n \geq M+2)
\end{split}
\end{equation}
where $I_{M+2}$ is and as yet arbitrary constant.

In order to obtain \eqref{recursion} and \eqref{sol} we used 
the first of \eqref{gapeq} for $n' \geq M=1$ and the second 
of \eqref{gapeqsimplified} for $n' \geq M+2$. It turns out we can actually do better. Using the fourth lines in \eqref{gapeqmupm}
and \eqref{gapeqmunm} and also the expressions for 
$\zeta_{R, I, M+1}$ listed in the first lines of those equations, and 
proceeding along the lines of Appendix \ref{recrel}, it is possible to show  that
\begin{equation}\label{solIe}
\begin{split}
I_n=I_{M+1}+4(n-(M+1)) ~~~(n \geq M+1)
\end{split}
\end{equation}
From \eqref{definition} we obtain\footnote{More precisely, the quadratic equation that determines
$\chi_n$ in terms of $I_n$ has two solutions
\begin{equation}\label{bothsol}
\begin{split}
\chi_n=\pm \frac{1}{\sqrt{2}|\lambda_F|b}\left( (c_F^2+2b(n+1)) \pm ((c_F^2+2b(n+1))^2-(\lambda_Fb)^2)^{1/2} \right)^{1/2}
\end{split}
\end{equation}
We have fixed the sign ambiguity in \eqref{bothsol} by matching
with perturbation theory around $\lambda_F \to 0$. It is 
possible that the second solution in \eqref{bothsol} represents
a legal solution of the theory, but one that is intrinsically 
non perturbative in nature. We do not consider this possibility 
in the current paper, but hope to return to it and other issues 
in future work.}
\begin{equation}
\begin{split}\label{chicolution}
\chi_n= \frac{1}{\sqrt{2}|\lambda_F|b}\left( (c_F^2+2b(n+1)) - ((c_F^2+2b(n+1))^2-(\lambda_Fb)^2)^{1/2} \right)^{1/2} ~~~~~~ (n \geq M+1)
\end{split}
\end{equation}
where we have traded $I_{M+1}$ for another arbitrary constant $c_F$.

The energy eigenvalues $\zeta^F_\pm(n)$ can be obtained from $\chi_n$ using \eqref{difzeta}. Using the gap equations (third  of \eqref{gapeqmupm} and
\eqref{gapeqmunm}, the 
RHS of \eqref{difzeta} evaluates to $2 \lambda_F b\chi_n$ for 
$n \geq  M+1$
i.e. 
\begin{equation} \label{difzetaf}
\zeta^+(n)- \zeta^-(n) = 2 \lambda_F b\chi_n ~~~(n \geq M+1)
\end{equation} 
and so we find 
\begin{equation}
\begin{split} \label{betsol}
& \zeta_\pm^F(n)=\frac{1}{2}(\pm 2 \lambda_F b \chi_n+\frac{1}{\chi_n}) \\
\implies & \zeta_\pm^F(n)=\left[ c_F^2+2b(n+1)\pm \lambda_F b \right]^{1/2}~~~(n \geq M+1)
\end{split}
\end{equation}

We pause to provided a physical interpretation of 
\eqref{betsol}. 
Recall that in the study of the free Dirac equation we employed two distinct (but related) labelling schemes for the energy levels of our system. The physically more transparent labelling 
was in terms of the variable $\nu$ (in the non relativistic limit this label mapped directly to the usual `which Landau Level' label). The second, algebraically more convenient 
scheme - the one also adopted all through this section so far - was to label energy levels by the variable $n$. $n$  is an 
algebraically convenient auxiliary label related to $\nu$ 
via \eqref{nundic} which we repeat here for convenience
\begin{equation}\label{nundicr}
\begin{split}
\nu=
\begin{cases}
&=n+ \frac{1}{2} +\frac{{\rm sgn}(m_F)}{2}~~~{\rm +ve} ~~{\rm energy}\\
&=n+ \frac{1}{2} -\frac{{\rm sgn}(m_F)}{2}
~~~~~~{\rm -ve} ~~{\rm energy}\\
\end{cases}
\end{split}
\end{equation}
Rewriting the last line of \eqref{betsol} in terms of 
the more physical variable $\nu$ we find (in the appropriate 
range of $\nu$, i.e. the range that follows from $n \geq M+2$ 
using \eqref{nundicr})
\begin{equation}\label{rmp}
\zeta_\pm^F(\nu)=\left[ c_F^2+2b(\nu+\frac{1}{2}) \mp 
2 b s(\lambda_F)  \right]^{1/2}
\end{equation}
where 
\begin{equation}\label{slam}
s(\lambda_F)= \frac{\left( {\rm sgn} (m_F) -\lambda_F \right)}{2}
\end{equation}
is the effective spin of excitations in the regular fermion theory at t' Hooft coupling $\lambda$. 
It is quite remarkable that \eqref{rmp} matches exactly with \eqref{epespms} - our naive guess for the spectrum of a theory 
of `free particles of spin $s(\lambda)$', provided we identify
$c_F^2$ as the squared effective mass our excitation.

Another aside: note also that it follows immediately from \eqref{betsol} that 
\begin{equation}\label{zeres}
\begin{split}
& \zeta_+^F(n)^2-\zeta_-^F(n-1)^2=2b(1+\lambda_F)~~~(n \geq M+2)\\
& \zeta_-^F(n)^2-\zeta_+^F(n-1)^2=2b(1-\lambda_F) ~~~(n \geq M+2)
\end{split}
\end{equation}
Note that the RHS of the first of \eqref{zeres} vanishes 
at $\lambda_F=-1$ while the RHS of the second of \eqref{zeres}
vanishes at $\lambda_F=+1$. Comparing with \eqref{nundic}, it
follows in particular that when $|\lambda_F|=1$ and $\lambda_F m_F>0$ then positive and negative energy levels with the same 
value of $\nu$ are equal and opposite (i.e. the spectrum  is symmetric in positive and negative energies when rewritten in terms of $\nu$). We will return to this point below.

{\bf Solution for $\zeta_\pm^F(n)$ for $n \leq M$ }

The gap equations are much easier to solve for $n \leq M$.
For these values of $n$ it follows from \eqref{gapeqmunm} and  \eqref{gapeqmupm} that $\Sigma_{R, I, n}$ and 
$\Sigma_{R, +, n}$ are both independent of $n$ in this range.  
It follows immediately that 
\begin{equation}\label{zetacon}
\begin{aligned}
\zeta_\pm^F(n)^2 & =(m_F+\Sigma_{R,I,n})^2-2i\sqrt{b(n+1)}(i \sqrt{b(n+1)}+\Sigma_{R,+,n+1})\\
& =(m_F+\Sigma_{R,I,0})^2-2i\sqrt{b} \Sigma_{R,+,1}+2b(n+1)\\
& =(m_F+\Sigma_{R,I,0})^2+2b(n+1)
\end{aligned}
\end{equation}
Note that in this range (and unlike for the case $n \geq M+1$) $\zeta^F_+(n)= \zeta^F_-(n)$.
We remind the reader that 
\begin{equation}\label{kid}
m_F+\Sigma_{R,I,0}=K_{R, I, 0}=\zeta_{ex}
\end{equation}
It follows that \eqref{zetacon} can be rewritten as 
\begin{equation}\label{filledeff}
\begin{aligned}
\zeta_\pm^F(n)^2 & =\zeta_{ex}^2+2b(n+1) ~~~~~~ \textit{For $n \leq M$} 
\end{aligned}
\end{equation}
\eqref{filledeff} is precisely the spectrum of a free fermion 
of squared mass $$c_{F*}^2=\zeta_{ex}^2$$ in magnetic field. Translating to 
the $\nu$ variable we once again find the formula \eqref{rmp} 
but this time with 
\begin{equation}\label{slamuf}
s(\lambda)= \frac{ {\rm sgn} (m_F)}{2}
\end{equation}
In other words excitations in the filled Landau levels behave 
like free particles that carry the `unrenormalized' spin listed in \eqref{slamuf} rather than the `renormalized' spin listed in 
\eqref{slam}. We do not yet have a physical explanation for this 
striking result. We hope to return to this point in future work.  

Note that it follows immediately from \eqref{filledeff} together
with the definition of $\chi_n$  that
\begin{equation} \label{chismall}
\begin{aligned}
\chi_n=\frac{1}{2\sqrt{(\zeta_{ex}^2+2b(n+1))}}~~~~~~ \textit{For $n \leq M$}
\end{aligned}
\end{equation} 

{\bf Sewing The solutions together}

Above we have found solutions for $\zeta_\pm^F$ separately 
for $n \geq M+1$ and $n \leq M$. The solution in the two 
different ranges have been presented in terms of two independent
integration constants $c_F$ and $c_{F*}$. Of course these 
integration constants are not really independent of each other; 
they are, in fact, easily related to each other as we now 
explain. Using the third lines in \eqref{gapeqmupm}
 and \eqref{gapeqmunm} and also the expressions for 
 $\zeta_{R, I, M}$ and  $\zeta_{R, I, M}$ listed in the first lines of those equations, and proceeding as in Appendix 
 \ref{recrel} it is possible to show that\footnote{The difference between positive and negative bands is in equation involving $\Sigma_{R,+,M+1},\Sigma_{R,+,M+2}$.} 
 \begin{equation} \label{injunc}
 \begin{aligned}
 \frac{1}{\chi_{M+1}^2}-\frac{1}{\chi_{M}^2}	=-(2\lambda_Fb)^2(\chi_{M+1}^2)+8b \pm 4 \lambda_F b
 \end{aligned}
 \end{equation}
 Here $'+'$ is for positive $M$ band and $'-'$ is for positive $M$ band. 
 Substituting in the expressions for $\chi_{M+1}$ from  \eqref{chicolution} and the analogous expression for 
 $\chi_M$ from \eqref{chismall}, we find immediately from 
 \eqref{injunc} that 
 \begin{equation}\label{junctioncondition}
 \begin{aligned}
 c_F^2= \zeta_{ex}^2 \pm \lambda_F b 
 \end{aligned}
 \end{equation}
 where we use the sign $\pm$ depending on whether $\mu$ lies in
 the positive/negative $M^{th}$ band.

 {\bf The gap equation for $\zeta_{ex}$}
 
 We have now evaluated every quantity of interest in terms of 
 a single unknown quantity $\zeta_{ex}$. $\zeta_{ex}$ may, in turn, 
 be evaluated using the last `junction' condition - which is equivalent 
 to using the equation 
 \begin{equation} \label{gapef}
 \begin{aligned}
 \frac{1}{\sqrt{b}} \Sigma_{R,I,0} & =-2 \lambda_F \sum_{k=M+1}^{\infty}\sqrt{b} \chi_k
 \end{aligned}
 \end{equation}
 The RHS of \eqref{gapef} is simplified using 
 \eqref{betsol} 
 $$\chi_k= \frac{(\zeta_+^F(k)-\zeta_-^F(k))}{\left(
 	(\zeta_+^F(k))^2-(\zeta_-^F(k))^2 \right)}
 =\frac{\zeta_+^F(k)-\zeta_-^F(k)}{2 \lambda_F b}$$
 The LHS of \eqref{gapeq} is simplified using 
 $$\Sigma_{R, I, 0}= K_{R,I, 0}- m_F= \zeta_{ex}-m_F$$
 We find 
 \begin{equation}\label{polemassmd}
 \begin{aligned}
 & m_F - \zeta_{ex}= \sum_{k=M+1}^{\infty}((\zeta_{ex}^2+2b(k+1) +\lambda_F b (1 \pm 1)  )^{1/2}-(c_F^2+2b(k+1) +\lambda_F b (-1 \pm 1)  )^{1/2})	\end{aligned}
 \end{equation}
 (where the sign $\pm$ applies to $\mu$ in the positive/negative $M^{th}$ 
 band.)
 More explicitly, when $\mu$ is in the $M^th$ positive band  
 \begin{equation}\label{polemassmdp}
 \begin{aligned}
 & m_F - \zeta_{ex}= \sum_{k=M+1}^{\infty}(\zeta_{ex}^2+2b(k+1) +2 \lambda_F b  )^{1/2}-(c_F^2+2b(k+1) )^{1/2} 	\end{aligned}
 \end{equation}
 while in the $M^{th}$ negative band 
 \begin{equation}\label{polemassmdn}
 \begin{aligned}
 & m_F - \zeta_{ex}= \sum_{k=M+1}^{\infty} (\zeta_{ex}^2+2b(k+1)   )^{1/2}-(c_F^2+2b(k+1) - 2 \lambda_F b )^{1/2} 	\end{aligned}
 \end{equation}

\subsubsection{$\mu$ in the positive/negative exceptional band}

It turns out that the final results for the case that $\mu$ lies in the the positive/negative exceptional band can be obtained from the results of 
the previous subsection by setting $M=-1$ in all final formulae. The 
derivation of these results closely parallel those presented in the 
previous subsection, and we will be brief in our presentation. 

The gap equations in this case are listed in \eqref{gapeqmupe} and 
\eqref{gapeqmune}. Using these equations we find 
\begin{equation}\label{recursionIex}
\begin{split}
I_{n+1}-I_{n}=4 ~~~(n \geq 0)
\end{split}
\end{equation}
It follows that 
\begin{equation}
\begin{split}\label{chicolutionex}
\chi_n= \frac{1}{\sqrt{2}|\lambda_F|b}\left( (c_F^2+2b(n+1)) - ((c_F^2+2b(n+1))^2-(\lambda_Fb)^2)^{1/2} \right)^{1/2} ~~~~~~ (n \geq 0)
\end{split}
\end{equation}
and
\begin{equation}
\begin{split} \label{betsolex}
& \zeta_\pm^F(n)=\frac{1}{2}(\pm 2 \lambda_F b \chi_n+\frac{1}{\chi_n}) \\
\implies & \zeta_\pm^F(n)=\left[ c_F^2+2b(n+1)\pm \lambda_F b \right]^{1/2}~~~(n \geq 0)
\end{split}
\end{equation}
Similar considerations as before gives
\begin{equation}\label{junctioncondition}
 \begin{aligned}
 c_F^2= \zeta_{ex}^2 \pm \lambda_F b 
 \end{aligned}
 \end{equation}
Where $'\pm'$ is taken for positive/negative bands. 

Once again our results can be rewritten in terms of the  
more physical variable $\nu$ labelling we find (in the appropriate 
range of $\nu$, i.e. the range that follows from $n \geq 0$ 
using \eqref{nundicr}) that \eqref{rmp} and \eqref{slam} apply. 
Once again we find agreement with  \eqref{epespms} - our naive guess for the spectrum of a theory of `free particles of spin $s(\lambda)$' once 
we identify $c_F^2$ as the squared effective mass our excitation. 

In this case there is no separate range (no analogue of the range $n \leq M$ )
Once again we find a gap equation for $\zeta_{ex}$ using the gap equation
\begin{equation} \label{gapefex}
\begin{aligned}
\frac{1}{\sqrt{b}} \Sigma_{R,I,0} & =-2 \lambda_F \sum_{k=0}^{\infty}\sqrt{b} \chi_k
\end{aligned}
\end{equation}
Simplifying the RHS and LHS of \eqref{gapefex} as before we obtain
\begin{equation}\label{polemassmdex}
\begin{aligned}
& m_F - \zeta_{ex}= \sum_{k=0}^{\infty}((\zeta_{ex}^2+2b(k+1) +\lambda_F b (1 \pm 1)  )^{1/2}-(c_F^2+2b(k+1) +\lambda_F b (-1 \pm 1)  )^{1/2})	\end{aligned}
\end{equation}
(where the sign $\pm$ applies to $\mu$ in the positive/negative exceptional
band.)
More explicitly, when $\mu$ is in the $M^th$ positive exceptional band
\begin{equation}\label{polemassmdpex}
\begin{aligned}
& m_F - \zeta_{ex}= \sum_{k=0}^{\infty}(\zeta_{ex}^2+2b(k+1) +2 \lambda_F b  )^{1/2}-(c_F^2+2b(k+1) )^{1/2} 	\end{aligned}
\end{equation}
while in the $M^{th}$ negative band 
\begin{equation}\label{polemassmdnex}
\begin{aligned}
& m_F - \zeta_{ex}= \sum_{k=0}^{\infty} (\zeta_{ex}^2+2b(k+1)   )^{1/2}-(c_F^2+2b(k+1) - 2 \lambda_F b )^{1/2} 	\end{aligned}
\end{equation}

\subsection{The $b \to 0$ limit and the regulated gap equation}

The gap equations \eqref{polemassmdp}, \eqref{polemassmdn}, \eqref{polemassmdex} and \eqref{polemassmdnex} - which we need to solve 
to determine $\zeta_{ex}$ and hence $c_F$ are not well defined as they 
stand, as the summations on the RHS of these equations are divergent. 
These gap equations need to be regulated and renormalized. In order 
to understand how this should be done in a physically sensible manner, it 
is useful to first consider the limit $b \to 0$. In this limit the 
gap equations presented in this paper reduce to the previously well 
studied gap equations for a massive fermion in the absence of a magnetic 
field. 

Let us consider, for instance, the gap equation \eqref{polemassmdp} which 
we reproduce here for convenience. 
 \begin{equation}\label{polemassmdpre}
\begin{aligned}
& m_F - \zeta_{ex}= \sum_{k=M+1}^{\infty}(\zeta_{ex}^2+2b(k+1) +2 \lambda_F b  )^{1/2}-(\zeta_{ex}^2+2b(k+1) )^{1/2} 	\end{aligned}
\end{equation}
In the limit $b \to 0$ we get a significant contribution to the summation 
on the RHS only for values of $k$ that scale like $k \sim \frac{1}{b}$. 
It is thus useful to work with the variable  $w=bk$ and also to work  
with a value of $\mu$ that is held fixed as $b$ is taken to infinity (so 
that we work in the  $M^{th}$ positive band with $\mu= \sqrt{ 2 M b + \zeta_{ex}^2}$. In other words we scale $M$ as 
$$ M = \frac{\mu^2-\zeta_{ex}^2}{2 b}$$
with $\mu$ and $\zeta_{ex}$ held fixed.  
\eqref{polemassmdpre} simplifies to  
 \begin{equation}\label{pmint}
\begin{aligned}
& m_F - \zeta_{ex}=  \lambda_F 
\int_ \frac{\mu^2-\zeta_{ex}^2}{2 }^{\infty} \frac{dw}{ \sqrt{\zeta_{ex}^2+2w}}
\end{aligned}
\end{equation}
If we make the change of variables $2w = p_s^2$ then \eqref{pmint}
turns into 
 \begin{equation}\label{pmintmod}
\begin{aligned}
m_F - \zeta_{ex}= & \lambda_F 
\int_{\sqrt{\mu^2-\zeta_{ex}^2}}^{\infty} \frac{p_s dp_s}{ \sqrt{\zeta_{ex}^2+p_s^2}}=   \lambda_F 
\int_{|{\vec p|}=\sqrt{\mu^2-\zeta_{ex}^2}}^{\infty} \frac{d^2 {\vec p}}{ (2 \pi) \sqrt{\zeta_{ex}^2+|{\vec p}|^2}}\\
&= \frac{\lambda_F}{2(2 \pi)^2} 
\int  \frac{d^3 p}{\zeta_{ex}^2+ p^2 }\\
\end{aligned}
\end{equation}
where the integral in the final line of \eqref{pmintmod} is taken over the 
contour $p_3=\omega= i \mu$. \footnote{The integral over $\omega=p_3$ on 
this contour produces the integral in the first line of \eqref{pmintmod}. 
Note that the integral over $\omega$ vanishes when $|\mu|^2 > p_s^2+\zeta_{ex}^2$. }
The important point here is that the integrals on the RHS of \eqref{pmintmod} 
are divergent; this is the continuum analogue of the divergence of the sum 
in \eqref{polemassmdpre}. As explained in \cite{Giombi:2011kc}, this divergence may be regulated and renormalized away by continuing  the integral over $p_3= \omega$ to an integral in $1-\epsilon$ dimensions. This regulation produces the 
integral
$$\int_{\sqrt{\mu^2-\zeta_{ex}^2}}^{\infty} \frac{p_s dp_s}{ \sqrt{\zeta_{ex}^2+p_s^2}^{1+\epsilon}}$$
As explained in \cite{Giombi:2011kc} this dimensionally regulated integral 
is easily evaluated; in-fact it evaluates to $|\mu|$ 
\footnote{In this subsection we have assumed that $|\mu|> |\zeta_{ex}|$; more 
	generally the integral evaluates to ${\rm max}(|\mu|, |\zeta_{ex}|)$}

Dimensionally regulating the continuum integral \eqref{pmintmod} is equivalent
to adding an appropriate counterterm to the field theory action that cancels 
the divergence in \eqref{pmint}. In this paper we wish to study the same quantum field theory - the theory defined with the counterterm that affects the 
dimensional regulation of \eqref{pmint} - in the presence of a background 
magnetic field. We can accomplish this by manipulating \eqref{polemassmdpre} 
as follows
\begin{equation}\label{ppaf}
\begin{aligned}
&m_F - \zeta_{ex}= \sum_{k=M+1}^{\infty}(\zeta_{ex}^2+2b(k+1) +2 \lambda_F b  )^{1/2}-(\zeta_{ex}^2+2b(k+1) )^{1/2} \\
&=\lim_{P \to \infty} \left[\sum_{k=M+1}^{P} \left( (\zeta_{ex}^2+2b(k+1) +2 \lambda_F b  )^{1/2}-(\zeta_{ex}^2+2b(k+1) )^{1/2} \right)  -  \lambda_F 
\int_ \frac{\mu^2-\zeta_{ex}^2}{2 }^{(P+1)b} \frac{dw}{ \sqrt{\zeta_{ex}^2+2w}}
\right] \\
&+ \lambda_F \int_{\sqrt{\mu^2-\zeta_{ex}^2}}^{\infty} \frac{p_s dp_s}{ \sqrt{\zeta_{ex}^2+p_s^2}^{1+\epsilon}}\\
&=\lim_{P \to \infty} \left[\sum_{k=M+1}^{P} \left( (\zeta_{ex}^2+2b(k+1) +2 \lambda_F b  )^{1/2}-(\zeta_{ex}^2+2b(k+1) )^{1/2} \right)  -  \lambda_F 
\int_\frac{a^2}{2}^{(P+1)b} \frac{dw}{ \sqrt{\zeta_{ex}^2+2w}}
\right] \\
&+ \lambda_F \int_{a}^{\infty} \frac{p_s dp_s}{ \sqrt{\zeta_{ex}^2+p_s^2}^{1+\epsilon}}\
	\end{aligned}
\end{equation}
Several comments are in order. First note that the limits $P \to \infty$ in the second and fourth lines of 
\eqref{ppaf} are well defined (i.e. finite); the divergence in the sums 
in these lines is cancelled by analogous divergence in the corresponding 
integrals. Next note that the integrals in the third and fifth lines of
\eqref{ppaf} are to be evaluated in a dimensionally regulated manner, i.e. are 
to be evaluated at a value of $\epsilon$ that is large enough to ensure the integral converges ($\epsilon >1$); the result of the integral is then continued to $\epsilon=0$. It follows that the integrals on the third and fifth lines of 
\eqref{ppaf} are also finite. \footnote{These integrals  can effectively be 
	evaluated as follows. Working directly at $\epsilon=0$ we evaluate the integral as a difference between the indefinite 
	integral evaluated at infinity and the same indefinite integral evaluated 
	at the lower limit. The procedure of dimensional regularization effectively instructs us to discard the (divergent) term at$p_s=\infty$.} 
Finally we have used the fact that the expression in the fourth and fifth lines 
of \eqref{ppaf} does not depend on the variable $a$ (this can be verified by 
differentiating the expression w.r.t. $a$). This explains why the expression 
on the fourth and fifth lines of \eqref{ppaf} equals the expression on the 2nd and 
3rd lines of the same equation. 

In this section we have, in particular, demonstrated that the 
gap equations presented in this paper reduce to the well known 
zero background field gap equations in the limit $b \to 0$, as 
expected in general grounds.

\subsection{Summary of final results for the 
fermion propagator} \label{sumfinp}

The poles in the Euclidean propagator occur at the values 
\begin{equation}\label{eucpoles}
\omega=-i\left(\zeta^+(n)-\mu \right)~ {\rm and}~
\omega= -i\left(-\zeta^n(n)-\mu \right)
\end{equation} 
 - corresponding to Lorentzian quasi particle states at energies 
 $\delta E= E-\mu$ at 
\begin{equation}\label{lorpoles}
\delta E= \left(\zeta^+(n)-\mu \right)~ {\rm and}~
\delta E= -\zeta^-(n)-\mu 
\end{equation}
In this subsection we summarize our results for $\zeta^F_\pm(n)$ for different cases.

\subsubsection{$\mu$ in the Positive/Negative M band}
When $\mu$ is in the positive M band, 
\begin{equation}
\begin{aligned}
 & \zeta_\pm^F(n)=\left[ \zeta_{ex}^2+2b(n+1)+ \lambda_F b\pm \lambda_F b \right]^{1/2}~~~(n \geq M+1)\\
 & \zeta_\pm^F(n)  =\left[\zeta_{ex}^2+2b(n+1)\right]^{1/2} ~~~~~~ ~~~~~~~~~~~~~~~~~(n \leq M)
\end{aligned}
\end{equation}
When $\mu$ is in the negative M band, 
\begin{equation}
\begin{aligned}
 & \zeta_\pm^F(n)=\left[ \zeta_{ex}^2+2b(n+1)- \lambda_F b\pm \lambda_F b \right]^{1/2}~~~(n \geq M+1)\\
 & \zeta_\pm^F(n)  =\left[\zeta_{ex}^2+2b(n+1)\right]^{1/2} ~~~~~~ ~~~~~~~~~~~~~~~~~(n \leq M)
\end{aligned}
\end{equation}
 $\zeta_{ex}$ is the solution to the equation 
\begin{equation}\label{ppafin}
\begin{aligned}
&m_F = \lambda_F \int_{a}^{\infty} \frac{p_s dp_s}{ \sqrt{\zeta_{ex}^2+p_s^2}^{1+\epsilon}} + \zeta_{ex} \\
&+\lim_{P \to \infty} \left[\sum_{k=M+1}^{P} \left( (\zeta_{ex}^2+2b(k+1) +2 \lambda_F b  )^{1/2}-(\zeta_{ex}^2+2b(k+1) )^{1/2} \right)  -  \lambda_F 
\int_\frac{a^2}{2}^{(P+1)b} \frac{dw}{ \sqrt{\zeta_{ex}^2+2w}}
\right] \\
\end{aligned}
\end{equation}
for $\mu$ in the positive band while 
\begin{equation}\label{ppafinm}
\begin{aligned}
&m_F =  \lambda_F \int_{a}^{\infty} \frac{p_s dp_s}{ \sqrt{\zeta_{ex}^2+p_s^2}^{1+\epsilon}} + \zeta_{ex}\\
&\lim_{P \to \infty} \left[\sum_{k=M+1}^{P} \left( (\zeta_{ex}^2+2b(k+1) )^{1/2}-(\zeta_{ex}^2+2b(k+1) - 2 \lambda_F b )^{1/2} \right)  -  \lambda_F 
\int_\frac{a^2}{2}^{(P+1)b} \frac{dw}{ \sqrt{\zeta_{ex}^2+2w}}
\right] \\
\end{aligned}
\end{equation}
for $\mu$ in the negative $M$ band. In both \eqref{ppafin} and 
\eqref{ppafinm}, $a$ is any convenient quantity and the integrals on the 
RHS of the first lines of these equations are the finite results for these 
integrals obtained in the  dimensional regularization scheme. Explicitly 
\begin{equation}\label{finexp} 
\begin{split}
 \int_{a}^{\infty} \frac{p_s dp_s}{ \sqrt{\zeta_{ex}^2+p_s^2}^{1+\epsilon}}= - \sqrt{\zeta_{ex}^2+ a^2}  ~~~ a \geq 0
\end{split}\end{equation} 

The full propagator is given by the expressions \eqref{alphaform} and 
\eqref{alphasol2} where 
\begin{equation}\label{sigm}
\Sigma_{R,I, 0}= \zeta_{ex}-m_F, 
\end{equation}  
$\Sigma_{R, I, n}$ may be evaluated from the first two lines of \eqref{gapeqmupm} and \eqref{gapeqmune}. \footnote{The infinite sums that occur in the second lines of these equations may be converted into finite sums by subtracting the expressions for $\Sigma_{R, I, n}$ from the expression for $\Sigma_{R, I, 0}$ whose explicit form we have presented in \eqref{sigm}. } 
and the expression for $\Sigma_{R, +,n}$ is listed subsequent lines of the same two equations.  \footnote{The RHS of the expressions that determine 
	$\Sigma_{R, +,n}$ depends on $K_{R, I, n}$. This quantity  is given in terms of $\Sigma_{R, I, n}$ - which we have already determined - in  \eqref{wheyu}. }

\subsubsection{$\mu$ in the positive/negative exceptional band}

When $\mu$ is in the positive M band, 
\begin{equation}
\begin{aligned}
 & \zeta_\pm^F(n)=\left[ \zeta_{ex}^2+2b(n+1)+ \lambda_F b\pm \lambda_F b \right]^{1/2}~~~(n \geq 0)
\end{aligned}
\end{equation}
When $\mu$ is in the negative M band, 
\begin{equation}
\begin{aligned}
 & \zeta_\pm^F(n)=\left[ \zeta_{ex}^2+2b(n+1)- \lambda_F b\pm \lambda_F b \right]^{1/2}~~~(n \geq 0)
\end{aligned}
\end{equation}
$\zeta_{ex}$ is the solution to the equation 
\begin{equation}\label{ppafin}
\begin{aligned}
&m_F = \lambda_F \int_{a}^{\infty} \frac{p_s dp_s}{ \sqrt{\zeta_{ex}^2+p_s^2}^{1+\epsilon}} + \zeta_{ex} \\
&+\lim_{P \to \infty} \left[\sum_{k=0}^{P} \left( (\zeta_{ex}^2+2b(k+1) +2 \lambda_F b  )^{1/2}-(\zeta_{ex}^2+2b(k+1) )^{1/2} \right)  -  \lambda_F 
\int_\frac{a^2}{2}^{(P+1)b} \frac{dw}{ \sqrt{\zeta_{ex}^2+2w}}
\right] \\
\end{aligned}
\end{equation}
for $\mu$ in the positive exceptional band while 
\begin{equation}\label{ppafinm}
\begin{aligned}
&m_F =  \lambda_F \int_{a}^{\infty} \frac{p_s dp_s}{ \sqrt{\zeta_{ex}^2+p_s^2}^{1+\epsilon}} + \zeta_{ex}\\
&+\lim_{P \to \infty} \left[\sum_{k=0}^{P} \left( (\zeta_{ex}^2+2b(k+1) )^{1/2}-(\zeta_{ex}^2+2b(k+1) - 2 \lambda_F b )^{1/2} \right)  -  \lambda_F 
\int_\frac{a^2}{2}^{(P+1)b} \frac{dw}{ \sqrt{\zeta_{ex}^2+2w}}
\right] \\
\end{aligned}
\end{equation}
for negative exceptional band.
Once again $a$ is any convenient quantity and \eqref{finexp} applies. 

Once again the full propagator is given by the expressions \eqref{alphaform} and 
\eqref{alphasol2}; the various quantities that occur in this expression are 
evaluated using \eqref{sigm}, \eqref{gapeqmupm} and \eqref{gapeqmune} as in the 
previous subsubsection. 

\subsection{Gap equation in bosonic variables in exceptional bands with some choices of signs} \label{spclb}

In this subsection we focus on the special case of $\mu$ in the 
positive/negative exceptional bands. We also assume that 
\begin{equation}\label{assumptions}
\begin{split}
&\zeta_{ex} \lambda_F>0\\
&\lambda_F<0 ~({\rm in ~+ve~ exceptional~ band})~~~
{\rm and}~~~\lambda_F>0~~~({\rm in~-ve ~exceptional ~band} )\\
\end{split}
\end{equation}
We will see below that the conditions \eqref{assumptions} lead 
to particularly simple gap equations in bosonic variables. 
We believe that the reason for this simplicity is that 
to ensure that the bosonic description is in an un Higgsed or 
uncondensed phase. 
 With these assumptions we have two cases that we consider separately. 

\subsubsection{ $\zeta_{ex}$, $\lambda_F$ all negative with 
$\mu$ in the positive exceptional band}

In this case  the duality map takes the 
form
\begin{equation}\label{duma}
\begin{split} 
&\lambda_F=\lambda_B-1, ~~~\lambda_B>0\\
&m_F=-\lambda_B m_B^{\rm cri},
\end{split}
\end{equation}
Substituting \eqref{duma} into \eqref{ppafin} turns the gap equation into 
\begin{equation}\label{ppafinno}
\begin{aligned}
& -\lambda_B m_B^{\rm cri} = (\lambda_B-1) \int_{a}^{\infty} \frac{p_s dp_s}{ \sqrt{\zeta_{ex}^2+p_s^2}^{1+\epsilon}}  \\
&+\lim_{P \to \infty} \bigg[-|\zeta_{ex}| + \sum_{k=0}^{P} \left( (\zeta_{ex}^2+2bk +2 \lambda_B b  )^{1/2}-(\zeta_{ex}^2+2b(k+1) )^{1/2} \right) \\
& +(1-\lambda_B)
\int_\frac{a^2}{2}^{(P+1)b} \frac{dw}{ \sqrt{\zeta_{ex}^2+2w}}
\bigg] \\
\end{aligned}
\end{equation}
Using 
\begin{equation}\label{rareq}
\begin{split}
&-|\zeta_{ex}| + \left(  \sum_{k=0}^{P} \left[ (\zeta_{ex}^2+2bk +2 \lambda_B b  )^{1/2}-(\zeta_{ex}^2+2b(k+1) )^{1/2} \right] \right)\\
&= \left( \sum_{k=0}^{P} \left[ (\zeta_{ex}^2+2bk +2 \lambda_B b  )^{1/2}-(\zeta_{ex}^2+2bk )^{1/2} \right] \right)   - (\zeta_{ex}^2+2b(P+1) )^{1/2}
\end{split}
\end{equation}
we can rewrite \eqref{ppafinno} as 
\begin{equation}\label{ppafinno}
\begin{aligned}
& -\lambda_B m_B^{\rm cri} = (\lambda_B-1) \int_{a}^{\infty} \frac{p_s dp_s}{ \sqrt{\zeta_{ex}^2+p_s^2}^{1+\epsilon}}  \\
&+\lim_{P \to \infty}
 \bigg[ \sum_{k=0}^P \left[ (\zeta_{ex}^2+2bk +2 \lambda_B b  )^{1/2}-(\zeta_{ex}^2+2bk )^{1/2} \right] \\
& - (\zeta_{ex}^2+2b(P+1) )^{1/2}  +(1-\lambda_B)
\int_\frac{a^2}{2}^{(P+1)b} \frac{dw}{ \sqrt{\zeta_{ex}^2+2w}}
\bigg] \\
\end{aligned}
\end{equation}
Using 
\begin{equation}\label{sein}
\int_a^{(P+1)b} \frac{dw}{ \sqrt{\zeta_{ex}^2+2w}}=
(\zeta_{ex}^2+2b(P+1) )^{1/2}-(\zeta_{ex}^2 +a^2 )^{1/2}
\end{equation}
and \eqref{finexp} we see that \eqref{ppafinno} can be 
re-expressed as 
\begin{equation}\label{ppafinnt}
\begin{aligned}
& -\lambda_B m_B^{\rm cri} = \lambda_B \int_{a}^{\infty} \frac{p_s dp_s}{ \sqrt{\zeta_{ex}^2+p_s^2}^{1+\epsilon}}  \\
&+\lim_{P \to \infty}
\bigg[ \sum_{k=0}^P \left[ (\zeta_{ex}^2+2bk +2 \lambda_B b  )^{1/2}-(\zeta_{ex}^2+2bk )^{1/2} \right]  -\lambda_B
\int_\frac{a^2}{2}^{(P+1)b} \frac{dw}{ \sqrt{\zeta_{ex}^2+2w}}
\bigg] \\
\end{aligned}
\end{equation}
Note that every term in \eqref{ppafinnt} is of order $\lambda_B$ 
at small $\lambda_B$. It follows that \eqref{ppafinnt} 
has a smooth $\lambda_B \to 0$ limit.

Finally we note that \eqref{ppafinnt} has an aestheically unpleasing 
feature. The summation in the second line of this equation is a 
discretised version of an integral with upper limit $w=Pb$. However 
the `subtraction integral' in the second line of \eqref{ppafinnt} 
has an upper limit $(P+1)b$. Note however that 
\begin{equation}\label{sqap}
\int_\frac{a^2}{2}^{(P+r)b} \frac{dw}{ \sqrt{\zeta_{ex}^2+2w}}
= 
\int_\frac{a^2}{2}^{(P+s)b} \frac{dw}{ \sqrt{\zeta_{ex}^2+2w}}
+ 
{\cal O}\left( \frac{1}{\sqrt{P}} \right) 
\end{equation}
where $r$ and $s$ are any numbers that are held fixed in the limit $P \to 0$. It follows that \eqref{ppafinnt} is equivalent to the the aesthetically 
more pleasing gap equation  
\begin{equation}\label{apep}
\begin{aligned}
& -\lambda_B m_B^{\rm cri} = \lambda_B \int_{a}^{\infty} \frac{p_s dp_s}{ \sqrt{\zeta_{ex}^2+p_s^2}^{1+\epsilon}}  \\
&+\lim_{P \to \infty}
\bigg[ \sum_{k=0}^P \left[ (\zeta_{ex}^2+2bk +2 \lambda_B b  )^{1/2}-(\zeta_{ex}^2+2bk )^{1/2} \right]  -\lambda_B
\int_\frac{a^2}{2}^{Pb} \frac{dw}{ \sqrt{\zeta_{ex}^2+2w}}
\bigg] \\
\end{aligned}
\end{equation}

\subsubsection{ $\zeta_{ex}$, $\lambda_F$ all positive with 
	$\mu$ in the negative exceptional band}

\begin{equation}\label{ppafinnt}
\begin{aligned}
& -\lambda_B m_B^{\rm cri} = \lambda_B \int_{a}^{\infty} \frac{p_s dp_s}{ \sqrt{\zeta_{ex}^2+p_s^2}^{1+\epsilon}}  \\
&+\lim_{P \to \infty}
\bigg[ \sum_{k=0}^P\left[ (\zeta_{ex}^2+2bk +2 \lambda_B b  )^{1/2}-(\zeta_{ex}^2+2bk )^{1/2} \right]  -\lambda_B
\int_\frac{a^2}{2}^{(P+1)b} \frac{dw}{ \sqrt{\zeta_{ex}^2+2w}}
\bigg] \\
\end{aligned}
\end{equation}

In this case  the duality map takes the 
form
\begin{equation}\label{dumab}
\begin{split} 
&\lambda_F= 1 +\lambda_B ~~~\lambda_B<0\\
&m_F=-\lambda_B m_B^{\rm cri},
\end{split}
\end{equation}
Substituting \eqref{dumab} into \eqref{ppafinm} we obtain
\begin{equation}\label{ppafinmb} 
\begin{aligned}
&-\lambda_B m_B^{\rm cri} = (1 + \lambda_B) \int_{a}^{\infty} \frac{p_s dp_s}{ \sqrt{\zeta_{ex}^2+p_s^2}^{1+\epsilon}} + \zeta_{ex}\\
&+\lim_{P \to \infty} \left[\sum_{k=0}^{P} \left( (\zeta_{ex}^2+2b(k+1) )^{1/2}-(\zeta_{ex}^2+2bk - 2 \lambda_B b )^{1/2} \right)  -  (1+\lambda_B) 
\int_\frac{a^2}{2}^{(P+1)b} \frac{dw}{ \sqrt{\zeta_{ex}^2+2w}}
\right] \\
\end{aligned}
\end{equation}
Rearranging terms as in the previous subsubsection we find 
\begin{equation}\label{ppafinmb} 
\begin{aligned}
&-\lambda_B m_B^{\rm cri} = (1 + \lambda_B) \int_{a}^{\infty} \frac{p_s dp_s}{ \sqrt{\zeta_{ex}^2+p_s^2}^{1+\epsilon}} \\
&+\lim_{P \to \infty} \bigg[\sum_{k=0}^{P} \left( (\zeta_{ex}^2+2bk )^{1/2}-(\zeta_{ex}^2+2bk - 2 \lambda_B b )^{1/2} \right) \\
& + (\zeta_{ex}^2+2b(P+1) )^{1/2} -  (1+\lambda_B) 
\int_\frac{a^2}{2}^{(P+1)b} \frac{dw}{ \sqrt{\zeta_{ex}^2+2w}}
\bigg] \\
\end{aligned}
\end{equation}
As in the previous subsubsection, this equation can be further
simplified to 
\begin{equation}\label{ppafinmc} 
\begin{aligned}
&-\lambda_B m_B^{\rm cri} = \lambda_B \int_{a}^{\infty} \frac{p_s dp_s}{ \sqrt{\zeta_{ex}^2+p_s^2}^{1+\epsilon}} \\
&+\lim_{P \to \infty} \bigg[\sum_{k=0}^{P} \left( (\zeta_{ex}^2+2bk )^{1/2}-(\zeta_{ex}^2+2bk - 2 \lambda_B b )^{1/2} \right) - \lambda_B
\int_\frac{a^2}{2}^{(P+1)b} \frac{dw}{ \sqrt{\zeta_{ex}^2+2w}}
\bigg] \\
\end{aligned}
\end{equation}
As in the previous subsection, every term in \eqref{ppafinmc} 
is of order $\lambda_B$ at small $\lambda_B$ and so the 
equation \eqref{ppafinmc} has a smooth $\lambda_B \to 0$ limit. 
As in the previous subsubsection, we may use \eqref{sqap} to recast this 
equation in the aesthetically more pleasing form 
\begin{equation}\label{apem} 
\begin{aligned}
&-\lambda_B m_B^{\rm cri} = \lambda_B \int_{a}^{\infty} \frac{p_s dp_s}{ \sqrt{\zeta_{ex}^2+p_s^2}^{1+\epsilon}} \\
&+\lim_{P \to \infty} \bigg[\sum_{k=0}^{P} \left( (\zeta_{ex}^2+2bk )^{1/2}-(\zeta_{ex}^2+2bk - 2 \lambda_B b )^{1/2} \right) - \lambda_B
\int_\frac{a^2}{2}^{Pb} \frac{dw}{ \sqrt{\zeta_{ex}^2+2w}}
\bigg] \\
\end{aligned}
\end{equation}

\subsection{The gap equations in bosonic variables in the 
	generic case} \label{brgc}

\subsubsection{$\mu$ in positive M band, $\lambda_F$ negative}

In this case \eqref{duma} applies. 
Using manipulations similar to those in the previous 
subsection, the gap equations in this case can be recast as 
\begin{equation}\label{ppafinntn}
\begin{aligned}
& -\lambda_B m_B^{\rm cri} = \lambda_B \int_{a}^{\infty} \frac{p_s dp_s}{ \sqrt{\zeta_{ex}^2+p_s^2}^{1+\epsilon}}  \\
&+\lim_{P \to \infty}
\bigg[ \sum_{k=M+1}^P \left[ (\zeta_{ex}^2+2bk +2 \lambda_B b  )^{1/2}-(\zeta_{ex}^2+2bk )^{1/2} \right]  -\lambda_B
\int_\frac{a^2}{2}^{(P+1)b} \frac{dw}{ \sqrt{\zeta_{ex}^2+2w}}
\bigg] \\
&+\zeta_{ex} + (\zeta_{ex}^2+2b(M+1) )^{1/2}
\end{aligned}
\end{equation}
As in the previous subsection we can use \eqref{sqap} to rewrite 
\eqref{ppafinntn} in the equivalent but aesthetically more pleasing form 
\begin{equation}\label{appp}
\begin{aligned}
& -\lambda_B m_B^{\rm cri} = \lambda_B \int_{a}^{\infty} \frac{p_s dp_s}{ \sqrt{\zeta_{ex}^2+p_s^2}^{1+\epsilon}}  \\
&+\lim_{P \to \infty}
\bigg[ \sum_{k=M+1}^P \left[ (\zeta_{ex}^2+2bk +2 \lambda_B b  )^{1/2}-(\zeta_{ex}^2+2bk )^{1/2} \right]  -\lambda_B
\int_\frac{a^2}{2}^{Pb} \frac{dw}{ \sqrt{\zeta_{ex}^2+2w}}
\bigg] \\
&+\zeta_{ex} + (\zeta_{ex}^2+2b(M+1) )^{1/2}
\end{aligned}
\end{equation}

This result \eqref{ppafinntn} at $M=-1$ applies to the positive 
exceptional band. In this case the last line of \eqref{ppafinntn} reduces to 
$$+\zeta_{ex} + |\zeta_{ex}|$$ 
and vanishes when $\zeta_{ex}$ is negative in agreement with 
\eqref{ppafinnt}. In every other case (i.e. if $M>-1$ or if 
$M=-1$ and $\zeta_{ex}$ is positive) the third line of 
\eqref{ppafinntn} is nonzero, and is, moreover, of order unity 
(rather than order $\lambda_B$) in the limit $\lambda_B \to 0$. 
It follows that in these cases the limit $\lambda_B \to 0$ of 
\eqref{ppafinntn} is not smooth.

\subsubsection{$\mu$ in negative M band, $\lambda_F$ positive}

In this case \eqref{dumab} applies. 
Using manipulations similar to those in the previous 
subsection, the gap equations can be shown to take the form 
\begin{equation}\label{ppafinntnb}
\begin{aligned}
& -\lambda_B m_B^{\rm cri} = \lambda_B \int_{a}^{\infty} \frac{p_s dp_s}{ \sqrt{\zeta_{ex}^2+p_s^2}^{1+\epsilon}}  \\
&+\lim_{P \to \infty}
\bigg[ \sum_{k=M+1}^P \left[ (\zeta_{ex}^2+2bk   )^{1/2}-(\zeta_{ex}^2+2bk -2 \lambda_B b)^{1/2} \right]  -\lambda_B
\int_\frac{a^2}{2}^{(P+1)b} \frac{dw}{ \sqrt{\zeta_{ex}^2+2w}}
\bigg] \\
&+\zeta_{ex} - (\zeta_{ex}^2+2b(M+1) )^{1/2}
\end{aligned}
\end{equation}
Again the result \eqref{ppafinntnb} can be rewritten (using \eqref{sqap})
as
\begin{equation}\label{aemm}
\begin{aligned}
& -\lambda_B m_B^{\rm cri} = \lambda_B \int_{a}^{\infty} \frac{p_s dp_s}{ \sqrt{\zeta_{ex}^2+p_s^2}^{1+\epsilon}}  \\
&+\lim_{P \to \infty}
\bigg[ \sum_{k=M+1}^P \left[ (\zeta_{ex}^2+2bk   )^{1/2}-(\zeta_{ex}^2+2bk -2 \lambda_B b)^{1/2} \right]  -\lambda_B
\int_\frac{a^2}{2}^{Pb} \frac{dw}{ \sqrt{\zeta_{ex}^2+2w}}
\bigg] \\
&+\zeta_{ex} - (\zeta_{ex}^2+2b(M+1) )^{1/2}
\end{aligned}
\end{equation}

This result \eqref{ppafinntnb} at $M=-1$ applies to the negative 
exceptional band. In this case the last line of \eqref{ppafinntnb} reduces to 
$$+\zeta_{ex} - |\zeta_{ex}|$$ 
and vanishes when $\zeta_{ex}$ is positive in agreement with 
\eqref{ppafinmc}. In every other case (i.e. if $M>-1$ or if 
$M=-1$ and $\zeta_{ex}$ is negative) the third line of 
\eqref{ppafinntnb} is nonzero, and is, moreover, of order unity 
(rather than order $\lambda_B$); it follows that the equation  
\eqref{ppafinntnb} does not have a smooth $\lambda_b \to 0$ limit. 

\subsubsection{$\mu$ in positive M band, $\lambda_F$ positive}

In this case the duality map takes the form \eqref{dumab}. Re-expressing \eqref{ppafin} in dual variables and using  
\begin{equation}\label{rearr}
\begin{split}  
&\sum_{k=M+1}^P 
\left( (\zeta_{ex}^2+2b(k+2) +2\lambda_B b  )^{1/2}-(\zeta_{ex}^2+2b(k+1) )^{1/2} \right)\\
& -(\lambda_B+1)
\int_\frac{a^2}{2}^{(P+1)b} \frac{dw}{ \sqrt{\zeta_{ex}^2+2w}}  +(\lambda_B+1) \int_{a}^{\infty} \frac{p_s dp_s}{ \sqrt{\zeta_{ex}^2+p_s^2}^{1+\epsilon}} \\
&=\left[\sum_{k=M+1}^P 
\left( (\zeta_{ex}^2+2b(k+2) +2\lambda_B b  )^{1/2}-(\zeta_{ex}^2+2b(k+2) )^{1/2} \right)-\lambda_B
\int_\frac{a^2}{2}^{(P+1)b} \frac{dw}{ \sqrt{\zeta_{ex}^2+2w}}  \right] \\
& + (\zeta_{ex}^2+2b(P+2) )^{1/2} 
-(\zeta_{ex}^2+2b(M+2) )^{1/2} \\
& -(\zeta_{ex}^2+2b(P+1) )^{1/2} +\lambda_B \int_{a}^{\infty} \frac{p_s dp_s}{ \sqrt{\zeta_{ex}^2+p_s^2}^{1+\epsilon}}
\end{split}
\end{equation} 
and \eqref{sqap} we find 
\begin{equation}\label{apppn}
\begin{aligned}
& -\lambda_B m_B^{\rm cri} = \lambda_B \int_{a}^{\infty} \frac{p_s dp_s}{ \sqrt{\zeta_{ex}^2+p_s^2}^{1+\epsilon}}  \\
&+\lim_{P \to \infty}
\bigg[ \sum_{k=M+3}^{P+2} \left[ (\zeta_{ex}^2+2bk +2 \lambda_B b  )^{1/2}-(\zeta_{ex}^2+2bk )^{1/2} \right]  -\lambda_B
\int_\frac{a^2}{2}^{(P+1)b} \frac{dw}{ \sqrt{\zeta_{ex}^2+2w}}
\bigg] \\
&+\zeta_{ex} - (\zeta_{ex}^2+2b(M+2) )^{1/2}\\
&  + (\zeta_{ex}^2+2b(P+2) )^{1/2}-(\zeta_{ex}^2+2b(P+1) )^{1/2}
\end{aligned}
\end{equation}

\subsubsection{$\mu$ in negative M band, $\lambda_F$ negative}

In this case the duality map takes the form \eqref{duma}. Reexpressing \eqref{ppafinm} in dual variables and using
\begin{equation}\label{lon} \begin{split}
&\sum_{k=M+1}^{P} \left( (\zeta_{ex}^2+2b(k+1) )^{1/2}-(\zeta_{ex}^2+2b(k+2) - 2 \lambda_B b )^{1/2} \right)\\
& -(\lambda_B-1)
\int_\frac{a^2}{2}^{(P+1)b} \frac{dw}{ \sqrt{\zeta_{ex}^2+2w}}  +(\lambda_B-1) \int_{a}^{\infty} \frac{p_s dp_s}{ \sqrt{\zeta_{ex}^2+p_s^2}^{1+\epsilon}}
\\=
&\sum_{k=M+1}^{P} \left[ (\zeta_{ex}^2+2b(k+2) )^{1/2}-(\zeta_{ex}^2+2b(k+2) - 2 \lambda_B b )^{1/2}  -\lambda_B
\int_\frac{a^2}{2}^{(P+1)b} \frac{dw}{ \sqrt{\zeta_{ex}^2+2w}} \right] \\
&+(\zeta_{ex}^2+2b(M+2) )^{1/2} - (\zeta_{ex}^2+2b(P+2) )^{1/2}\\
& +\lambda_B \int_{a}^{\infty} \frac{p_s dp_s}{ \sqrt{\zeta_{ex}^2+p_s^2}^{1+\epsilon}}+(\zeta_{ex}^2+2b(P+1) )^{1/2}
\end{split} 
\end{equation}
along with \eqref{sqap}, the gap equation becomes 
\begin{equation}\label{aempn}
\begin{aligned}
& -\lambda_B m_B^{\rm cri} = \lambda_B \int_{a}^{\infty} \frac{p_s dp_s}{ \sqrt{\zeta_{ex}^2+p_s^2}^{1+\epsilon}}  \\
&+\lim_{P \to \infty}
\bigg[ \sum_{k=M+3}^{P+2} \left[ (\zeta_{ex}^2+2bk   )^{1/2}-(\zeta_{ex}^2+2bk -2 \lambda_B b)^{1/2} \right]  -\lambda_B
\int_\frac{a^2}{2}^{(P+1)b} \frac{dw}{ \sqrt{\zeta_{ex}^2+2w}}
\bigg] \\
&+\zeta_{ex} + (\zeta_{ex}^2+2b(M+2) )^{1/2}\\
& - (\zeta_{ex}^2+2b(P+2) )^{1/2}
\end{aligned}
\end{equation}

\subsection{Solving for $\zeta_{ex}$}

Consider, for instance, the positive M band gap equation \eqref{ppafin} which we reproduce here for convenience
\begin{equation}\label{ppafinre}
\begin{aligned}
&m_F = \lambda_F \int_{a}^{\infty} \frac{p_s dp_s}{ \sqrt{\zeta_{ex}^2+p_s^2}^{1+\epsilon}} + \zeta_{ex} \\
&+\lim_{P \to \infty} \left[\sum_{k=M+1}^{P} \left( (\zeta_{ex}^2+2b(k+1) +2 \lambda_F b  )^{1/2}-(\zeta_{ex}^2+2b(k+1) )^{1/2} \right)  -  \lambda_F 
\int_\frac{a^2}{2}^{(P+1)b} \frac{dw}{ \sqrt{\zeta_{ex}^2+2w}}
\right] \\
\end{aligned}
\end{equation}
Making the choice $a=0$ we evaluate the first integral
\eqref{ppafinre} to obtain the simplified equation
\begin{equation}\label{proco}
\begin{aligned}
&m_F = \zeta_{ex} - \lambda_F |\zeta_{ex}| -
\sum_{k=0}^{M} \left( (\zeta_{ex}^2+2b(k+1) +2 \lambda_F b  )^{1/2}-(\zeta_{ex}^2+2b(k+1) )^{1/2} \right) \\
&+\lim_{P \to \infty} \left[\sum_{k=0}^{P} \left( (\zeta_{ex}^2+2b(k+1) +2 \lambda_F b  )^{1/2}-(\zeta_{ex}^2+2b(k+1) )^{1/2} \right)  -  \lambda_F 
\int_0^{(P+1)b} \frac{dw}{ \sqrt{\zeta_{ex}^2+2w}}
\right] \\
&~~~~ \equiv f(\zeta_{ex}, \lambda_F)
\end{aligned}
\end{equation}

\subsubsection{Existence of solutions}
The second and third lines of \eqref{proco} are both of order 
${\cal O}(1/|\zeta_{ex}|)$ at large $|\zeta_{ex}|$\footnote{Here we are measuring all dimensionful qualities in the units of magnetic field.}. It follows 
that the RHS of \eqref{proco} behaves like $(1-\lambda_F)|\zeta_{ex}|$ as $\zeta_{ex} \to \infty$, but 
behaves like $-(1+\lambda_F)|\zeta_{ex}|$ as $\zeta_{ex} \to -\infty$. As the RHS of \eqref{proco} is a continuous function 
of $\zeta_{ex}$, it follows that it takes every real value as 
$\zeta_{ex}$ varies from $-\infty$ to $\infty$, except possibly the extreme value $\lambda_F=\pm 1$.\footnote{Recall that $|\lambda_F|\leq 1$.} As a consequence
\eqref{proco} has at least one real solution (for the variable 
$\zeta_{ex}$) for every choice of $m_F$ provided $|\lambda_F|<1$. It is easily verified that this conclusion continues to hold for negative bands 
and also for exceptional bands. 

\subsubsection{Uniqueness of solutions}

Consider the equation \eqref{proco} at any fixed $\lambda_F$. 
We have already seen that the RHS of this equation increases from $-\infty$ to $\infty$ as $\zeta_{ex}$ varies from $-\infty$ to $\infty$. If this increase occurs monotonically, it follows 
that \eqref{proco} has a unique solution for every value of 
$m_F$. If, on the other hand, the RHS of \eqref{proco} is 
not a monotonic function of $\zeta_{ex}$ then there exists 
a range of $m_F$ over which there exist three or more solutions 
to the equation \eqref{proco}. 

\begin{figure}[h!]
	\centering
	\includegraphics[width=5in,height=4in]{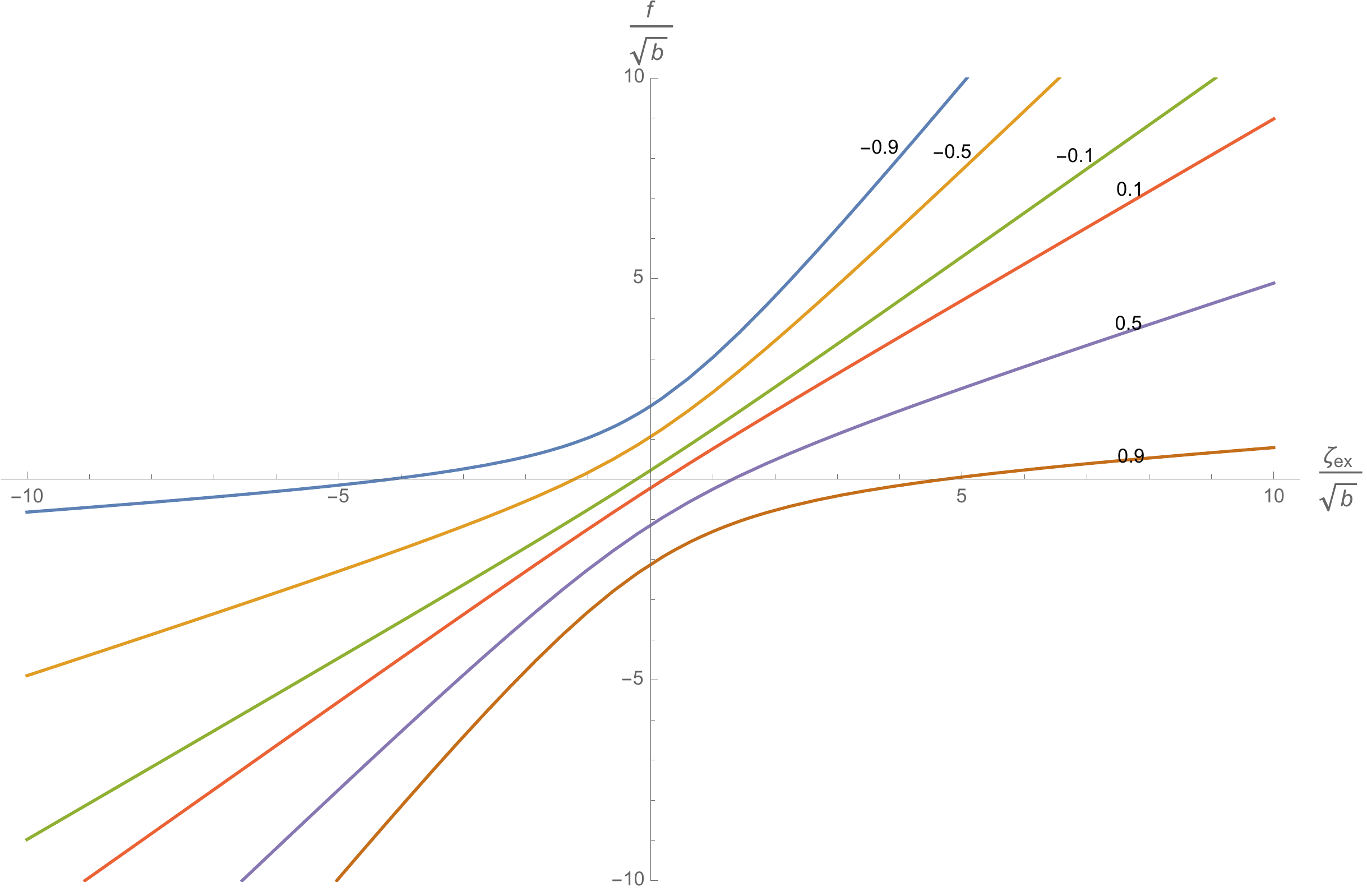}
	\caption{Here we have plotted $f$ defined in \eqref{proco} for different values of $\lambda_F$ that are mentioned by the side of each curve. This qualitative behaviour is valid for both positive and negative bands including exceptional ones.}
	\label{Uniqueness}
\end{figure}

When $\lambda_F$ is small, the second and third lines of 
\eqref{proco} are both of order $\lambda_F$. In this limit 
the RHS of \eqref{proco} is approximately given by 
$\zeta_{ex}$ and so is monotonic. It follows that the solutions 
of \eqref{proco} are unique at small enough $\lambda_F$. 
Below we will also argue that the solutions of \eqref{proco} 
are also unique when $\lambda_F$ is in the vicinity of $\pm 1$
We do not know whether this property (monotonicity of the 
RHS of \eqref{proco} and consequent uniqueness of the solutions 
to the gap equation) persists at arbitrary values of $\lambda_F$. 
Preliminary numerical experiments suggest that the uniqueness 
of solutions may persist for all values of $\lambda_F$ and 
all values of $M$ (see Fig \ref{Uniqueness}). This question deserves 
a more careful numerical study which we defer to future work. 
Again all the conclusions of this subsection apply also 
to negative and exceptional bands.

\subsubsection{Perturbation Theory in small $\lambda_F$}

We have argued above that \eqref{proco} has a unique solution 
(at least) at small enough $\lambda_F$. In a perturbative expansion in $\lambda_F$ this solution takes the form 
\begin{equation}\label{zetapert}
\zeta_{ex,+,M}= m_F + \sum_{n=1}^\infty \lambda_F^n \zeta^n_{ex,+,M}
\end{equation}
It is not difficult to obtain explicit expressions for the 
coefficients $\zeta^n_{ex,+,M}$ at small values of $n$. The expressions are less than completely illuminating as they are given in terms of (completely explicit) infinite summations. 
For this reason we present explicit formulae only for 
$\zeta^1_{ex,+,M}$ below
\begin{equation}
\begin{aligned} \label{olc}
&\zeta_{ex, +,M}^1= |m_F|  +\sum_{n=0}^{M} \frac{1}{\sqrt{m_F^2+2b(n+1)}}\\
 & - \lim_{P \to \infty} \left[ \sum_{n=0}^{P} \frac{1}{\sqrt{m_F^2+2b(n+1)}}-\int_0^{b(P+1)} dw \frac{1}{\sqrt{m_F^2+2bn}} \right]\\&
\end{aligned}
\end{equation}
Note that the complicated part of the solution (the second line 
of \eqref{olc}) is independent of $M$. This fact allows us 
to reach a simple explicit conclusion about the energy width of Landau levels, as we now explain.  

Let us first assume that physical interpretation of $\mu$ lying in the $M^{th}$ positive energy gap is that all Landau levels with $n \leq M$ are completely filled. When this 
is the case, the formula for the energy of the Landau level 
at $n=M$ is given by $E_{+,M}=\zeta_+^F(M)$ where 
\begin{equation}\label{zetp}
(E_{+,M}^{\rm up})^2 =\zeta_{ex,M}^2+2b(M+1)
\end{equation}
The energy $E_{+,M}^{\rm up}$ reported above may physically be thought of as the energy of the last `electron' whose addition completely filled up the Landau Level with $n=M$. 

On the other hand when $\mu$ lies in the $M-1$ positive level,
the energy of the Landau level with $n=M$ is given by 
\begin{equation}\label{othen}
(E_{+,M}^{\rm low})^2 =\zeta_{ex,M-1}^2+2b(M+1) + 2 \lambda_Fb
\end{equation}
Physically, $E_{+,M}^{\rm low}$ is the energy of the first `electron' added to the Landau level with $n=M$. 

Notice that 
\begin{equation}\label{energydiff}
(E_{+,M}^{\rm up})^2-(E_{+,M}^{\rm low})^2=
\zeta_{ex,M}^2-\zeta_{ex,M-1}^2 -2 \lambda_Fb
\end{equation}
It follows from \eqref{olc} that 
\begin{equation}\label{energydifffp}
\Delta_{+}(M)=(E_{+,M}^{\rm up})^2-(E_{+,M}^{\rm low})^2= - 2 \lambda_F b 
\left( 1 - \frac{m_F}{\sqrt{m_F^2+2b(M+1)}} \right)  +\mathcal{O}(\lambda_F^2)
\end{equation}
Similar considerations for negative $M$ band shows that  \footnote{Note that in this case $E_{-,M}=-\zeta_-^F(M)$
\begin{equation}\label{zetp}
(E_{-,M}^{\rm up})^2 =\zeta_{ex,M-1}^2+2b(M+1)- 2 \lambda_Fb,~~~ (E_{-,M}^{\rm low})^2 =\zeta_{ex,M}^2+2b(M+1) 
\end{equation}}
\begin{equation}\label{energydifffn}
\begin{aligned}
 & \zeta_{ex, -,M}^1= |m_F|  +\sum_{n=0}^{M} \frac{1}{\sqrt{m_F^2+2b(n+1)}}\\
 & - \lim_{P \to \infty} \left[ \sum_{n=0}^{P} \frac{1}{\sqrt{m_F^2+2b(n+1)}}-\int_0^{b(P+1)} dw \frac{1}{\sqrt{m_F^2+2bn}} \right]\\
& \Delta_{-}(M)=(E_{-,M}^{\rm up})^2-(E_{-,M}^{\rm low})^2= - 2 \lambda_F b 
\left( 1 + \frac{m_F}{\sqrt{m_F^2+2b(M+1)}} \right)  +\mathcal{O}(\lambda_F^2)
\end{aligned}
\end{equation}
On the other hand for chemical potential in both positive and negative exceptional band it follows that
\begin{equation}
\begin{aligned} \label{olcex}
&\zeta_{ex, \pm}^1= |m_F|  
 & - \lim_{P \to \infty} \left[ \sum_{n=0}^{P} \frac{1}{\sqrt{m_F^2+2b(n+1)}}-\int_0^{b(P+1)} dw \frac{1}{\sqrt{m_F^2+2bn}} \right]\\&
\end{aligned}
\end{equation}
Therefore exceptional band does not split to the first order in $\lambda_F$,i.e.,
\begin{equation}\label{energydifffex}
\Delta_{ex}= \mathcal{O}(\lambda_F^2)
\end{equation}

\subsubsection{Perturbation Theory in small $\lambda_B$} \label{bpc}

In this subsection we will move to the opposite end of parameter space 
from the previous section. We study the gap equations in the limit that $|\lambda_F|$ is near to its largest allowed value, i.e. unity. In this 
limit the dual t' Hooft coupling, $\lambda_B$ is small. In this section 
we study $\zeta_{ex}$ at leading order in $\lambda_B$ perturbation theory.
In this regime of parameters the bosonic theory is weakly coupled 
and so it is most natural to view the gap equations in bosonic 
variables. Motivated by this observation, in this section we take  $\lambda_B$ to zero holding $m_B^{{\rm cri}}$ fixed. In this limit 
the gap equations we wish to solve are presented in \eqref{appp}, \eqref{aemm}, \eqref{apppn}, and \eqref{aempn} respectively. 

In order to initiate our discussion let us consider, for instance, 
the equation \eqref{appp}. Notice that, in the limit under study 
in this section, every term in  the first two lines of \eqref{appp} 
has an explicit factor of $\lambda_B$. On the other hand the 
the terms in the third line of \eqref{appp} appear with no power 
of $\lambda_B$. In order to find a solution to this equation at small 
$\lambda_B$ we are left with two options. The first option is that 
$\zeta_{ex}$ stays finite in the limit $\lambda_B \to 0$. The second 
option is that $\zeta_B$ scales to infinity as $\lambda_B$ is taken 
to zero. Let us consider each of these options in turn. 

Option one can only work if the terms in the third line of \eqref{appp} cancel against each other. This can only happen when $M=-1$ and 
also when $\zeta_{ex}$ is negative. In this special case we can 
simply set $\lambda_B=0$ in \eqref{appp} and that equation reduces to  
\begin{equation} \label{nrc}
\begin{aligned}
&- m_B^{\rm cri} = -|\zeta_{ex}| 
&+\lim_{P \to \infty} \bigg[\sum_{k=0}^{P} \frac{ b}{(\zeta_{ex}^2+2bk )^{1/2}}   - 
\int_\frac{a^2}{2}^{Pb} \frac{dw}{ \sqrt{\zeta_{ex}^2+2w}}
\bigg] \\
\end{aligned}
\end{equation}

As a slight aside we pause to note that a very similar discussion applies also to the study of the equation \eqref{aemm}. For that equation as well there exist solutions to the gap equation with $\zeta_{ex}$ held fixed as $\lambda_B$ is taken to zero only when $M=-1$ and when, in this case, $\zeta_{ex}$ is positive. 
Indeed the final equation for $\zeta_{ex}$ in this case reduces 
exactly to \eqref{nrc} (with no modification). Below we will present 
numerical evidence that suggests that \eqref{nrc} does indeed have 
a (likely unique) solution for $|\zeta_{ex}|$ for every value of 
$m_B^{{\rm cri}}$. 

Note that the condition for existence of `option one' solutions of 
\eqref{appp} and \eqref{aemm} can be presented in a uniform manner. These
solutions exist only for $M=-1$ and also provided
\begin{equation}\label{all}
\zeta_{ex} \lambda_B<0
\end{equation}
Recall also that the equation \eqref{appp} only 
applies provided that
\begin{equation} \label{tc}
\eta \lambda_B>0
\end{equation}
( here the variable $\eta=\pm 1$ is defined to be positive or negative 
depending on whether $\mu$ lies  in the positive/negative exceptional band) 
so \eqref{tc} is effectively a third condition for the existence 
of these solutions. 
 
Let us now return to the consideration of the second option for 
solutions of \eqref{appp}, i.e. the option that we find a solution 
with $\zeta_{ex}$ blowing up as $\lambda_B \to 0$. In this limit 
every term in the second line of \eqref{appp} vanishes faster than 
$\lambda_B$. The LHS of the first line of \eqref{appp} goes to 
zero like $\lambda_B$. On the other hand the RHS of the first line of \eqref{appp} is of order $\lambda_B |\zeta_{ex}|$. Finally each of the 
terms in the third line of \eqref{appp} are of order $|\zeta_{ex}|$.
It follows that the two terms on the third line of \eqref{appp} are 
dominant in the small $\lambda_B$ limit. The only way our gap equation 
can have a solution is if these two terms cancel against each other 
upto a residue of order $\lambda_B |\zeta_{ex}|$. Clearly such a cancellation can only happen if $\zeta_{ex}$ is negative. Assuming this 
condition is met, the difference between the two terms on the third 
line of \eqref{appp} is of order $1/\sqrt{|\zeta_{ex}|}$. This difference
can cancel against the term of order $\lambda_B |\zeta_{ex}|$ only 
if $\zeta_{ex}$ is of order $1/\sqrt{|\lambda_B|}$, more specifically 
if 
\begin{equation}\label{forsol}
\zeta_{ex}=\frac{\zeta_{ex}^{-}}{\sqrt{|\lambda_B}|}+ \mathcal{O}(1)
\end{equation}

The discussion presented of the second option of solutions - i.e. solutions
of the gap equation for which $\zeta_{ex}$ diverges as $\lambda_B$ 
is taken to zero - is easily generalized to the the study of the 
gap equations \eqref{aemm} , \eqref{apppn}, and \eqref{aempn}. In 
each case we find that solutions of this form occur only when 
\eqref{all} is obeyed, and in each case the solutions we seek take 
the form \eqref{forsol}. 

It is a simple matter to plug \eqref{forsol} into the gap equations 
\eqref{appp}, \eqref{aemm}, \eqref{apppn} and \eqref{aempn} and to solve 
for $\zeta_{ex}^{-}$. Provided $M\neq -1$, in every case we find that  
\begin{equation}\label{zetaB}
\begin{aligned}
\zeta^-_{ex}=-\rm{sgn}(\lambda_B) \left[b \left(M+\frac{3}{2}\mp\frac{\rm{sgn}(\lambda_B )}{2} \right)    \right]^{1/2}
\end{aligned}
\end{equation}
Where sign $\mp$ is for positive/negative Mth band.

It follows that for $M \neq -1$ the width of 
the $M^{th}$ energy state is given by 
\begin{equation}\label{energydiffB}
\Delta_{\pm}(M)=\pm \frac{b}{\sqrt{|\lambda_B|}}
\end{equation}
Note that this (divergent) width is independent 
of $M$. 

In the special case $M=-1$, i.e $\mu$ is in positive/negative exceptional band but does not satisfy the condition in \eqref{tc} (this option 
only arises in the study of \eqref{apppn} and \eqref{aempn}), we find
\begin{equation}\label{zetaBex}
 \begin{aligned}
  \zeta_{ex}^{-}=-\rm{sgn}(\lambda_B)b^{1/2} 
 \end{aligned}
\end{equation}

\subsection{Numerical interpolation between these
two regimes} \label{numint}

Although we have been able to solve analytically 
for $\zeta_{ex}$ only at small $\lambda_F$ 
or small $\lambda_B$, it is not difficult to 
numerically solve the gap equations for $\zeta_{ex}$ at any given value of parameters. 
We have carried out this exercise for a range of 
different values of parameters.  

In this section we briefly present some of our results 
with the help of a few graphs. The numerical results of this 
section have all been unsophisticatedly obtained  using the numerical equation solving routine on Mathematica. We have not attempted to seriously  estimate the errors in our computations, and have presented no error bars on our graphs.

\begin{figure}[!h]
	\begin{subfigure}{0.5\textwidth}
	\includegraphics[width=3in,height=2.5in]{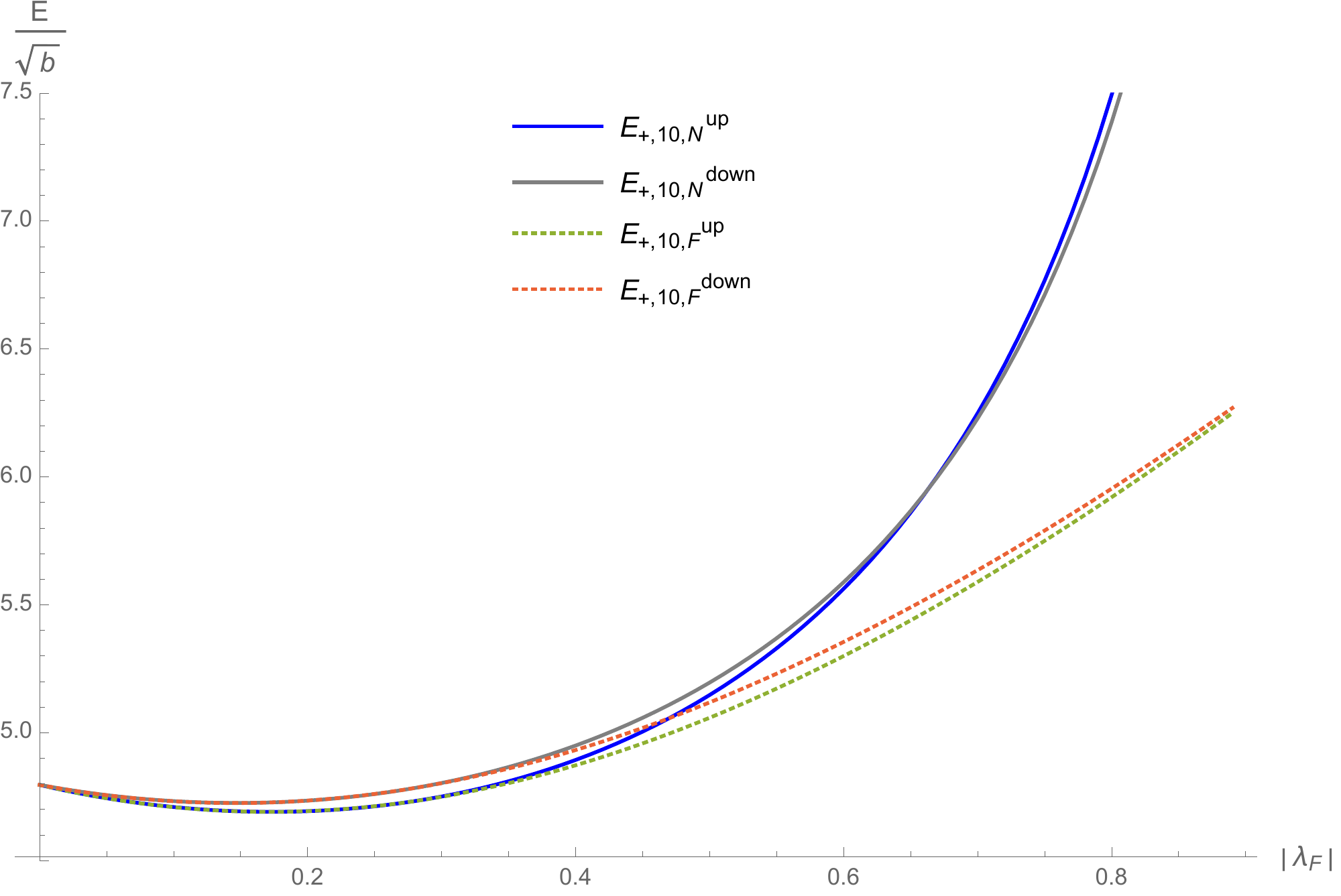}
		\caption{Dashed lines are from fermionic perturbation theory. Solid lines are from numerical analysis.}
		\label{A1}
	\end{subfigure}\hspace{15pt}
	~~~~~~~~
\begin{subfigure}{0.5\textwidth}
	\includegraphics[width=3in,height=2.5in]{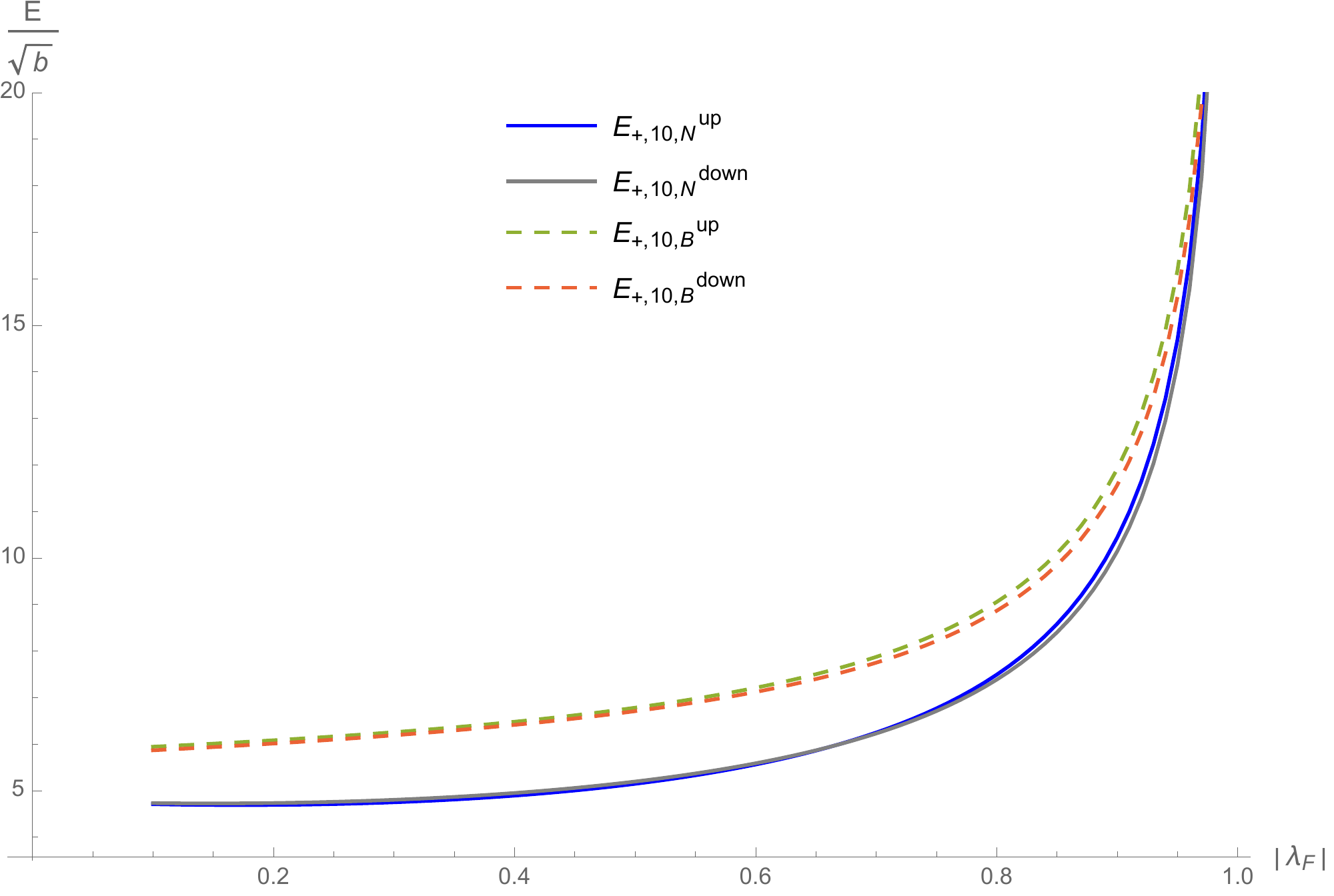}
	\caption{Dashed lines are from bosonic perturbation theory. Solid lines are from numerical analysis..}
	\label{B1}
\end{subfigure}\hspace{15pt}
\caption{Upper and lower critical chemical potentials of the $M=10$ positive energy band as  function of $\lambda_F$ for $\rm{sgn}( b\lambda_F)=1$ at $m_B^{cri}=-\sqrt{|b|}$.}
\label{graph1}
\end{figure}

Let us start by noting that the width of the $M^{th}$ positive level, $\Delta_+(M)$ has the same sign as $-\lambda_F b$ at small $\lambda_F$ (see \eqref{energydifffp}) but has is positive 
at $|\lambda_F|$ near unity (see  \eqref{energydiffB}).It follows that this width must change sign at a finite value $\lambda_F$. 
We illustrate how this happens in  Fig \ref{graph1}
where we separately plot $E^{\rm up}_{+10}$ 
and $E^{\rm low}_{+10}$ at the fixed value of 
$m_B^{{\rm cri}}=-\sqrt{|b|}$, as a function of 
$\lambda_F$.  In subfigure \ref{A1} we compare 
our numerical results (subscript $N$ on the graphs)
with the results of small $\lambda_F$ perturbation 
theory (subscript F on the graphs). Note that 
the agreement is very good at small values of 
$\lambda_F$. Note also that the numerical curves for $E^{\rm up}_{+10}$ and $E^{\rm low}_{+10}$ 
do indeed cross at a value of $\lambda_F$ of 
order 0.7. In sub Figure \ref{B1} we have plotted 
the same quantities at larger values of $\lambda_F$
and have compared the results with those of 
small $\lambda_B$ perturbation theory (subscript 
$B$ on the graphs.) Note again the excellent 
agreement at small $|\lambda_B|$.  

\begin{figure}[!h]
	\begin{subfigure}{0.5\textwidth}
	\includegraphics[width=3in,height=2.5in]{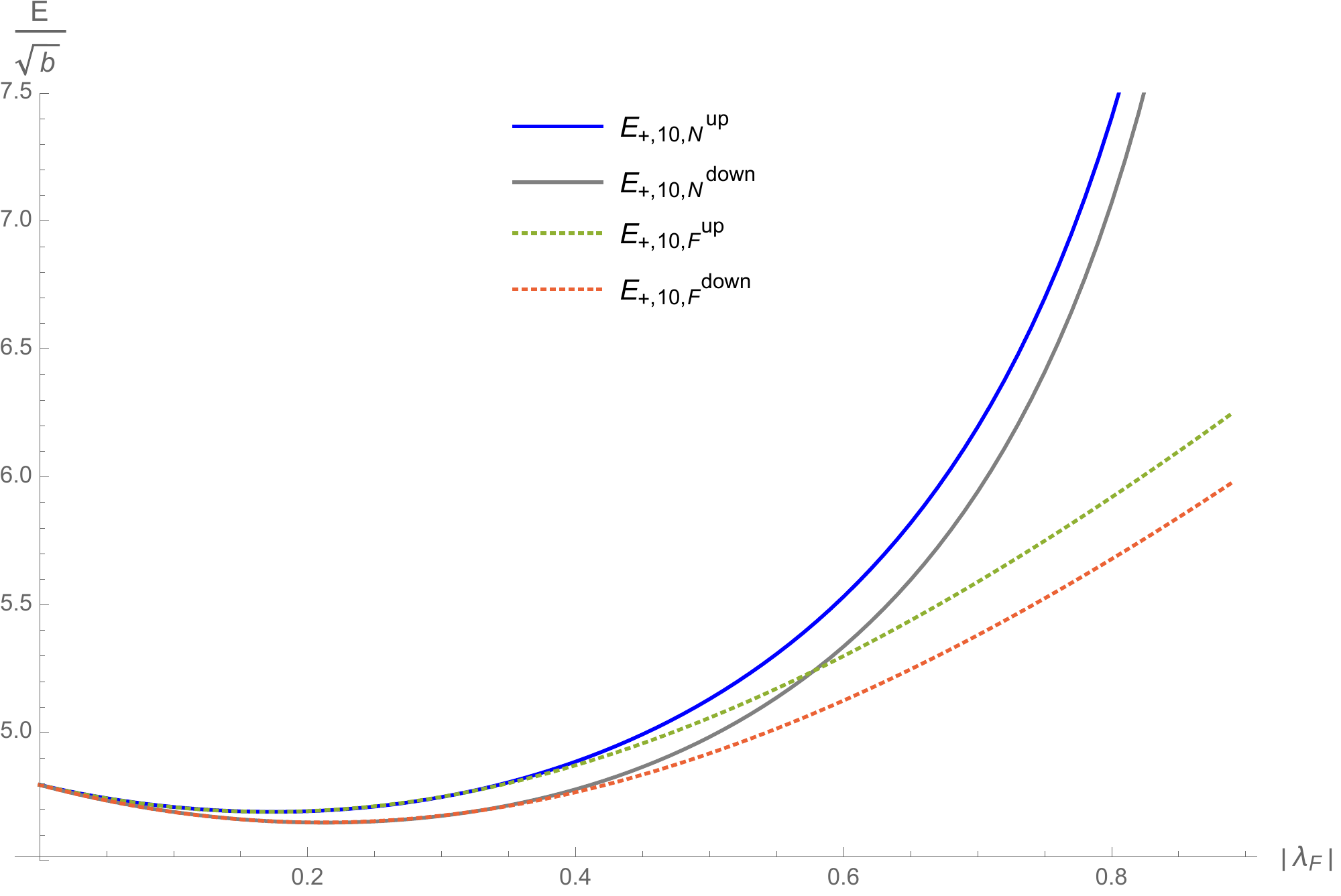}
		\caption{Dashed lines are from fermionic perturbation theory. Solid lines are from numerical analysis.}
		\label{A1}
	\end{subfigure}\hspace{15pt}
	~~~~~~~~
\begin{subfigure}{0.5\textwidth}
	\includegraphics[width=3in,height=2.5in]{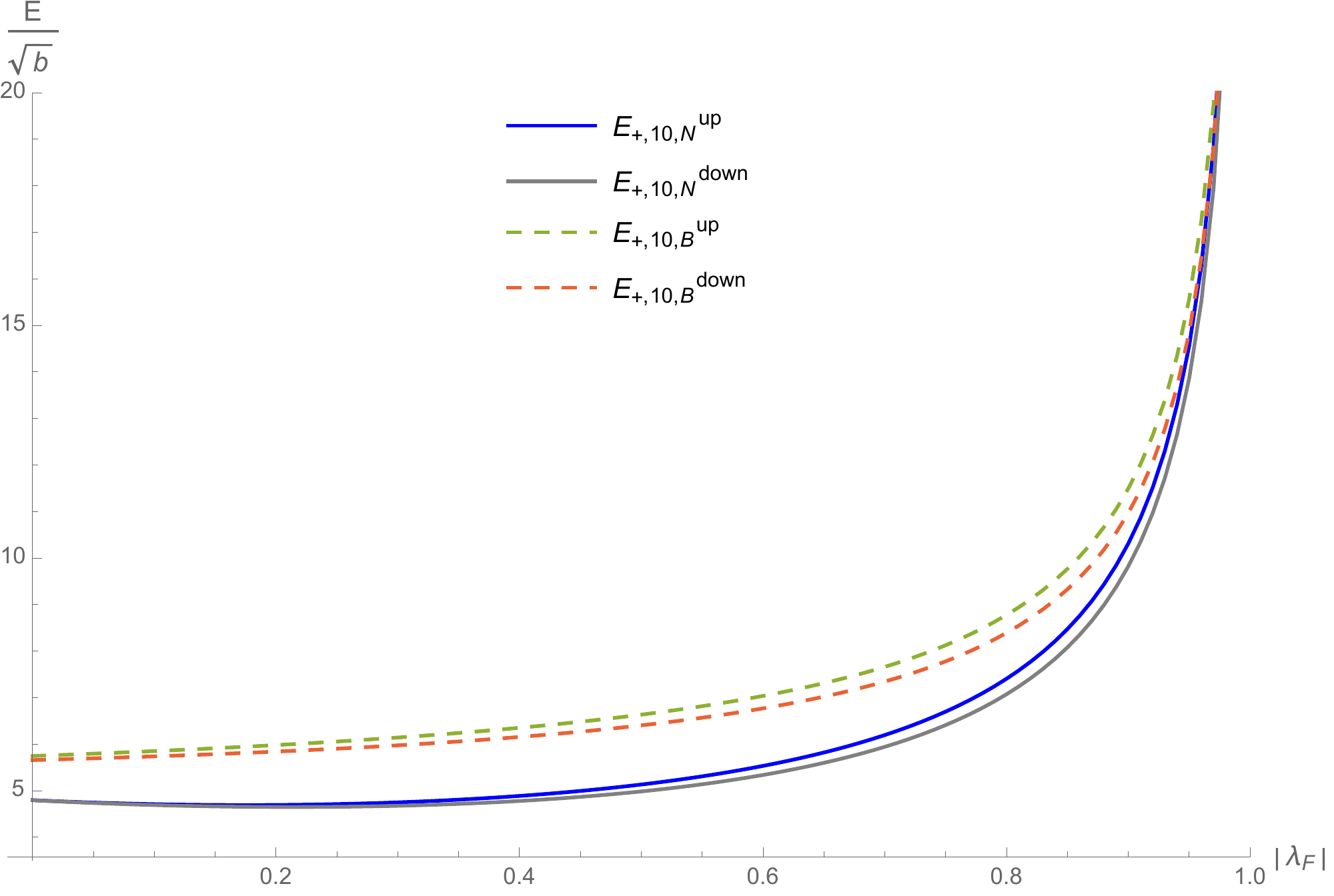}
	\caption{Dashed lines are from bosonic perturbation theory. Solid lines are from numerical analysis.}
	\label{B1}
\end{subfigure}\hspace{15pt}
\caption{Upper and lower critical chemical potentials for the $M=10$ positive energy band function of $\lambda_F$ for $\rm{sgn}( b\lambda_F)=-1$, at fixed $m_B^{cri}=-\sqrt{|b|}$.}
\label{grapho}
\end{figure}

In Fig. \ref{grapho} we once again plot
$E^{\rm up}_{+10}$  and $E^{\rm low}_{+10}$ at 
$m_B^{{\rm cri}}=-\sqrt{|b|}$, as a function of 
$\lambda_F$, but this time for $\lambda_F$ negative. Note that in this case 
$E^{\rm up}_{+10}>E^{\rm low}_{+10}$ for all 
values of $\lambda_F$; the curves never cross. 
Once again note the excellent agreement 
between the numerical curves and perturbation 
theory in its regime of validity. 

\begin{figure}[!h]
	\begin{subfigure}{0.5\textwidth}
	\includegraphics[width=3in,height=2.5in]{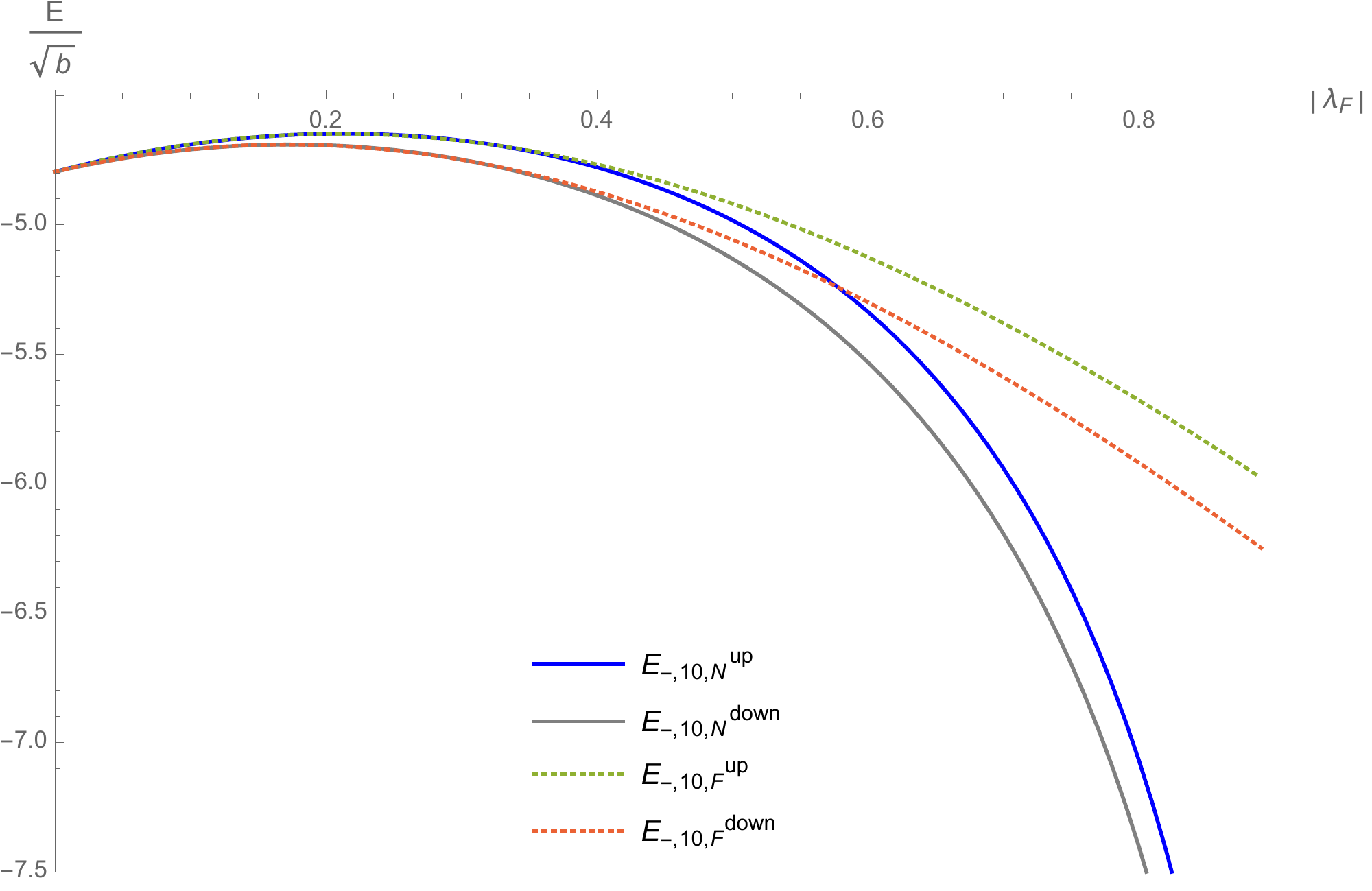}
		\caption{Dashed lines are from fermionic perturbation theory. Solid lines are from numerical analysis.}
		\label{A1}
	\end{subfigure}\hspace{15pt}
	~~~~~~~~
\begin{subfigure}{0.5\textwidth}
	\includegraphics[width=3in,height=2.5in]{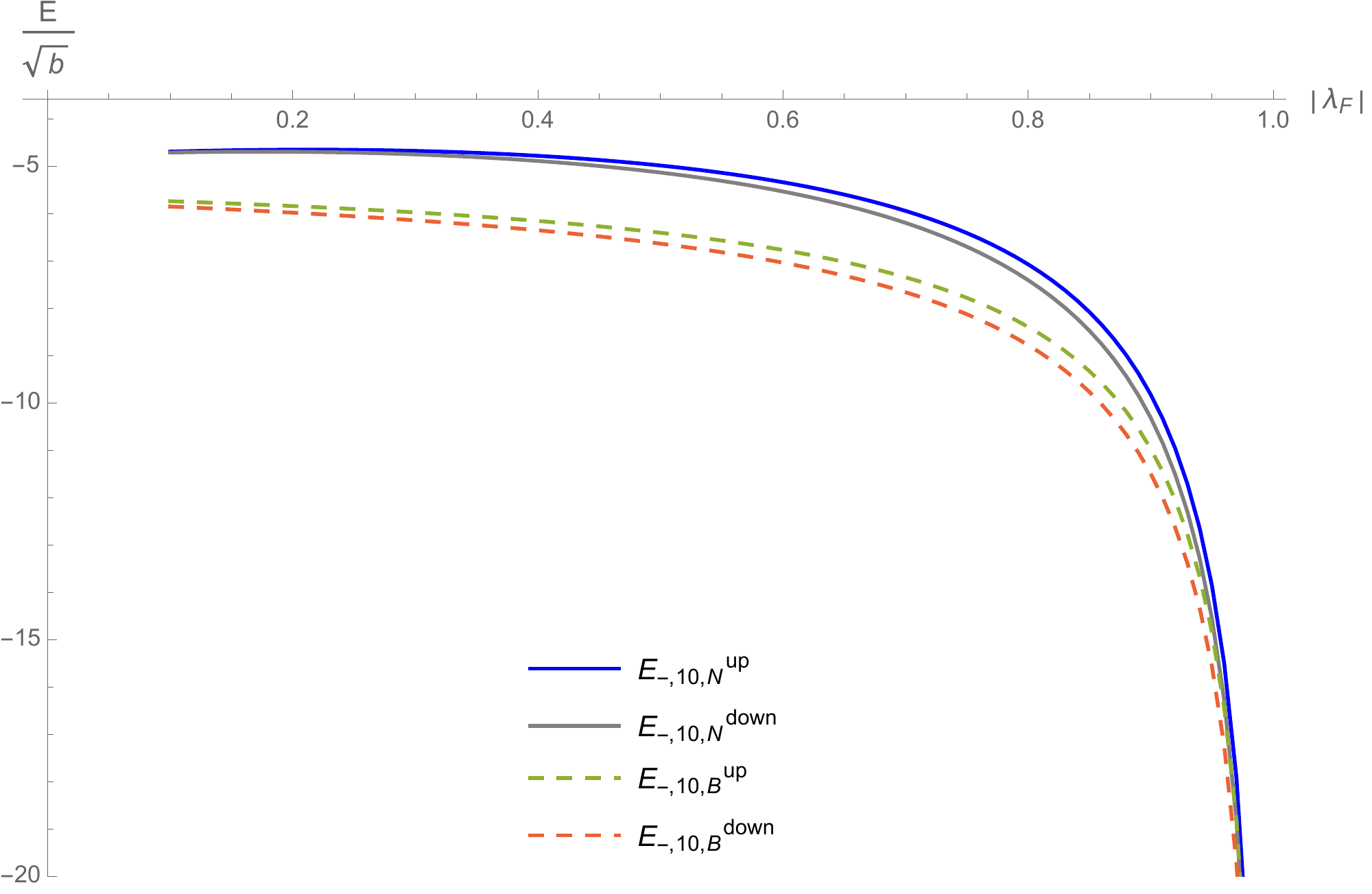}
	\caption{Dashed lines are from bosonic perturbation theory. Solid lines are from numerical analysis.}
	\label{B1}
\end{subfigure}\hspace{15pt}
\caption{Critical chemical potentials for the 
$M=10$ negative energy band as a function of $\lambda_F$ for $\rm{sgn}( b\lambda_F)=1$ at 
$m_B^{cri}=-\sqrt{|b|}$.}
\label{grapht}
\end{figure}

\begin{figure}[!h]
	\begin{subfigure}{0.5\textwidth}
	\includegraphics[width=3in,height=2.5in]{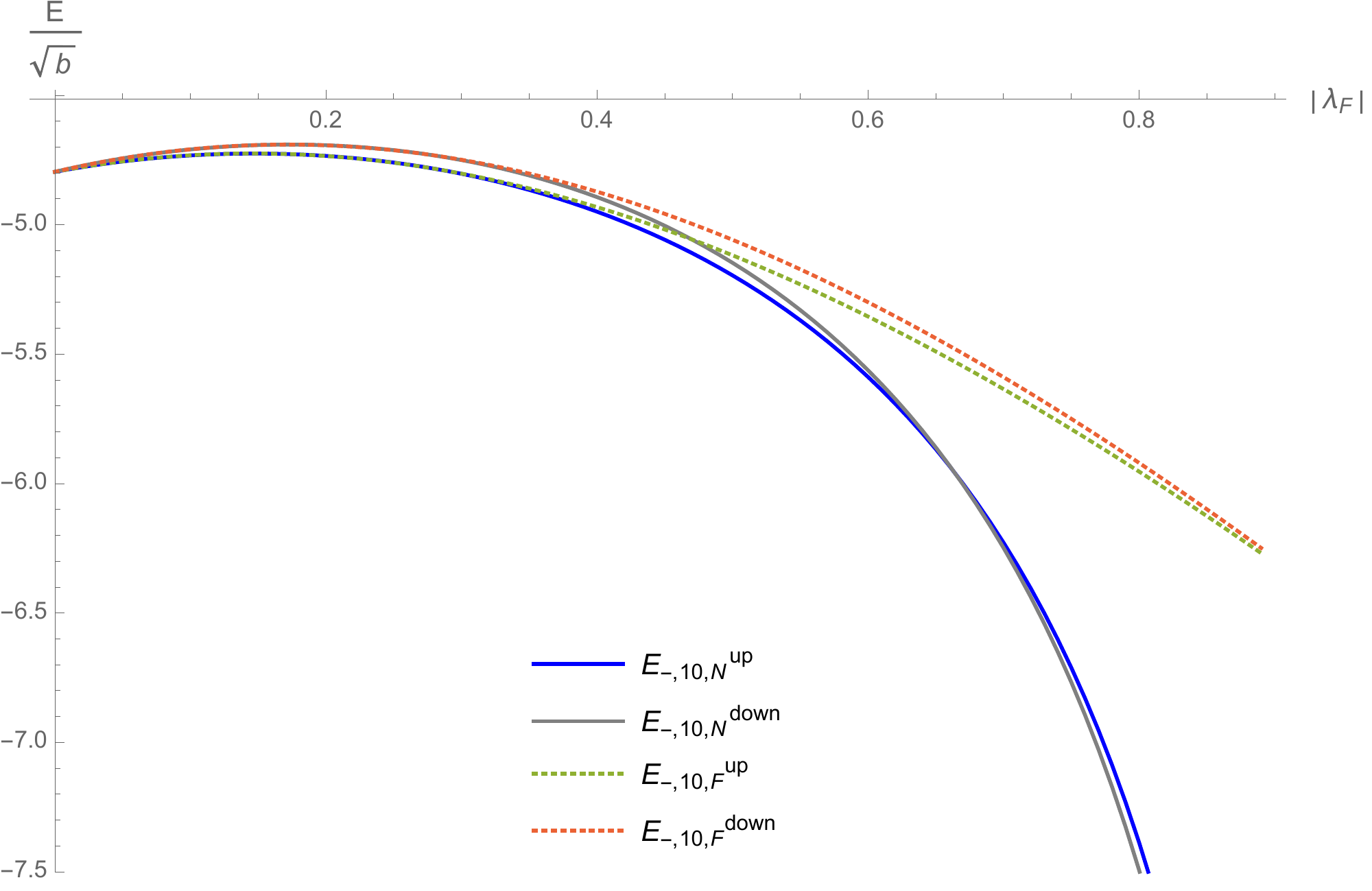}
		\caption{Dashed lines are from fermionic perturbation theory. Solid lines are from numerical analysis.}
		\label{A1}
	\end{subfigure}\hspace{15pt}
	~~~~~~~~
\begin{subfigure}{0.5\textwidth}
	\includegraphics[width=3in,height=2.5in]{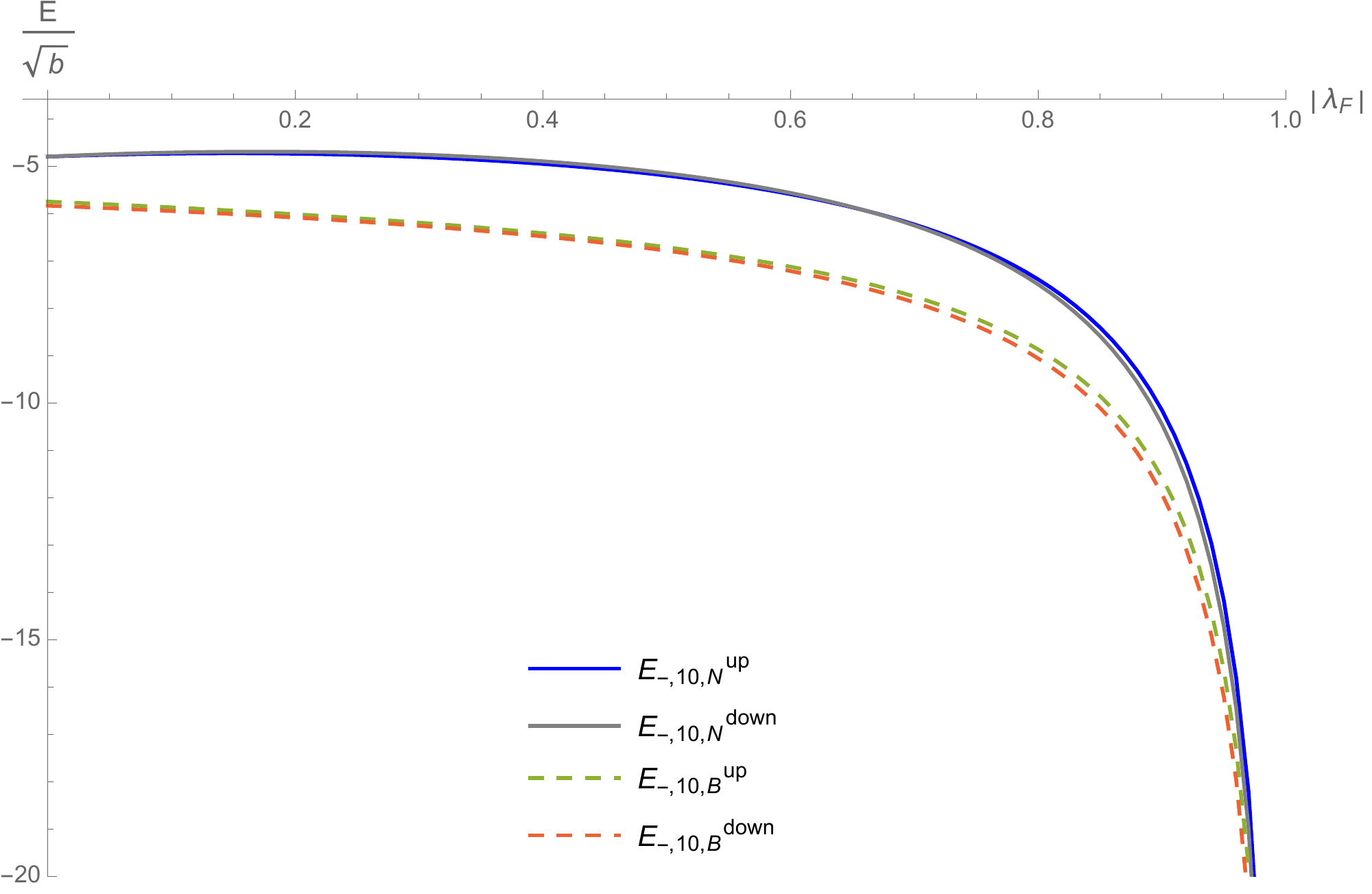}
	\caption{Dashed lines are from bosonic perturbation theory. Solid lines are from numerical analysis.}
	\label{B1}
\end{subfigure}\hspace{15pt}
\caption{The critical chemical potentials for 
the $M=10$ negative energy band as a function of $\lambda_F$ for $\rm{sgn}( b\lambda_F)=-1$ at fixed $m_B^{cri}=-\sqrt{|b|}$.}
\label{graphth}
\end{figure}

In Figure \ref{grapht} and Fig \ref{graphth} 
we turn to the study 
of negative energy bands. In these two graphs we plot $E^{\rm up}_{-10}$ and $E^{\rm low}_{-10}$ (note that $E^{\rm up}_{-10}<E^{\rm low}_{-10}$ but 
$|E^{\rm up}_{-10}|>|E^{\rm low}_{-10}|$) as a 
function of $\lambda_F$ respectively for positive 
and negative values of $\lambda_F b$. As 
could have been anticipated by comparing 
\eqref{energydifffn} and \eqref{energydiffB}, 
in this case the curves cross when $\lambda_F b$
is negative (Fig \ref{graphth} ) but not when it is positive (Fig. \ref{grapht}). Note again the 
excellent agreement with perturbation theory
in the appropriate parameter ranges.

\vspace{30pt}

\begin{figure}[!h]
	\begin{subfigure}{0.5\textwidth}
	\includegraphics[width=3in,height=2.5in]{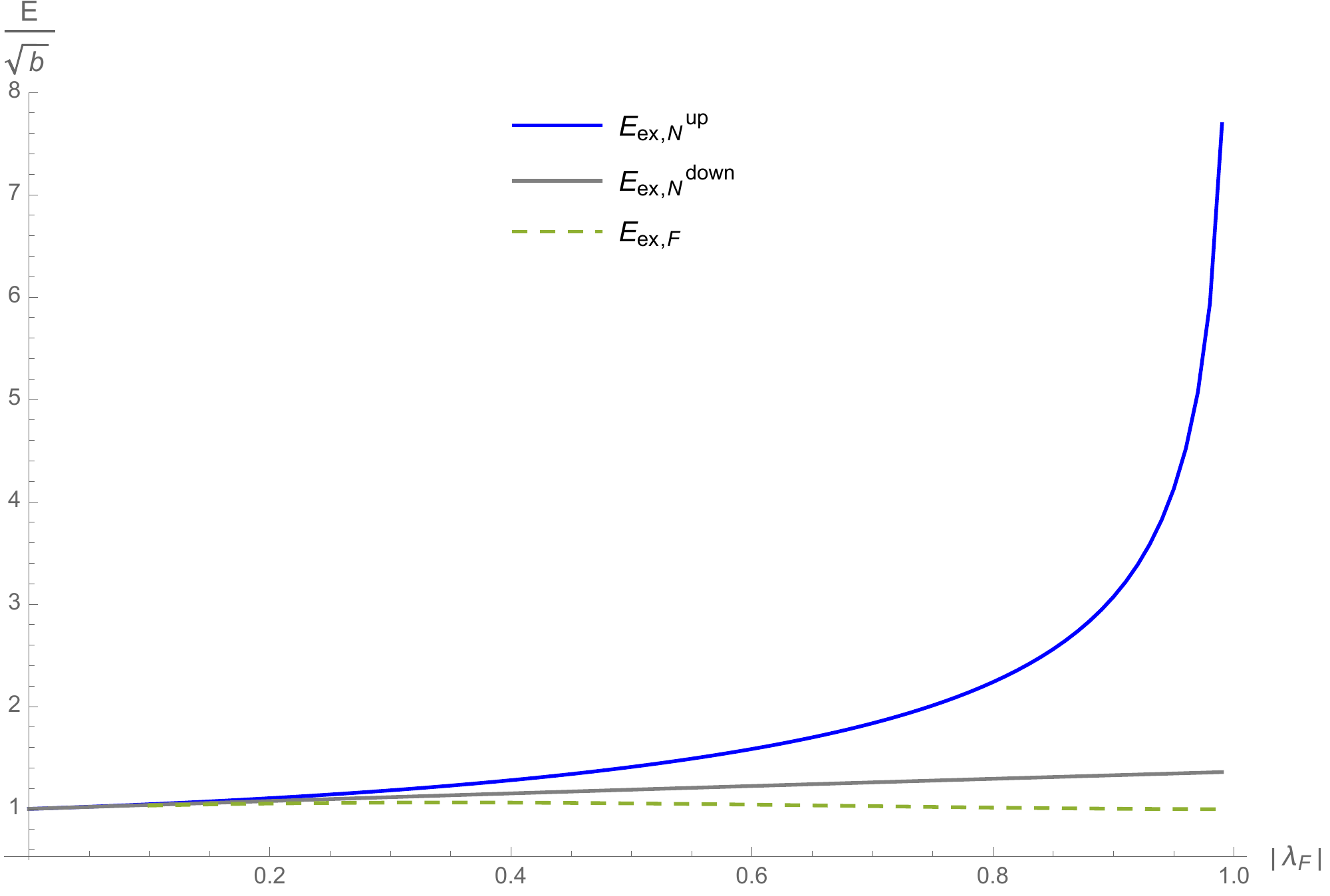}
		\caption{Dashed lines are from fermionic perturbation theory. Solid lines are from numerical analysis.}
		\label{A1}
	\end{subfigure}\hspace{15pt}
	~~~~~~~~
\begin{subfigure}{0.5\textwidth}
	\includegraphics[width=3in,height=2.5in]{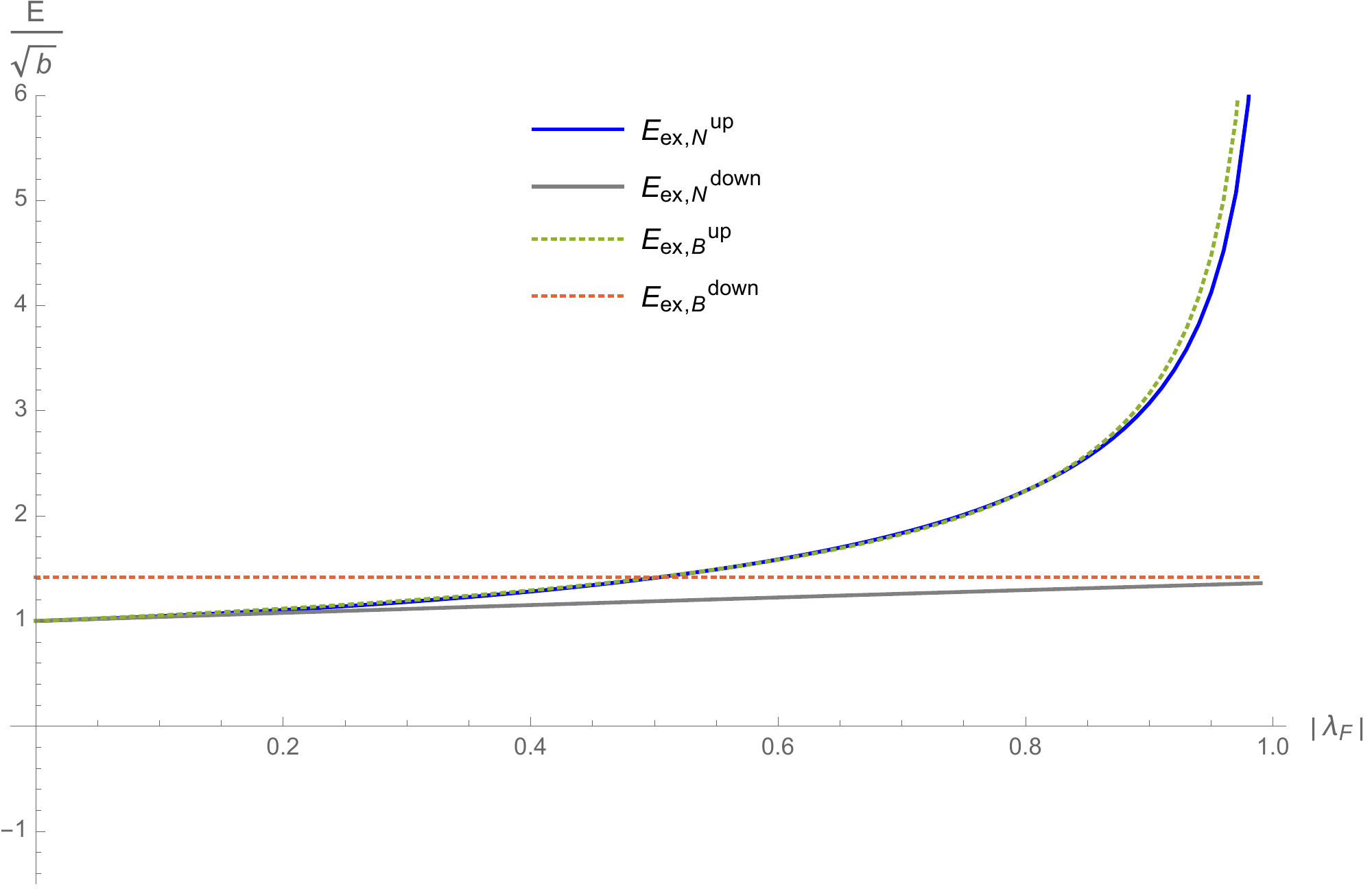}
	\caption{Dashed lines are from bosonic perturbation theory. Solid lines are from numerical analysis.}
	\label{B1}
\end{subfigure}\hspace{15pt}
\caption{Exceptional energy band as a function of $\lambda_F$ for $\rm{sgn}( b\lambda_F)=1$, $\rm{sgn}(m_F\lambda_F)= 1$ at $m_B^{cri}=
	\sqrt{|b|}$.}
\label{grapheo}
\end{figure}

\begin{figure}[!h]
	\begin{subfigure}{0.5\textwidth}
	\includegraphics[width=3in,height=2.5in]{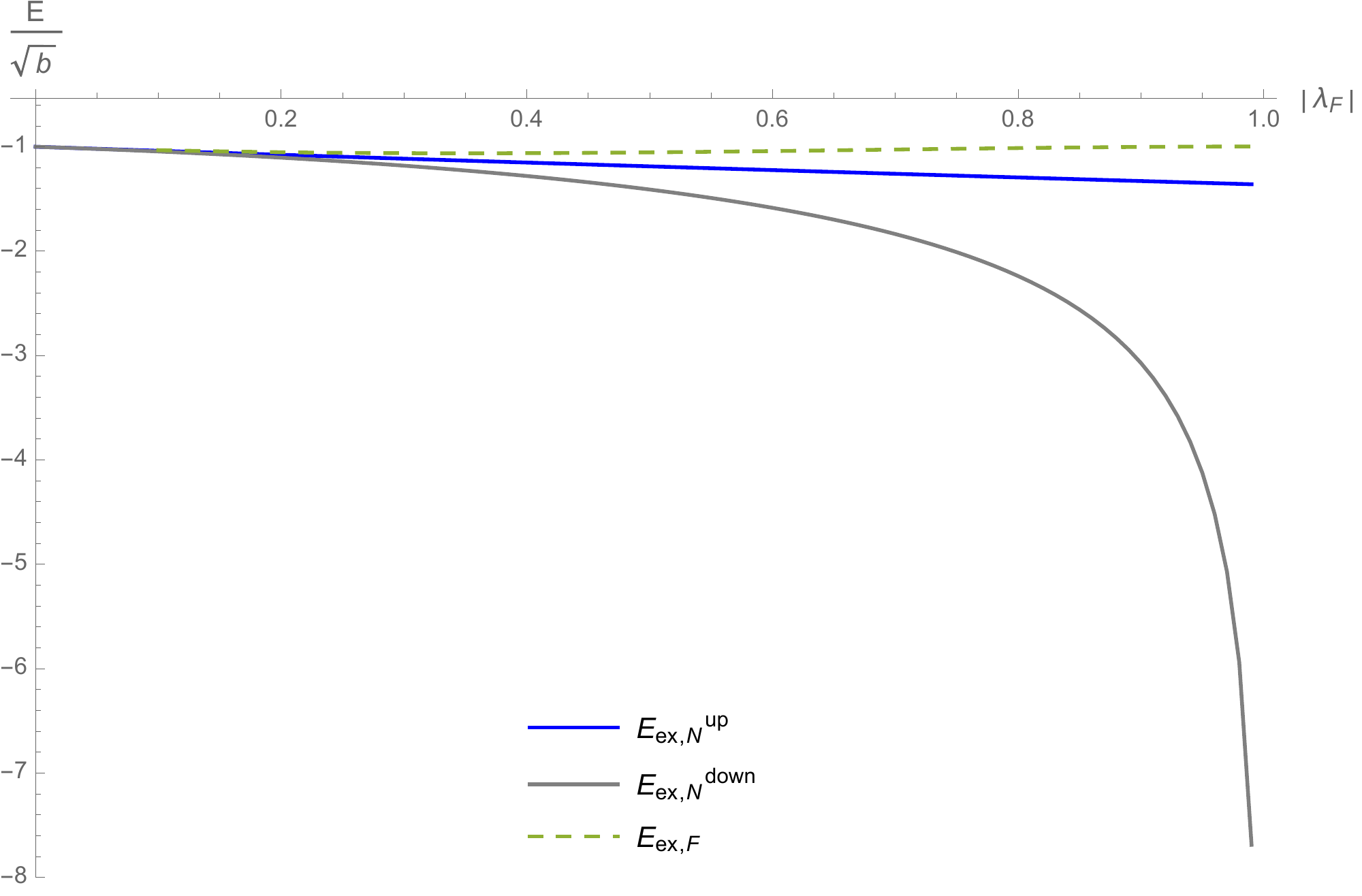}
		\caption{Dashed lines are from fermionic perturbation theory. Solid lines are from numerical analysis.}
		\label{A1}
	\end{subfigure}\hspace{15pt}
	~~~~~~~~
\begin{subfigure}{0.5\textwidth}
	\includegraphics[width=3in,height=2.5in]{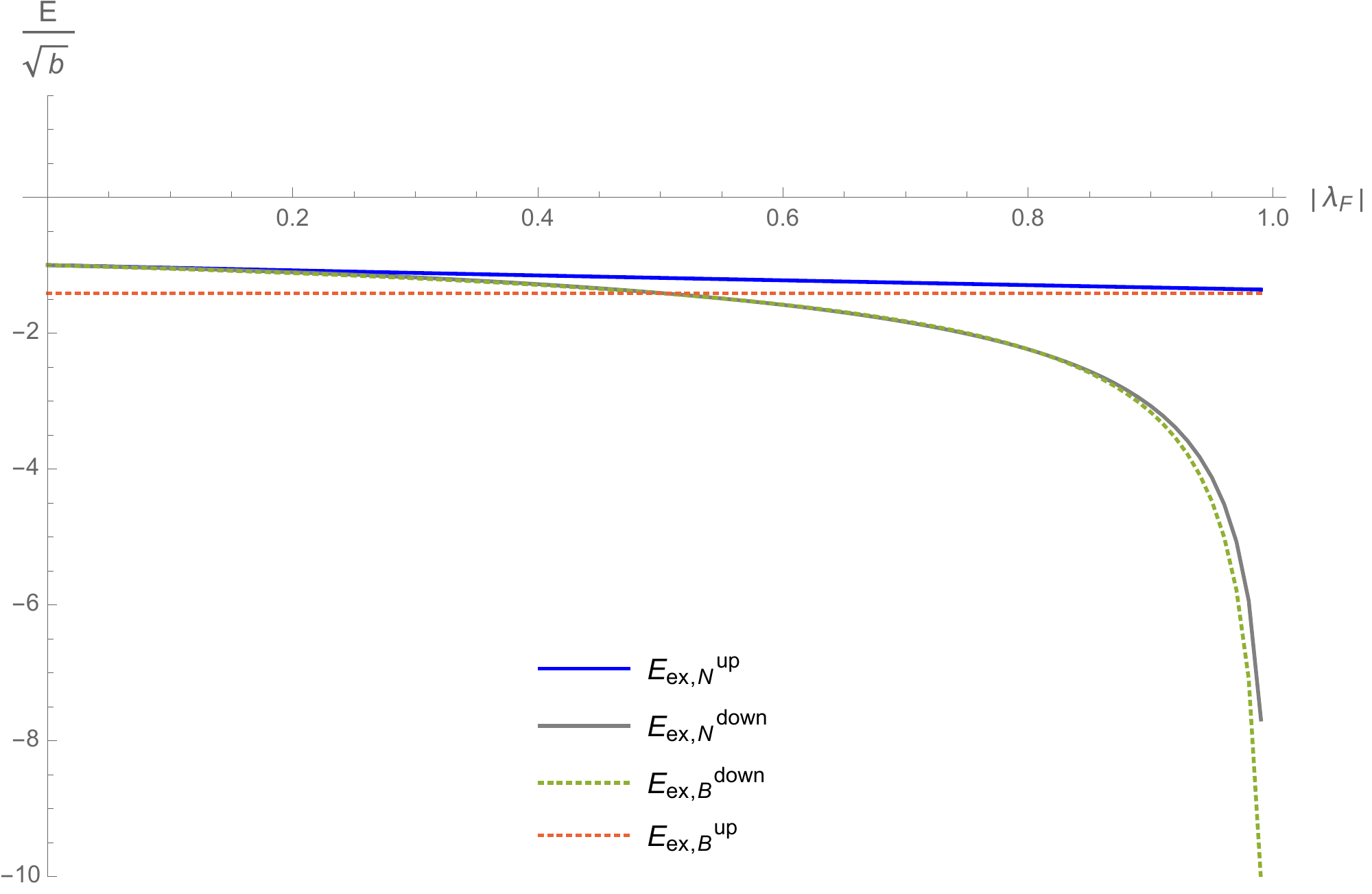}
	\caption{Dashed lines are from bosonic perturbation theory. Solid lines are from numerical analysis.}
	\label{B1}
\end{subfigure}\hspace{15pt}
\caption{Exceptional energy band as a function of $\lambda_F$ for $\rm{sgn}( b\lambda_F)=-1$, $\rm{sgn}(m_F\lambda_F)= 1$ at $m_B^{cri}=\sqrt{|b|}$.}
\label{graphet}
\end{figure}

\begin{figure}[!h]
	\begin{subfigure}{0.5\textwidth}
	\includegraphics[width=3in,height=2.5in]{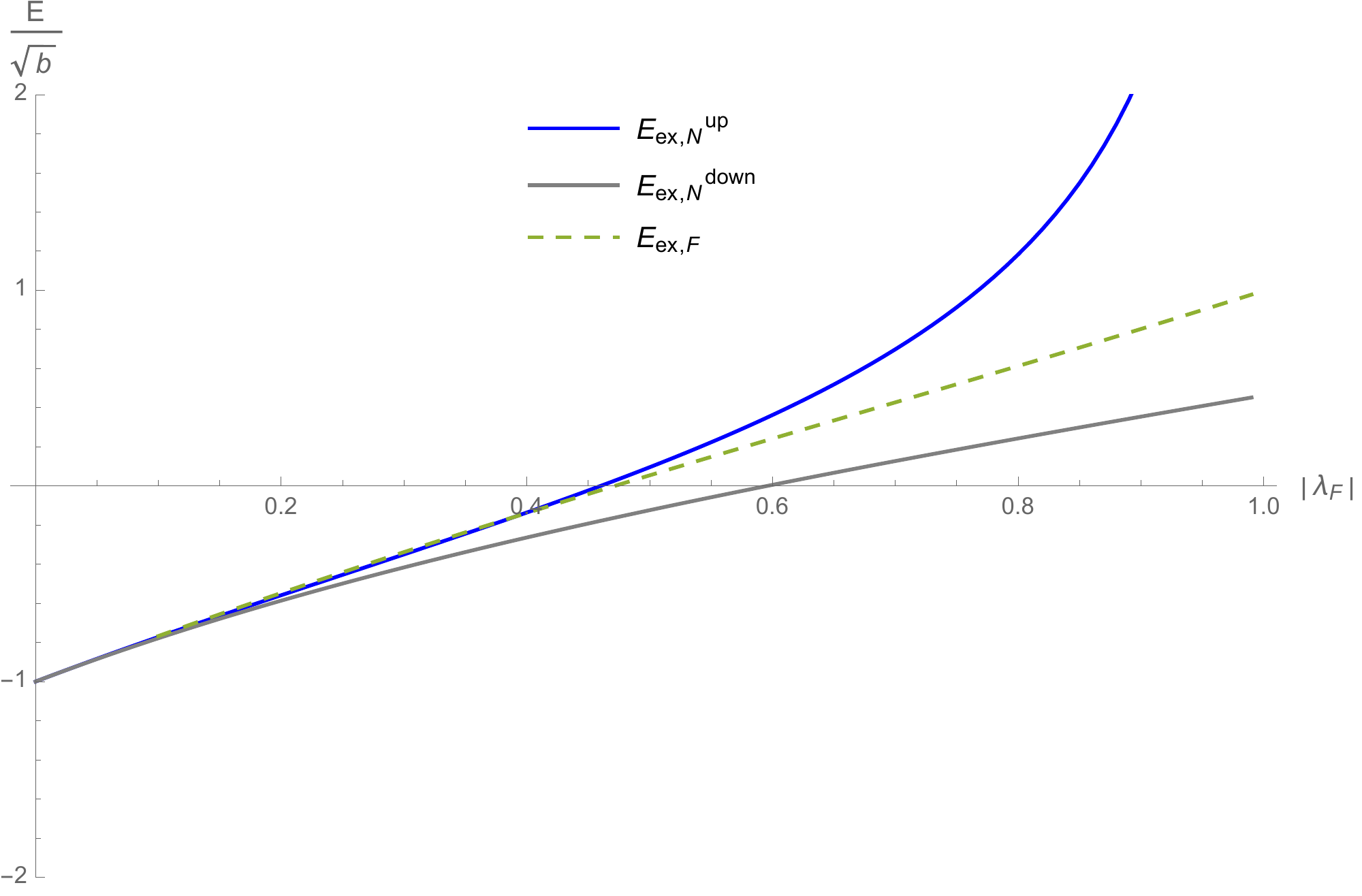}
		\caption{Dashed lines are from fermionic perturbation theory. Solid lines are from numerical analysis.}
		\label{A1}
	\end{subfigure}\hspace{15pt}
	~~~~~~~~
\begin{subfigure}{0.5\textwidth}
	\includegraphics[width=3in,height=2.5in]{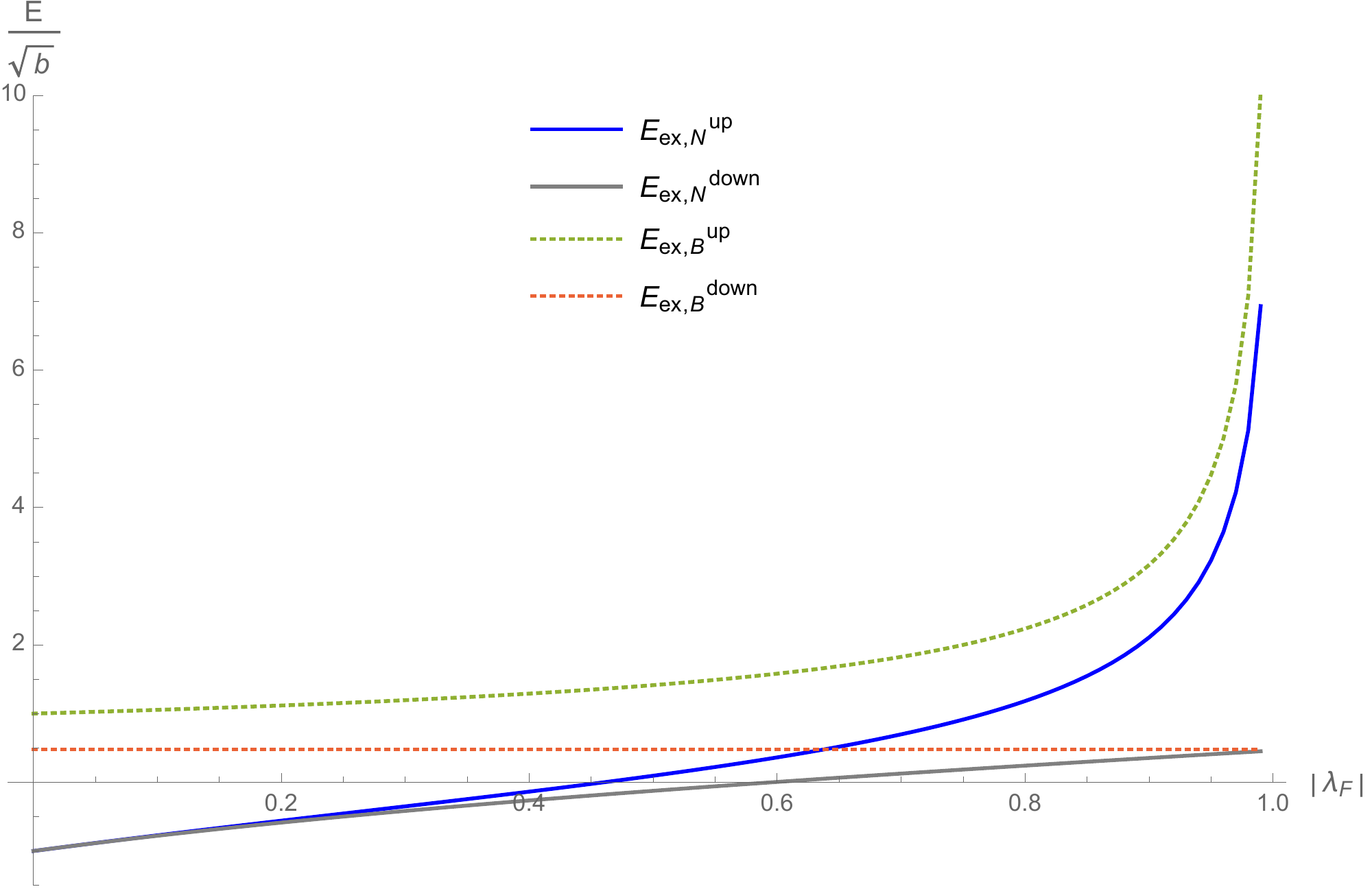}
	\caption{Dashed lines are from bosonic perturbation theory. Solid lines are from numerical analysis.}
	\label{B1}
\end{subfigure}\hspace{15pt}
\caption{Exceptional energy band as a function of $\lambda_F$ for $\rm{sgn}( b\lambda_F)=1$, $\rm{sgn}(m_F\lambda_F)= -1$ at 
	$m_B^{cri}=-\sqrt{|b|}$.}
\label{grapheth}
\end{figure}

\begin{figure}[!h]
	\begin{subfigure}{0.5\textwidth}
	\includegraphics[width=3in,height=2.5in]{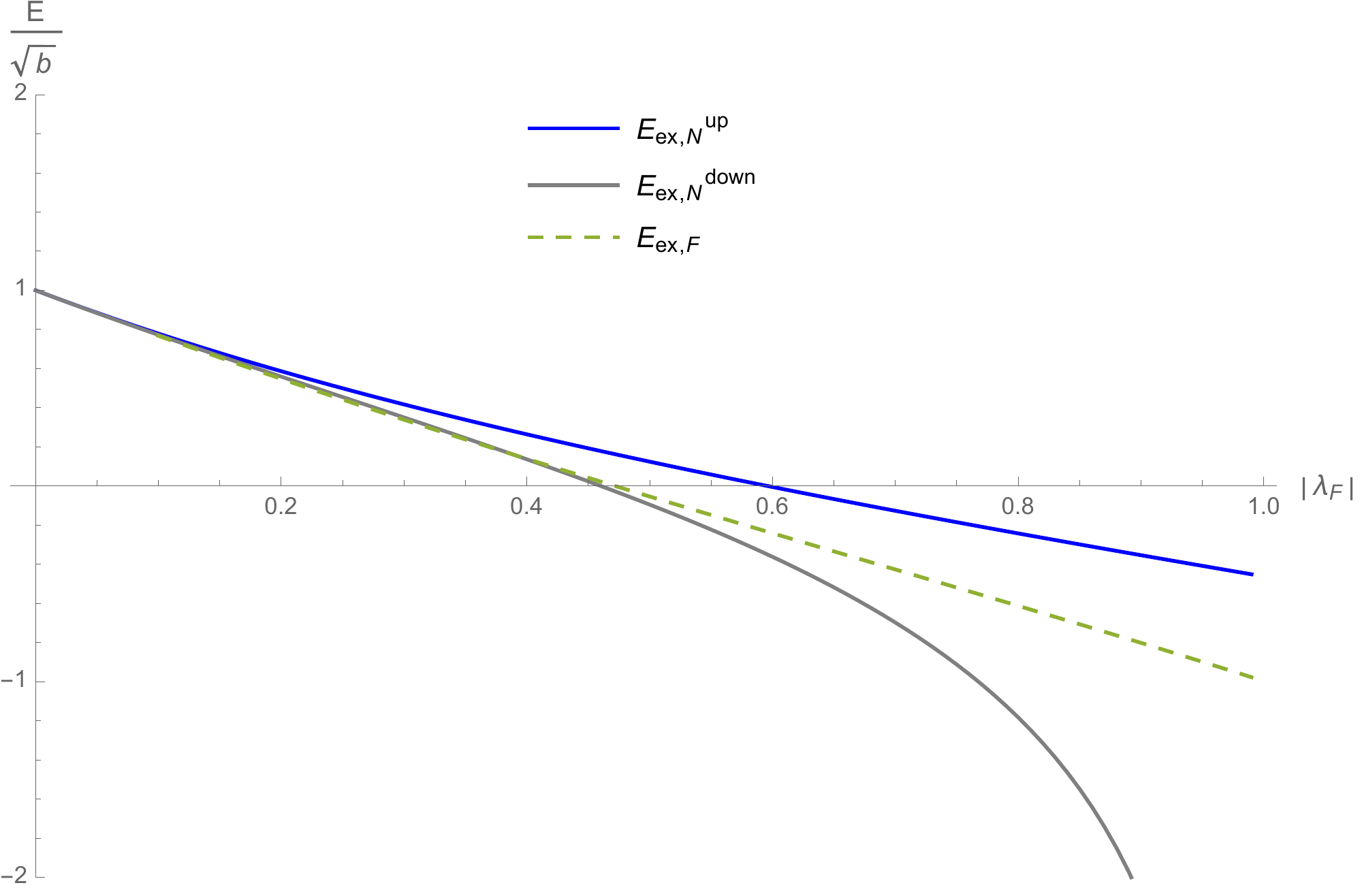}
		\caption{Dashed lines are from fermionic perturbation theory. Solid lines are from numerical analysis.}
		\label{A1}
	\end{subfigure}\hspace{15pt}
	~~~~~~~~
\begin{subfigure}{0.5\textwidth}
	\includegraphics[width=3in,height=2.5in]{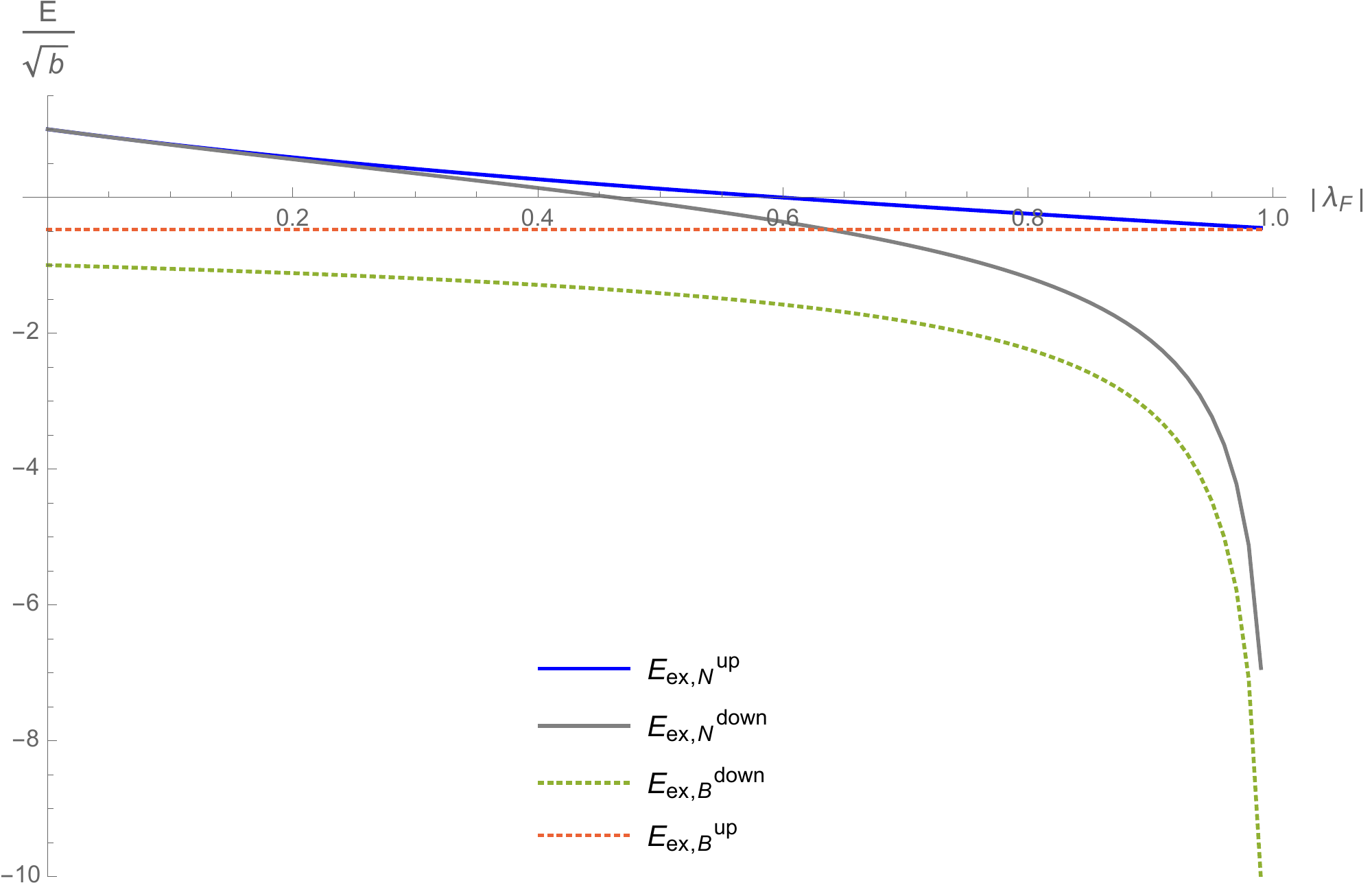}
	\caption{Dashed lines are from bosonic perturbation theory. Solid lines are from numerical analysis.}
	\label{B1}
\end{subfigure}\hspace{15pt}
\caption{Exceptional energy band as a function of $\lambda_F$ for $\rm{sgn}( b\lambda_F)=-1$, $\rm{sgn}(m_F\lambda_F)= -1$ at 
	$m_B^{cri}=-\sqrt{|b|}$.}
\label{graphef}
\end{figure}

Finally we turn to a study of the exceptional 
bands. In Figures \ref{grapheo}, \ref{graphet}, 
\ref{grapheth} and \ref{graphef} respectively
we plot the critical energies $E_{ex}^{\rm up}$ a
and $E_{ex}^{\rm down}$ of the exceptional 
energy band the following four cases. First when ${\rm sgn}(\lambda_F b)=1$ and ${\rm sgn}(M_F \lambda_F)=1$. Second when ${\rm sgn}(\lambda_F b)=-1$ and ${\rm sgn}(M_F \lambda_F)=1$. Third 
when ${\rm sgn}(\lambda_F b)=1$ and ${\rm sgn}(M_F \lambda_F)=-1$. Fourth when ${\rm sgn}(\lambda_F b)=-1$ and ${\rm sgn}(M_F \lambda_F)=-1$. Note 
again the reasonable agreement with perturbation 
theory. Note also that in every case that 
$E_{ex}^{\rm down}$ is positive at small $|\lambda_B|$, it remains finite as $|\lambda_B|\to 0$, while in every case that $E_{ex}^{\rm up}$ is
negative at small $\lambda_B$, it remains 
finite as $|\lambda_B|\to 0$.  Note also that 
the curves cut the x axis in the last two cases.

\vspace{30pt}

\vspace{30pt}

\clearpage

\section{Regular bosons coupled with Chern Simons gauge field} \label{regbo}

In the previous section we have presented an 
exhaustive analysis of Greens function and the 
single particle eigen energies for the regular 
fermion theory in a magnetic field background, 
both with and without a chemical potential. 
As we have reviewed in the introduction, however, 
the regular fermion theory is conjectured to be 
dual to the theory of critical bosons. As a 
check both of this conjectured duality - but also 
of the fermionic results of the previous 
section - in this section we study the propagator 
and single particle energy eigenstates of the 
critical boson theory in the background of 
a global symmetry magnetic field. For the reason 
explained in the introduction, in this section (i.e. the relative difficult of solving the gap 
equations in the presence of a Bose condensate) 
in this section we work only at zero chemical potential, leaving the generalization to nonzero 
chemical potential to future work.

\subsection{Gap equations}
We consider following Euclidean action for a fundamental boson coupled with Chern Simons gauge field
\begin{equation}\label{ActionScalar}
	\begin{aligned}
		S_B= \frac{i\kappa_B}{4 \pi}\int Tr(AdA-\frac{2i}{3}A^3)+\int \left(\tilde{D}^\mu \bar{\phi}\tilde{D}_\mu \phi+m_B^2 \bar{\phi}\phi + \frac{4 \pi b_4}{k} (\bar{\phi}\phi )^2 + \frac{4 \pi^2 x_6}{k^2} (\bar{\phi}\phi )^3\right)
	\end{aligned}
\end{equation}
Here $A$ is a $U(N_B)$ connection, and we have turned on a suitably normalized background (topological) $U(1)$ gauge field $a_\mu$. In the large $N_B$ limit under study the fluctuations of the $U(1)$ part of the gauge field  are subdominant in a $\frac{1}{N_B}$ expansion, and so the $U(1)$ part of the $U(N_B)$ gauge field can effectively be replaced by its saddle point value. This saddle 
point value is obtained from the equation of motion 
of the $U(1)$ gauge field in the presence of the 
source coupling to the topological current. Ignoring the matter contribution to this equation 
of motion (which we know to be suppressed in the 
large $N_B$ limit) and with a suitable choice of 
normalization for $a$ we find that the equation of motion simply identifies the $U(1)$ part of the 
dynamical gauge field with $a_\mu$. When the 
dust settles, to the accuracy we work in this 
paper, the bosons can be thought of as charged 
under the gauge group $SU(N_B)$ and the covariant 
derivative that appears in \eqref{ActionScalar}
may be thought of as acting as 
\begin{equation} \label{Ddef}
\tilde{D}_\mu \phi  = \left( \partial_\mu -i A_\mu  -i a_\mu \right) 
\phi
\end{equation}
in close analogy with the previous section. 

Below we study the theory leading order in large $N$ approximation, but all order in t 'Hooft coupling $\lambda_B=\frac{N_B}{\kappa_B}$. The saddle point equations in $A_-=0$ gauge (See Appendix \ref{bosonicgapeq} for details) take the form
\begin{equation}\label{bgapeq1}
	\begin{aligned}
		(-D_x^2+m_B^2)\alpha(x,y)+\int d^3z \Sigma(x,z)\alpha(z,y)=\delta^{(3)}(x-y)
	\end{aligned}
\end{equation}
where $D_\mu$ is the `background covariant derivative' 
\begin{equation}
	D_\mu \phi  = \left( \partial_\mu   -i a_\mu \right) \phi 
\end{equation}

\begin{figure}[!h]
	\begin{subfigure}{0.5\textwidth}
	\includegraphics[width=3in,height=1.5in]{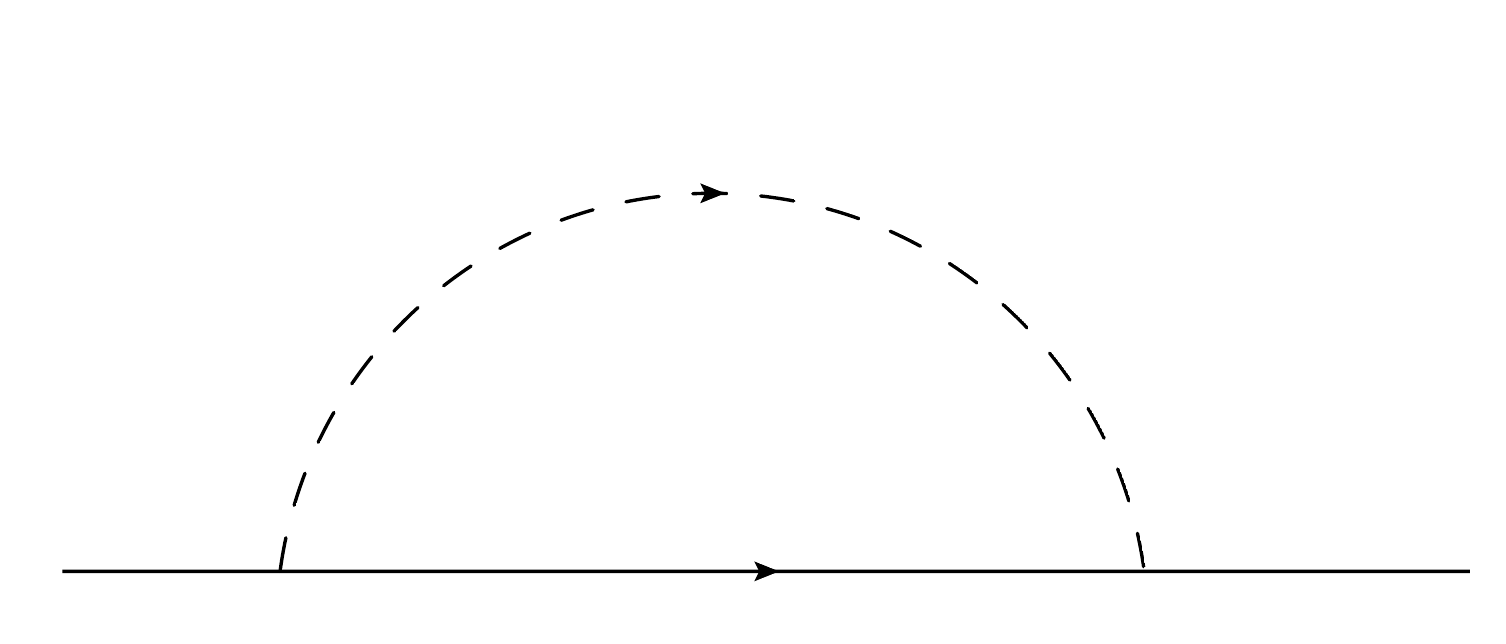}
		\caption{Contribution of this diagram is presented in \eqref{bgapeq21}. Solid lines
		are scalar propagators while dashed lines 
	are gauge boson propagators.}
		\label{A1}
	\end{subfigure}\hspace{15pt}
	~~~~~~~~
\begin{subfigure}{0.5\textwidth}
	\includegraphics[width=3in,height=1.5in]{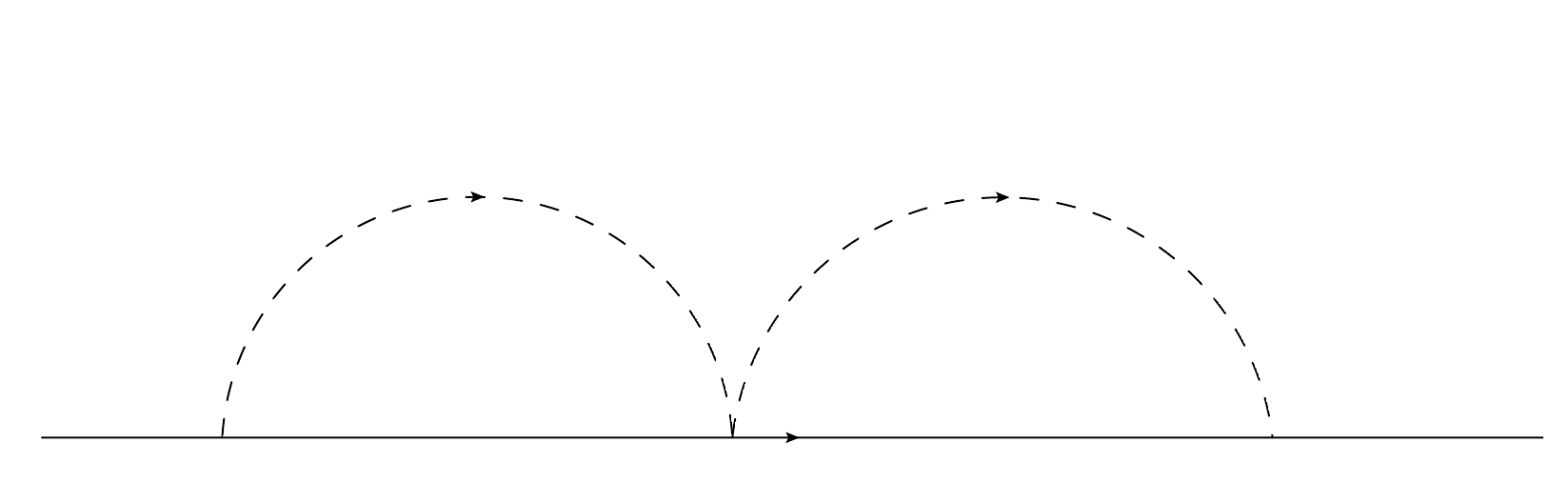}
	\caption{Contribution of this diagram is presented in \eqref{bgapeq22A}. Solid lines 
    scalar field propagators while dotted lines 
are gauge boson propagators.}
	\label{B1}
\end{subfigure}\hspace{15pt}
\caption{The `one loop' and one of the `two loop' 
graphs that contribute to the bosonic self energy. 
Note that the contribution of the one loop graph vanishes in the absence of a magnetic field. }
\label{Fgraph1}
\end{figure}

\begin{figure}[!h]
	\begin{subfigure}{0.5\textwidth}
	\includegraphics[width=3in,height=1.5in]{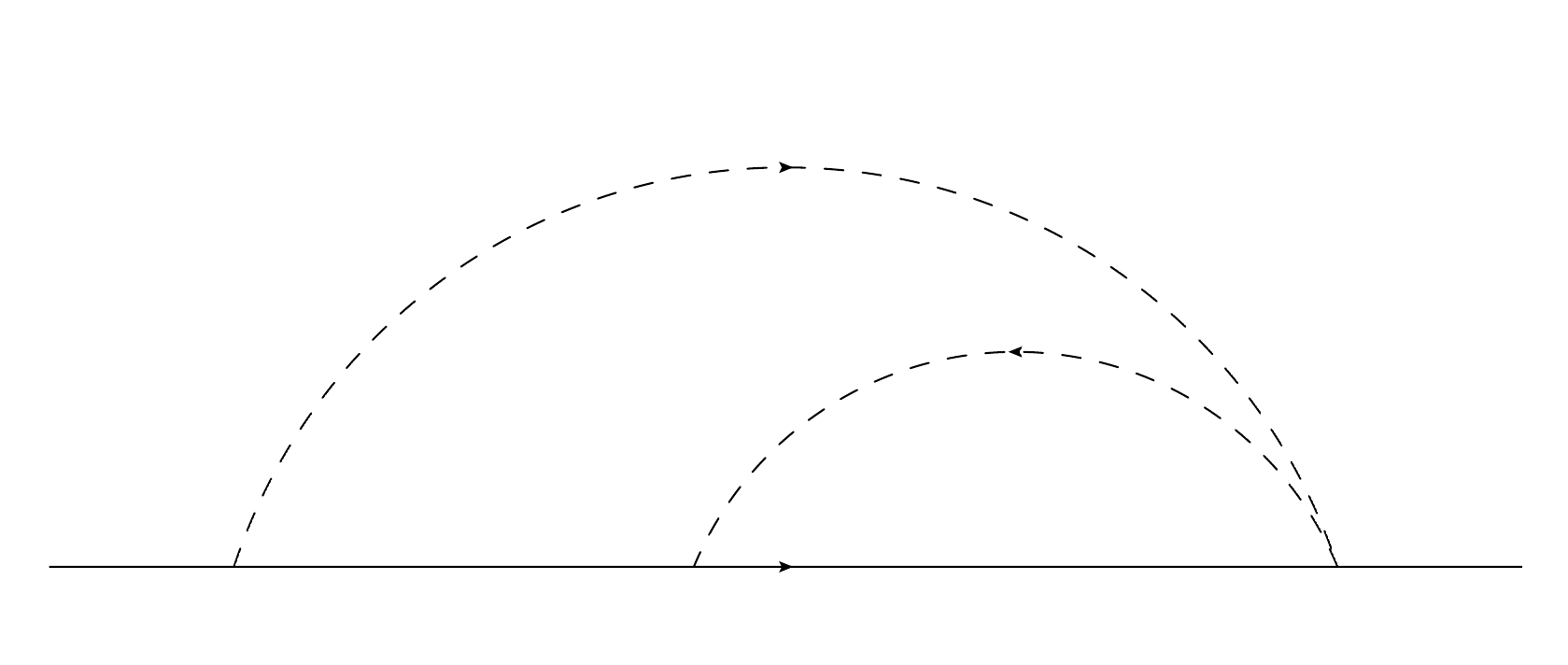}
		\caption{Contribution of this diagram is presented in \eqref{bgapeq22B}. Solid lines
			are scalar propagators while dashed lines 
			are gauge boson propagators.}
		\label{A1}
	\end{subfigure}\hspace{15pt}
	~~~~~~~~
\begin{subfigure}{0.5\textwidth}
	\includegraphics[width=3in,height=1.5in]{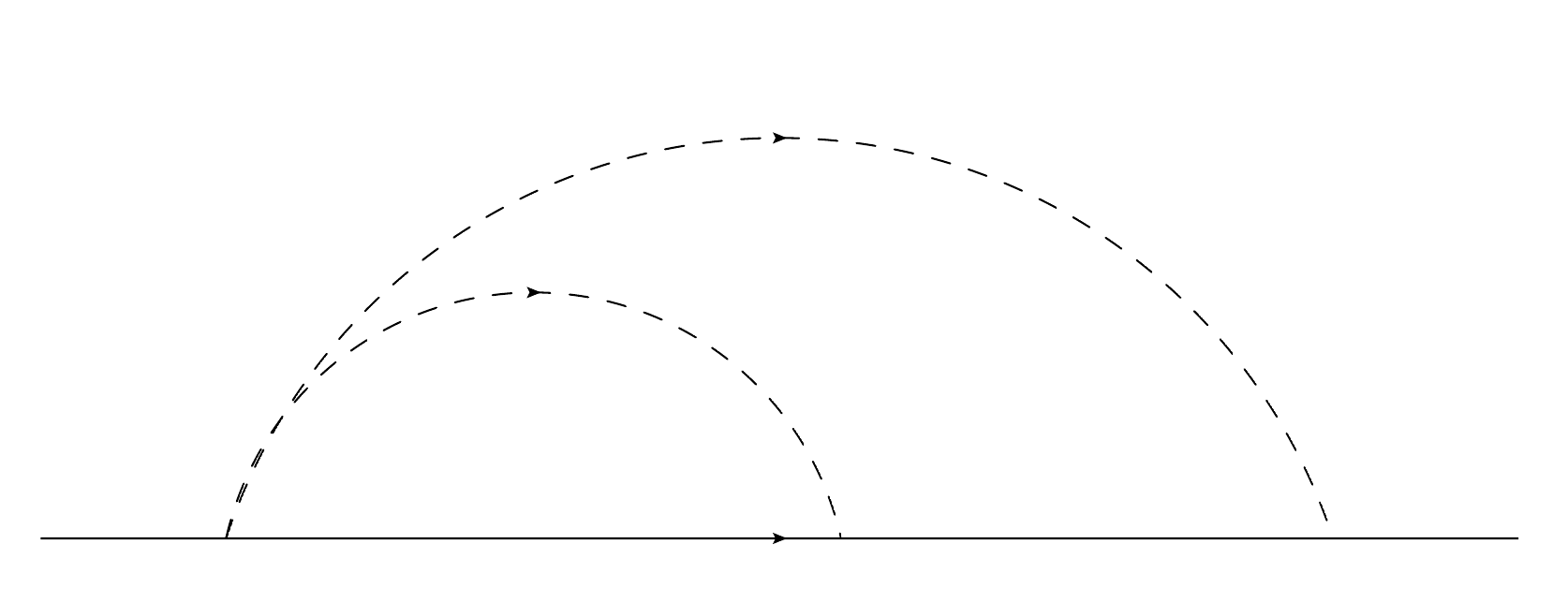}
	\caption{Contribution of this diagram is presented in \eqref{bgapeq22B'}. Solid lines 
	are scalar propagators while dotted lines 
are gauge boson propagators}
	\label{B1}
\end{subfigure}\hspace{15pt}
\caption{The remaining two `two loop' graphs that 
contribute to the bosonic self energy. The two graphs above contribute identically in the $b=0$ 
limit, but are distinct at nonzero $b$. While each of the graphs drawn above is individually complicated, their sum turns out to be relatively simple.}
\label{Fgraph2}
\end{figure}

As in the fermionic theory, the self energy $\Sigma(x,y)$ is the sum of a set of one 
(and in this case also two) loop diagrams graphs 
built using the exact propagator $\alpha$. 
We find  
\begin{equation}\label{bgapeq2}
	\begin{aligned}
		\Sigma(x,y) &=\Sigma_{1}(x_T,x_R)+\Sigma_{2A}(x_T,x_R)+\Sigma_{2B}(x_T,x_R)+\Sigma_{2B'}(x_T,x_R)\\
	\end{aligned}
\end{equation}
where the quantities $\Sigma_1$, $\Sigma_{2A}$ 
$\Sigma_{2B}$ and $\Sigma_{2C}$ are  the expressions computed by the Feynman graphs 
presented in Fig \ref{Fgraph1}, \ref{Fgraph2}.

The coordinates $x_T$ and $x_R$ that appear in 
\eqref{bgapeq2} refer respectively to centre of 
mass and relative coordinates
\begin{equation}\label{relcm}
	x_T=x+y,~~~ x_R=x-y
\end{equation}

The explicit expressions for
 $\Sigma_1$, $\Sigma_{2A}$ 
$\Sigma_{2B}$ and $\Sigma_{2C}$ 
take the form

\begin{equation}\label{bgapeq21}
	\begin{aligned}
		\Sigma_{1}(x_T,x_R) &=8 \pi \lambda_B (b_4 \alpha(x_T,x_R)\delta^{(3)}(x_R)+i \partial_{x_T^-} (\partial_{x_R^3}(G(x_T,x_R)\alpha(x_T,x_R) ) \\& ~~~~~~~~~~~~~ ~~~~~~~~~~~~~~~~ ~~~~~~~~~~~~ + G(x_T,x_R)\partial_{x_R^3}\alpha(x_T,x_R)))\\
	\end{aligned}
\end{equation}

\begin{equation}\label{bgapeq22A}
	\begin{aligned}
		\Sigma_{2A}(x_T,x_R) &=4 \pi^2 \lambda_B^2(x_6\alpha(x_T,x_R)^2\delta^{(3)}(x_R)+ \alpha(x_T,x_R)^2 \delta^{(3)}(x_R)\\& +4i \partial_{x_R^-} \int d^3z \ \alpha(x_{1,T},x_{1,R})\delta^{(3)}(x_{1,R})G(x_{2,T},x_{2,R})\alpha(x_{2,T},x_{2,R})\\
		&-4(\partial_{x_R^-}+\partial_{x_T^-})(\partial_{x_R^-}-\partial_{x_T^-})\int  d^3z \ \alpha(x_{1,T},x_{1,R})G(x_{1,T},x_{1,R})\alpha(x_{2,T},x_{2,R})G(x_{2,T},x_{2,R}))\\
	\end{aligned}
\end{equation}

\begin{equation}\label{bgapeq22B}
	\begin{aligned}
		\Sigma_{2B}(x_T,x_R) &=4 \pi^2 \lambda_B^2(x_6\alpha(x_T,x_R)^2\delta^{(3)}(x_R)\\
		&+4(\partial_{x_R^-}+\partial_{x_T^-})(\partial_{x_R^-}+\partial_{x_T^-})\int  d^3z \ \alpha(x_{1,T},x_{1,R})G(x_{T},x_{R})\alpha(x_{2,T},x_{2,R})G(x_{2,T},x_{2,R})\\
		&-6(\partial_{x_R^-}+\partial_{x_T^-})\int  d^3z \ \alpha(x_{1,T},x_{1,R})\partial_{x_R^-}G(x_{T},x_{R})\alpha(x_{2,T},x_{2,R})G(x_{2,T},x_{2,R})\\
		&-2(\partial_{x_R^-}+\partial_{x_T^-})\int  d^3z \ \alpha(x_{1,T},x_{1,R})G(x_{T},x_{R})\alpha(x_{2,T},x_{2,R})\partial_{x_{2,R}^-}G(x_{2,T},x_{2,R})\\
		&-8(\partial_{x_R^-}+\partial_{x_T^-})\int  d^3z \ \partial_{x_{1,T}^-}\alpha(x_{1,T},x_{1,R})G(x_{T},x_{R})\alpha(x_{2,T},x_{2,R})G(x_{2,T},x_{2,R})\\
		&+4\int  d^3z \ \partial_{x_{1,T}^-}\alpha(x_{1,T},x_{1,R})\partial_{x_R^-}G(x_{T},x_{R})\alpha(x_{2,T},x_{2,R})G(x_{2,T},x_{2,R})\\
		&+2\int  d^3z \ \alpha(x_{1,T},x_{1,R})\partial^2_{x_R^-}G(x_{T},x_{R})\alpha(x_{2,T},x_{2,R})G(x_{2,T},x_{2,R})\\
		&+\int  d^3z \ \alpha(x_{1,T},x_{1,R})\partial_{x_R^-}G(x_{T},x_{R})\alpha(x_{2,T},x_{2,R}) \partial_{x_{2,R}^-}G(x_{2,T},x_{2,R}))
	\end{aligned}
\end{equation}

\begin{equation}\label{bgapeq22B'}
	\begin{aligned}
				\Sigma_{2B'}(x_T,x_R) &=4 \pi^2 \lambda_B^2(x_6\alpha(x_T,x_R)^2\delta^{(3)}(x_R)\\
		&+4(\partial_{x_R^-}-\partial_{x_T^-})(\partial_{x_R^-}+\partial_{x_T^-})\int  d^3z \ \alpha(x_{1,T},x_{1,R})G(x_{1,T},x_{1,R})\alpha(x_{2,T},x_{2,R})G(x_{T},x_{R})\\
		&-2(\partial_{x_R^-}+\partial_{x_T^-})\int  d^3z \ \alpha(x_{1,T},x_{1,R})G(x_{1,T},x_{1,R})\alpha(x_{2,T},x_{2,R})\partial_{x_R^-}G(x_{T},x_{R})\\
		&-4(\partial_{x_R^-}-\partial_{x_T^-})\int  d^3z \ \alpha(x_{1,T},x_{1,R})G(x_{1,T},x_{1,R})\alpha(x_{2,T},x_{2,R})\partial_{x_R^-}G(x_{T},x_{R})\\
		&-2(\partial_{x_R^-}-\partial_{x_T^-})\int  d^3z \ \alpha(x_{1,T},x_{1,R})\partial_{x_{1,R}^-}G(x_{1,T},x_{1,R})\alpha(x_{2,T},x_{2,R})G(x_{T},x_{R})\\
		&-8(\partial_{x_R^-}-\partial_{x_T^-})\int  d^3z \ \partial_{x_{1,T}^-}\alpha(x_{1,T},x_{1,R})G(x_{1,T},x_{1,R})\alpha(x_{2,T},x_{2,R})G(x_{T},x_{R})\\
		&+4\int  d^3z \ \partial_{x_{1,T}^-}\alpha(x_{1,T},x_{1,R})\partial_{x_R^-}G(x_{T},x_{R})\alpha(x_{2,T},x_{2,R})G(x_{1,T},x_{1,R})\\
		&+2\int  d^3z \ \alpha(x_{1,T},x_{1,R})\partial^2_{x_R^-}G(x_{T},x_{R})\alpha(x_{2,T},x_{2,R})G(x_{1,T},x_{1,R})\\
		&+\int  d^3z \ \alpha(x_{1,T},x_{1,R})\partial_{x_R^-}G(x_{T},x_{R})\alpha(x_{2,T},x_{2,R}) \partial_{x_{2,R}^-}G(x_{1,T},x_{1,R}))
	\end{aligned}
\end{equation}

In the integral expressions above $z$ represents
the position of the interaction vertex which is 
integrated over all of spacetime. In each of the expressions above we have used 
the following notation (see Appendix \ref{conventions})
\begin{equation}
\begin{aligned}
x_{1,T}=x+z, ~~~x_{1,R}=x-z,~~~x_{2,T}=z+y,~~~ x_{2,R}=z-y
\end{aligned}
\end{equation}

\subsection{Constant magnetic field}
 
As in the previous section, we wish to study our  theory in the presence of a background gauge field of the special form 
 \begin{equation} 
a_j = -b\frac{\epsilon_{ji} x^i} {2}. 
\end{equation}
It follows from the discussion of section 
\ref{ReducedVariables} that 
\begin{equation}\label{form}
\alpha(x, y)=e^{-i \frac{b}{2} \left(x^1 y^2 -x^2 y^1 
	\right) } \alpha_R(x-y), ~~~~
\Sigma(x,y) =e^{-i \frac{b}{2} \left(x^1 y^2 -x^2 y^1 
	\right) } \Sigma_R(x-y)
\end{equation}
where the functions $\alpha_R(x-y)$ and 
$\Sigma_R(x-y)$ are both simultaneously 
rotationally and translationally invariant. 
One difference with the previous section is 
that, in this case, $\Sigma_R(x-y)$ and 
$\alpha_R(x-y)$ are both index free scalar 
functions. It follows that $\Sigma_R(x-y)$ and 
$\alpha_R(x-y)$ admit a much simpler expansion in terms of the functions $e_{m,n}$
than in the previous section. Explicitly we 
have 
\begin{equation} \label{decompo}
\begin{aligned}
&\alpha_R(x)= 
\int_{-\infty}^\infty \frac{dw}{2\pi} e^{i\omega t}
\alpha_R(\omega,\vec{x}), ~~~~~~ \alpha_R(\omega,\vec{x})=\sum_{n=0}^{\infty}\alpha_n(\omega) e_{n,n}(\vec{x}) \\
&\Sigma_R(x)= 
\int_{-\infty}^\infty \frac{dw}{2\pi} e^{i\omega t}
\Sigma(\omega,\vec{x}), ~~~~~~ \Sigma_R(\omega,\vec{x})=\sum_{n=0}^{\infty}\Sigma_n(\omega) e_{n,n}(\vec{x}). 
\end{aligned}
\end{equation}
Note that the RHS of has terms of at most first 
order in $x_R^3$. It follows from this observation
that the contribution of $\Sigma_1$ to the 
quantity $\Sigma_(\omega, {\vec x})$ defned in 
\eqref{decompo}  is a polynomial of first degree
in $\omega$. On the other hand the expressions 
on the RHS of \eqref{bgapeq22A}, \eqref{bgapeq22B}
and \eqref{bgapeq22B'} each have no derivatives
in $x^3_R$; it follows that the contributions of the $\Sigma_{2A}$, $\Sigma_{2B}$ and $\Sigma_{2B'}$ 
to $\Sigma_(\omega, {\vec x})$ are all independent 
of $\omega$. In summary $\Sigma_(\omega, {\vec x})$
is a polynomial of unit degree in $\omega$. 
It follows that the second expansion on the 
second line of \eqref{decompo} is completely 
specified by two expansion coefficients (the 
coefficient of the constant and the coefficient 
of the term proportional to $\omega$) at every 
value of $n$. We find it convenient to 
parameterize these two coefficients in terms 
of the coefficients $\zeta^B_+(n)$ and 
$\zeta^B_-(n)$ defined by 
\begin{equation}\label{sigiz}
 \Sigma_n(\omega)=-i \omega(\zeta_{+}^B(n)-\zeta_{-}^B(n))+\zeta_{+}^B(n)\zeta_{-}^B(n)-b(2n+1)-m_B^2
 \end{equation}
 The form of the parameterization 
 \eqref{sigiz} ensures that the gap equation \eqref{bgapeq1} has the following simple 
 `solution' in terms of  $\zeta^B_+(n)$ and 
 $\zeta^B_-(n)$ 
  \begin{equation}\label{bgapeq1f}
	\begin{aligned}
		& \alpha_R(\omega,\vec{x})=\sum_{n=0}^{\infty} \frac{1}{(\omega-i \zeta_{+}^B(n))(\omega+i \zeta_{-}^B(n))} e_{n,n}(\vec{x})\\
	\end{aligned}
\end{equation}
The second gap equation \eqref{bgapeq2} - after a considerable amount of algebraic manipulation  
can be massaged into the form   (see Appendix for details)
\begin{equation}\label{bgapeq2f}
	\begin{aligned}
		\Sigma_R(\omega,\vec{x})&=\Sigma_{R,0}(\omega,\vec{x})+\Sigma_{R,1}(\omega,\vec{x})+\Sigma_{R,2}(\omega,\vec{x})\\
		\Sigma_{R,0}(\omega,\vec{x})&=8 \pi \lambda_B b_4 \delta^{(2)}(\vec{x}) \int \frac{d \omega'}{2 \pi} \alpha_R( \omega',\vec{0})+4 \pi^2 \lambda_B^2(1+3x_6) \left(\int \frac{d \omega'}{2 \pi} \alpha_R( \omega',\vec{0}) \right)^2  \delta^{(2)}(\vec{x})\\
		\Sigma_{R,1}(\omega,\vec{x})&=-2\lambda_B b \left(i \omega \int \frac{d \omega'}{2 \pi} \alpha_R( \omega',\vec{x})+ \int \frac{d \omega'}{2 \pi}i \omega'  \alpha_R( \omega',\vec{x})\right)\\
		\Sigma_{R,2}(\omega,\vec{x})&=-(2\lambda_B b)^2 \left(\int \frac{d \omega_1'}{2 \pi}\frac{d \omega_2'}{2 \pi} \alpha_R(\omega_1',\vec{x}) *_b \alpha_R(\omega_2',\vec{x}) \right)
	\end{aligned}
\end{equation}
where $*_b$ is the twisted convolution defined 
in \eqref{stp}. 

In the rest of this section we will recast the  
gap equations \eqref{bgapeq1f} and \eqref{bgapeq2f}
into equations for $\zeta^B_{\pm}(n)$ and then 
proceed to solve these equations. 

\subsection{Solving the gap equations}
 In this subsection we study the gap equations obtained in previous section. We proceed by plugging $\alpha$ from \eqref{bgapeq1f} to RHS of \eqref{bgapeq2f} and performing $\omega$ integrals. To simplify the star product appearing in \eqref{bgapeq2f} we use property presented in \eqref{mrules} of the basis elements $e_{m,n}$. 
 
 As we have explained above, of the three terms 
 on the RHS of the first line of \eqref{bgapeq2f}, 
 only the term $\Sigma_{R,1}$ has a contribution 
 proportional to $\omega$. This term - i.e. 
 the term proportional to $\omega$ - arises from 
 the first integral on the RHS of third line 
 in \eqref{bgapeq2f}. The integral in question 
is easily verified to be proportional to
$\frac{1}{\zeta^B_+(n)+ \zeta^B_-(n)}$. Equating 
this integral to the coefficient of $\omega$ 
in \eqref{sigiz} we obtain the equation
 \begin{equation}\label{eq1b}
 	\begin{aligned}
 		\zeta_{+}^B(n)^2-\zeta_{-}^B(n)^2=2 \lambda_B b
 	\end{aligned}
 \end{equation}
 
Evaluating the remaining integrals in \eqref{bgapeq2f} gives us an expression for the $\omega$ independent part of $\Sigma_n(\omega)$. 
Collecting terms and comparing equating with 
the $\omega$ independent part of \eqref{sigiz} 
we find the second equation 
\begin{equation}\label{eq2b}
	\begin{aligned}
		\zeta_{+}^B(n)^2+\zeta_{-}^B(n)^2=2 \left( c_B^2+b(2n+1) \right)
	\end{aligned}
\end{equation}
where we have defined the quantity 
\begin{equation}\label{cb}
	\begin{aligned}
		c_B^2=m_B^2+8 \pi \lambda_B b_4 \int \frac{d \omega'}{2 \pi} \alpha_R( \omega',\vec{0})+4 \pi^2 \lambda_B^2(1+3x_6) \left(\int \frac{d \omega'}{2 \pi} \alpha_R( \omega',\vec{0}) \right)^2  
			\end{aligned}
\end{equation}
\eqref{eq1b}, \eqref{eq2b} are readily solved to give
\begin{equation}\label{polesboson}
	\begin{aligned}
				\zeta_{\pm}^B(n)^2=c_B^2+b(2n+1)\pm \lambda_B b
	\end{aligned}
\end{equation}
The quantity $c_B$ is now determined by plugging 
\eqref{bgapeq1f} and \eqref{polesboson} into 
\eqref{cb} and solving the resultant equation 
for $c_B$. 

\subsection{Critical scaling limit and the duality}

In order to match with the fermionic results of 
the previous section we are interested in the 
theory \eqref{ActionScalar} in the so called critical boson scaling limit defined by 
scaling $m_B^2$ and $b_4$ to $\infty$ while keeping the following quantity fixed
\begin{equation}
	m_B^{cri}=\frac{m_B^2}{2 \lambda_B b_4}
\end{equation}
In this limit equation \eqref{cb} simplifies to 
\begin{equation}
	4 \pi \lambda_B \int \frac{d \omega'}{2 \pi} \alpha_R( \omega',\vec{0})=-m_B^{cri} \lambda_B
\end{equation}
When this condition is maintained $c_B$ defined above has a finite value in the scaling limit under consideration. Using \eqref{bgapeq1f} and \eqref{eq1b} this equation can be simplified to
\begin{equation}\label{gapeqfinal}
	\begin{aligned}
		\sum_{n=0}^{\infty} ( \zeta_{+}^B(n)-\zeta_{-}^B(n) )=-m_B^{cri} \lambda_B
	\end{aligned}
\end{equation}
The equation \eqref{gapeqfinal} is formal as 
the summation on LHS of this equation is divergent. 
As in the previous section, this divergence 
should be regulated by adding the same counterterm
to the theory that one adds to renormalize the 
corresponding divergence in the absence of a 
magnetic field. As in the previous subsection this 
is achieved by adding and subtracting the 
relevant integral. Our gap equation turns into 

\begin{equation}\label{apeb}
\begin{aligned}
& -\lambda_B m_B^{\rm cri} = \lambda_B \int_{a}^{\infty} \frac{p_s dp_s}{ \sqrt{c_B^2 +p_s^2}^{1+\epsilon}}  \\
&+\lim_{P \to \infty}
\bigg[ \sum_{k=0}^P \left[ c_B^2+ b(2k+1) + \lambda_B b  )^{1/2}-(c_B^2+b(2k+1) - \lambda_B b )^{1/2} \right]  -\lambda_B
\int_\frac{a^2}{2}^{Pb} \frac{dw}{ \sqrt{\zeta_{ex}^2+2w}}
\bigg] \\
\end{aligned}
\end{equation}
where the integral on the first line is to be 
evaluated in the dimensional regulation scheme and  
can be evaluated using  \eqref{finexp}. 

\subsection{Matching under duality}

We have demonstrated above that the critical 
boson theory at zero chemical potential has 
a propagator, whose zeroes predict a spectrum 
of single particle energies at (Lorentzian) 
\begin{equation}\label{omemat}
E= \pm \zeta^B_\pm(n)
\end{equation} 
where $\zeta^B_\pm(n)$ are given by 
\eqref{polesboson} where $c_B$ is required to 
solve the gap equation \eqref{apeb}. 

We now wish to compare the result \eqref{omemat}
with our results for the spectrum of single particle
energies in the fermionic theory of the previous 
section. The variable $n$ that appears in \eqref{omemat} can simply be identified with the Landau level label $\nu$ used, for instance in \eqref{mit}. Comparing \eqref{polesboson} with 
\eqref{rmp}, we see that the spectrum of 
bosonic and fermionic single particle energies 
agrees exactly provided we make the identification
\begin{equation}\label{chvar}
c_B^2=c_F^2= \zeta_{ex}^2 \pm \lambda_F b
\end{equation}
(where the sign $\pm$ in \eqref{chvar} applies 
in the Fermionic positive/negative exceptional bands). 

In order to complete the check that the zero 
chemical potential bosonic single particle 
energy spectrum agrees with the corresponding 
single particle spectrum of fermionic energies, 
we must verify that the gap equation \eqref{apeb} 
reduces to the corresponding Fermionic gap
equation under the variable change \eqref{chvar}. 

Recall that we have performed the bosonic computation in the case that $m_B^{{\rm cri}}$ 
is positive (this is why were were able to 
perform our computation without turning on a 
bosonic condensate). Using the relationship 
$$m_F = - \lambda_B m_B^{{\rm cri}}$$ 
it follows that we are working in the case that $\lambda_F$ and $m_F$ have the same sign. We 
have also performed our bosonic computation assuming that the chemical potential lies between the positive and negative Landau $\nu=0$ Landau Levels. When ${\rm sgn}(m_F) >0$ this condition 
puts us in the fermionic `negative exceptional 
band'. On the other hand when 
${\rm sgn}(m_F) <0$ this condition puts us in the 
fermionic positive exceptional band. 

In summary, duality predicts that the bosonic gap 
equation should match (under the map \eqref{chvar} ) with the fermionic gap equation in the negative exceptional band when $m_F$ and $\lambda_F$ are both positive, and the fermionic gap equation
in the positive exceptional band when 
$m_F$ and $\lambda_F$ are both positive. 

It is easy to verify that the prediction of the 
previous paragraph is indeed borne out 
provided we assume that $\zeta_{ex}$ has the same 
sign as $m_F$ at every value of $\lambda_F$. 
Under this assumption, the conditions of the previous paragraph put us in the special 
case studied in section \ref{spclb}. It is 
easy to verify that \eqref{apeb} reduces to 
\eqref{apep} (using \eqref{chvar}) when 
$m_F$, $\lambda_F$ and $\zeta_{ex}$ are all 
positive, but reduces to \eqref{apem} when 
$m_F$, $\lambda_F$ and $\zeta_{ex}$ are all 
negative. 

In order to complete our demonstration of the matching of gap equations under duality, we 
need to demonstrate that our assumption about 
the sign of $\zeta_{ex}$ is correct. We do not, 
unfortunately, have an analytic proof of this 
fact. The numerical evidence presented in the 
previous section, however, appears to back this 
claim. 

It thus appears that all our results are in 
perfect agreement with the predictions of duality, 
although a watertight demonstration of this 
fact would need an analytic proof of the assumption
we have made earlier in this section for the sign
of $\zeta_{ex}$.

\section{Future directions}

Working at leading order in the large $N$ limit, 
in this paper we have computed the exact (i.e. 
all orders in the t' Hooft coupling) fermion - 
fermion propagator of the Chern Simons gauged 
regular fermion theory in the fundamental 
representation. A key structural feature of this 
propagator - whenever we have reliably been 
able to compute it - is that its only non 
analyticities in frequency space are a set of 
poles. While the propagator that we have computed 
is not itself gauge invariant, its poles capture 
the physical one particle excitations of the 
theory. In the free limit $\lambda_F \to 0$ the poles of our propagator lie at the locations of Landau levels. Our exact determination of the 
poles of our propagator allows us to track the 
change in the energies of Landau levels as 
a function of the t'Hooft coupling. The final 
result for the energies of unfilled Landau 
Levels is satisfyingly simple; the spectrum 
of single particle energies takes the simple 
form one might naively have guessed for  a system of free relativistic spin $s$ particles propagating in a magnetic field (see around \eqref{epesp})
with $s$ given by  \eqref{slam}, the true 
spin of the anyonic excitations in our theory. 

The large $N$ limit ensures that our propagator has only poles (and not also cuts) at all values of $\lambda_F$; cuts in the propagator all come from diagrams with holes or boundaries (in the sense of the t' Hooftian counting of large $N$ Feynman diagrams) and so are suppressed at large $N$. 
In this sense our large $N$ limit may be the 
nearest one can come to defining a theory 
of non interacting relativistic spin $s$ particles. 
The simple structure of the energies of 
unfilled Landau Levels is a further illustration 
of this fact. 

We find our  results for 
filled Landau levels, and especially for Landau 
levels in the process of being filled, particularly
interesting.
As we have explained in the introduction, in the 
interacting theory (and unlike in the free theory) the minimum chemical potential at which a Landau
Level is completely filled differs from the 
maximum chemical potential at which the same 
Landau level is completely empty. The first 
chemical potential is the one at which the 
pole in the propagator - corresponding to this 
Landau level - lies just below our integration 
contour in $\omega$ (frequency) space, while the second chemical potential is the one at which 
the same pole lies just above the same integration 
contour. We call the difference between the 
first chemical potential and the second chemical 
potential the width of the corresponding 
Landau level. This width can be either positive
or negative. 

A careful analysis of the physics within the 
width of any Landau level may turn out to be one 
of the most important research directions that 
follow from the work presented in this paper. 
We pause to discuss this point at some length. 

Let us first consider the case of Landau levels 
with a positive width. In this case the analysis of 
this paper yields  no valid solutions to the gap equation when $\mu$ lies within the width of any particular Landau level. At the technical level 
this happens because a pole in the complex $\omega$ plane hits the integration contour either from below (at the top of the band) or from above (at the bottom of the band). It is tempting to 
obtain new solutions to the gap equations at  intermediate values of $\mu$ by adopting the following prescription. 
We could allow the  pole corresponding to a particular Landau to lie precisely on the integration contour, and then split up this pole 
in an ad hoc manner so that 
a fraction $f$ of the pole lies under the contour 
and the remaining fraction $1-f$ of the pole lies 
above the contour (the special case $f= \frac{1}{2}$ defines the principle value 
prescription). Intuitively this corresponds to 
filling up part of the Landau Level under study
while leaving the rest of it unfilled. If we adopt 
this prescription then $\zeta_{ex}^2$ and 
and hence the value of $\mu$ corresponding to 
this situation \footnote{This is the value of 
$\mu$ at which the Landau level pole of interest 
lies on the contour.} will depend on $f$. As we 
vary $f$ from unity to zero, this prescription 
presumably  gives us a solution to the gap equation for values of $\mu$ that lies between 
the lower and upper end of the Landau Level width. 
This (natural but so far ad hoc) prescription 
would give us a result for the fermion propagator 
within a Landau level and would potentially allow us to study its physics. The discussion of this 
paragraph provides a natural first guess for the 
nature of dynamics within a particular Landau level, one that should certainly be pursued 
in future work. 

While the prescription developed in the 
previous paragraph is simple and appealing 
it may not be completely correct. 
It is possible that the correct propagator 
(when $\mu$ lies within the width of the Landau 
Level) has a more complicated singularities 
than poles even in the strict Large N limit
implying  that fermion 
dynamics within the Landau band is not single 
particle in nature, even in the strict large 
$N$ limit. At the technical level a conclusion of 
this sort would follow if, for instance, 
infrared divergences within a Landau band turn out
to complicate the naive large $N$ expansion, and 
cause naively sub-leading graphs to actually 
contribute at leading order in large $N$. 
The classic work of Sung-Sik Lee \cite{Lee:2009epi} 
provides a precedent for a similar effect in 
a vaguely similar context. As dynamics 
at chemical potentials that lie within a 
Landau level is of great interest - for instance
for the study of the quantum hall effect - 
we feel that these issues deserve careful 
investigation in future work. 

We have so far discussed the case in which 
a given Landau Level has a positive width. 
In certain ranges of parameters (though never 
at large enough values of $|\lambda_F|$) we find 
that the width of our Landau Levels are sometimes
negative. At the physical level this is not 
particularly strange - it simply means that the 
energy needed to add the last fermion into an 
almost filled band is lower than the energy 
needed to add the first Fermion into the otherwise
empty band. We would expect this to happen
whenever the effective interaction between fermions is attractive rather than repulsive.

At the technical level, whenever we have a negative 
Landau level width, for every $\mu$ that lies 
within this width we have two solutions 
for the fermion propagator. This may, at first
seem like a better situation to be in than 
the case of positive width (in which case we had 
no solution for the propagator at intermediate 
values of $\mu$) but we believe this is not really 
the case. The reason for this is that the two 
solutions that we do have in this case correspond
to the solution for a fully filled or a completely 
empty Landau level. For physical reasons we 
are really interested in studying a partially 
filled Landau level. \footnote{When the width of the 
Landau level is negative, we have already seen that
there is a range of $\mu$ in which there exist 
atleast two saddle point solutions. For the sake of argument let us assume the solutions we already know are the dominant ones for thermodynamical purposes. 
As $\mu$ is gradually increased, we would expect our theory to undergo a first 
order  phase transition from the
saddle in which the level is completely empty to the saddle in which it is completely filled. The charge density will increase discontinuously 
across this phase trasition. At fixed intermediate values of charge  (rather than fixed chemical potential, i.e. in the `microcanonical' ensemble) the system could well undergo a phase separation (``half water, half steam'') that breaks translational invariance. It would be useful to replace these speculations with quantitive
results from an honest evaluation of the partition function. We hope to return
 to this issue soon. We thank D. Tong for related discussions.}

 Hence in this case as well 
it is natural to seek a third solution, for 
every value of $\mu$ within the band, for the 
fermion propagator. As in the discussion above, 
a naive solution of this form could be generated 
by splitting the pole on the propagator into a part
above and a part below. As in the discussion above 
this naive solution may - or may not - capture the 
true nature of the propagator. This question 
deserves further study. 

Apart from the (intensely interesting but potentially tricky) study of the dynamics 
within a broadened Landau band, the work in 
this paper should be generalized in several ways.
First, it should be straightforward to 
use the results of this paper to compute 
the thermodynamical free energy of our system 
at zero temperature but arbitrary chemical 
potential. A concrete outcome of such a 
generalization would be a formula for the 
charge of our system as a function of the 
chemical potential. 

Next, it may be possible, and would be 
interesting,  to generalize the study of this paper to finite temperature. 

Next, we have already partially verified that the 
results of this paper are consistent with 
the predictions of the level rank type 
bosonisation duality conjectured between the 
Chern Simnons gauged regular fermion theory and the critical boson theory. We performed this 
verification by directly solving the Chern 
Simons gauged critical boson theory at zero 
chemical potential and verifying that our final 
results for single particle free energies matches 
those of the the fermionic computation. 
It would be very useful to complete this 
verification by performing the needed bosonic 
computation at finite chemical potential, 
perhaps using the methods of \cite{Choudhury:2018iwf,Dey:2018ykx}. 

The computations mentioned in the last paragraph 
are interesting in their own right for a 
qualitative reason. It is well known that free 
bosons unstable to unchecked condensation in 
a large enough chemical potential background. 
Interaction effects presumably cure this 
instability. It is of great physical interest 
to understand in detail precisely how  this happens. In the results presented in this paper 
we have already seen signs of strange behaviour
for filled Landau levels in the limit $\lambda_B \to 0$. In this limit the width of, for instance,  the first Landau level diverges. It is very important to semi quantitatively understand 
this phenomenon in qualitative physical terms. 
We leave this for future work. 

Another related issue concerns the charge of a filled Landau level. In the free fermion theory 
the charge density of a single filled Landau level is given by $ \frac{b N_F}{2 \pi}$. While we have not studied this issue, it is certainly possible 
that the charge of a filled Landau level is not renormalized as a function of $\lambda_F$. 
If this turns out to be the case, the charge 
density of a filled Landau level 
can be rewritten as $\frac{b N_B|\lambda_F|}{2 \pi |\lambda_B|}$ and so translates to a charge density
that diverges, in units of $N_B$, as $\lambda_B$ 
is taken to zero. If the (admittedly unsubstantiated) guesses of this paragraph turn 
out to be even qualitatively correct, they suggest
that the occupation number of any particular bosonic
state (in a filled Landau level) diverges as 
$\lambda_B \to 0$. It should not be difficult - and would be very interesting - to perform computations 
that verify this  guess. It would 
also be interesting to reproduce this behaviour 
starting from the bosonic theory, and so understand
it qualitatively in bosonic terms. We hope to 
return to this issue in future work. 

Finally we find it intriguing that the leading large $N$ gap equation in the presence of a background magnetic field, for the fundamental fermion (and fundamental boson) propagators, differ from the corresponding gap equations
at zero magnetic field in only two ways. First all 
derivatives are replaced by background gauge covariant 
derivatives. Second that products of the propagator 
and the self energy are replaced by star products. 
It would be interesting to investigate whether 
a similar replacement rule applies also for the 
Schwinger Dyson equations for higher (e.g 4 point) 
correlators of the fermions/bosons, and also whether
a similar replacement rule persists also at sub-leading 
orders in $\frac{1}{N}$. We also leave this 
question to future work. 

\acknowledgments

We would like to thank A. Vishwanath for 
a discussion that initiated this project. 
We would like to thank  A. Dey, K. Damle,  R. Gopakumar, T. Hartman,  S. Jain,   L. Janagal, G. Mandal, N. Prabhakar, D. Radicevic,  R. Sen Sharma, A. Strominger, G. Torraba,  V. Tripathy and S. Trivedi and especially A. Gadde and N. Seiberg for useful discussions. We would also like to thank
A. Dey, G. Mandal, N. Prabhakar, N. Seiberg, A. Stronminger,  D. Tong, S. Wadia and A. Vishwanath 
for useful comments on a preliminary draft of this 
manuscript. The work of both authors was supported by the Infosys Endowment for the study of the Quantum Structure of Spacetime. Finally we would both
like to acknowledge our debt to the steady support of the people of India for research in the basic sciences.

\appendix

\section{Conventions}\label{conventions}
In this appendix we summarize the conventions that are used in this note
\begin{equation}
 \begin{split}
 & ~~~~~x^{\pm}=\frac{x^1\pm ix^2}{\sqrt{2}}, ~ x^{\pm}=\frac{r}{\sqrt{2}}e^{\pm i \phi}, p^{\pm}=\frac{p^1\pm ip^2}{\sqrt{2}}, ~ p^{\pm}=\frac{p_s}{\sqrt{2}}e^{\pm i \theta}\\
  & g_{+-}=g_{-+}=g_{33}=1, ~
\epsilon_{123}=\epsilon^{123}=1, ~ \epsilon_{+-3}=-\epsilon^{+-3}=i, \epsilon_{ij}=\epsilon_{ij3}\\
&~\gamma^1=\sigma_{x} = \left( \begin{array}{cc} 0 & 1 \\ 1 & 0 \end{array} \right), \;\;\;
  \gamma^2=\sigma_{y} = \left( \begin{array}{cc} 0 & -i \\ i & 0 \end{array} \right), \;\;\;
  \gamma^3=\sigma_{z} = \left( \begin{array}{cc} 1 & 0 \\ 0 & -1 \end{array} \right)\\
 & \Tr(T^A T^B)=\frac{1}{2}\delta_{A,B},~ \sum_A (T^A)_{m}^{n} (T^A)_{p}^{q}=\frac{1}{2}\delta_{m}^q \delta_{p}^n, ~ \gamma^\mu \gamma^\nu=i \epsilon^{\mu \nu \rho}\gamma_\rho+g^{\mu\nu}
 \end{split}
\end{equation}
Conventions for Fourier transformations are
\begin{equation}
 \begin{split}
  f(x)=\int \frac{d^3P}{(2 \pi)^3} e^{i P.x}f(P), \ ~ f(x,y)=\int \frac{d^3P}{(2 \pi)^3} \frac{d^3q}{(2 \pi)^3} e^{i (\frac{P}{2}+q).x+i (\frac{P}{2}-q).y}f(P,q)
 \end{split}
\end{equation}
We will often find it useful to use following set of coordinates
\begin{equation}
	\begin{aligned}
		x_T=x+y,x_R=x-y
	\end{aligned}
\end{equation}
In this new set of coordinates metric
\begin{equation}
	\begin{aligned}
		ds^2=dx^2+dy^2=\frac{1}{2}(dx_T^2+dx_R^2)
	\end{aligned}
\end{equation}
is different by a factor and as a result of that $x_{T,R}^{\pm}=2x_{T,R,\mp} $. Further note that variable change from $(x,y)$ to $(x_T/2,x_R)$ does not involve any non-trivial Jacobin. 
We have following map between derivatives and polynomials
\begin{equation}
	\begin{aligned}
		iq_\mu f(P,q)=\int d^3x d^3y \frac{\partial}{\partial x_R^\mu}f(x,y)e^{-i (\frac{P}{2}+q).x-i (\frac{P}{2}-q).y}\\
		\frac{iP_\mu}{2} f(P,q)=\int d^3x d^3y \frac{\partial}{\partial x_T^\mu}f(x,y)e^{-i (\frac{P}{2}+q).x-i (\frac{P}{2}-q).y}
	\end{aligned}
\end{equation}
We will be dealing with functions of the form $$ f(x,y)=e^{-i \frac{b}{2} \left(x^1 y^2 -x^2 y^1 \right) } f_R(x,y)=e^{\frac{b}{4} \left(x_R^+ x_T^- -x_R^- x_T^+ \right) } f_R(x,y)$$
On functions of this form
\begin{equation}
	\begin{aligned}
	&	\frac{\partial}{\partial x_R^-}f(x_T,x_R)=(- \frac{b}{2} x_{T,-}+\frac{\partial}{\partial x_R^-})f_R(x_T,x_R)\\
	&	\frac{\partial}{\partial x_T^-}f(x_T,x_R)=(+ \frac{b}{2} x_{T,-}+\frac{\partial}{\partial x_T^-})f_R(x_T,x_R)
	\end{aligned}
\end{equation}

\section{Constraints of gauge invariance}
In this paper we choose to work in the gauge $g_r$. 
In this gauge $\alpha(x,y)$ is rotationally invariant, i.e.,
\begin{equation}\label{rotation}
 \begin{split}
  (x_1 \frac{\partial}{\partial x^2}-x_2 \frac{\partial}{\partial x^1} + y_1 \frac{\partial}{\partial y^2}-y_2 \frac{\partial}{\partial y^1})\alpha(x,y)+[S_{12},\alpha(x,y)]=0
  \end{split}
\end{equation}
Where $S_{12}$ is the contribution of the spin to angular momentum. It is not present for the bosons, i.e., for bosons $$S_{12}=0$$ and for fermions  $$S_{12}=\frac{1}{4}[\gamma_1,\gamma_2]=\frac{i}{2}\gamma_3$$
It also follows from \eqref{gaugetransf} and \eqref{gto} that 
$$ e^{i\frac{b x^1 x^2}{2}} \alpha(x,y) 
e^{-i\frac{b y^1 y^2}{2}} =f_1(x_1-y_1,x_2,x_3,y_2,y_3) $$
and 
$$ e^{-i\frac{b x^1 x^2}{2}} \alpha(x,y) 
e^{i\frac{b y^1 y^2}{2}}=f_2(x_1,y_1,x_2-y_2,x_3,y_3)$$
Eliminating $\alpha$ from above two equations are seeking consistency with infinitesimal translation in 1,2 direction solves $f_1,f_2$ upto a function that is translationally invariant in 1,2 directions.  Demanding translational invariance in 3 direction as well we get
\begin{equation}\label{uss}
\alpha(x, y)=e^{-i \frac{b}{2} \left(x^1 y^2 -x^2 y^1 
	\right) } \alpha_R(x-y)
\end{equation}
Plugging this into equation \eqref{rotation} determines rotational charge of (various components of $2 \times 2$ matrix in case of fermion) $\alpha$.

Analogous statements apply for $\Sigma(x,y)$ and we won't repeat them here.

\section{Spectrum of free particles in a magnetic field}

\subsection{Boson}\label{bAppendix}
In this subsection we study consider the non-relativistic Schrodinger equation for a spin-less particle of mass $|M|$ in a uniform
magnetic field $b$ described by Hamiltonian
\begin{equation}
H= - \frac{(\nabla_i - i a_i)^2}{2|M|}
\end{equation}
We work in the in a rotationally invariant gauge 
\begin{equation} 
a_j = -b\frac{\epsilon_{ji} x^i} {2}
\end{equation}
Eigen functions of the hamiltonian satisfies
\begin{equation}\label{eignevalueEQ}
 \begin{aligned}
  H\phi=E\phi
 \end{aligned}
\end{equation}
We work in polar coordinates $(r, \phi)$. We seek a solution of the form
\begin{equation}
 \begin{aligned}
  \phi=\frac{e^{il\phi }}{\sqrt{2\pi}×}R(u), ~ u = \frac{|b|}{2}r^2
 \end{aligned}
\end{equation}
The condition that this describes a particle of spin zero determines $l=0,\pm1,\pm2,..$
Plugging this form of $\phi$ into \eqref{eignevalueEQ}, it follows that $R(u)$ satisfies
\begin{equation}
 \begin{aligned}
  uR''(u)+R'(u)+\left(-\frac{1}{4}u+\beta-\frac{m^2}{4 u}\right)R(u)=0
 \end{aligned}
\end{equation}
where we have defined
\begin{equation}
 \begin{aligned}
  m=-\rm{sgn}(b)l, ~ \beta=\frac{E}{\omega_H}-\frac{m}{2}, ~ \omega_H=\frac{|b|}{|M|}
 \end{aligned}
\end{equation}
The equation for $R$ can be transformed into confluent hypergeometric equation for $\tilde{R}$ by following change of variables
\begin{equation}
 \begin{aligned}
  R(u)=e^{-u/2}u^{|m|/2}\tilde{R}(u)
 \end{aligned}
\end{equation}
We search for solutions that are normalisable by following inner product
\begin{equation}
 \begin{aligned}
  (\phi_1,\phi_2)=\int d^2x \phi_1(x)^*\phi_2(x)
 \end{aligned}
\end{equation}
For those solutions 
\begin{equation}
 \begin{aligned}
  \tilde{R}(u) \propto L_n^{|m|}(u)
 \end{aligned}
\end{equation}
where $n=0,1,2,..$ is a non-negative integer that determines energy to be
\begin{equation}
 \begin{aligned}
  E=\omega_H \left(n+\frac{|m|}{2}+\frac{m}{2}+\frac{1}{2} \right)  \end{aligned}
\end{equation}
Corresponding orthonormal eigen functions are given by\footnote{$L_n^l$ is associated Laguerre polynomial. For non-negative integer $n$, $L_n^l(x)$ is 
a polynomial in $x$ of degree $n$  normalised as $ L_n^l(0)=\frac{\Gamma(n+l+1)}{\Gamma(n+1) \Gamma(l+1)}$. For negative integer $n$,  $ L_n^l(x)$ is taken to be zero by definition. }
\begin{equation}
 \begin{aligned}
  \phi=& i^{|m|}\left[\frac{|b|}{2\pi} \frac{n!}{(|m|+n)!} \right]^{1/2} e^{-i\rm{sgn}(b)m\phi }e^{-u/2}u^{|m|/2}L_n^{|m|}(u)\\
   \end{aligned}
\end{equation}
Below we rewrite the expressions using following map
\begin{equation}
	\begin{aligned}
		m=-\rm{sgn}(b)l,~ n=\nu-\frac{|l|}{2}+\rm{sgn}(b)\frac{l}{2}
	\end{aligned}
\end{equation}
In these variables
\begin{equation}\label{bosonspectrum}
 \begin{aligned}
  \phi  =& i^{|l|}\left[\frac{|b|}{2\pi} \frac{(\nu-\frac{|l|}{2}+\rm{sgn}(b)\frac{l}{2})!}{(\nu+\frac{|l|}{2}+\rm{sgn}(b)\frac{l}{2})!} \right]^{1/2} e^{il\phi }e^{-u/2}u^{|l|/2}L_{\nu-\frac{|l|}{2}+\rm{sgn}(b)\frac{l}{2}}^{|l|}(u),
  ~ E=\omega_H \left(\nu+\frac{1}{2} \right)
 \end{aligned}
\end{equation}

\subsection{Fermion}
In this subsection we study spectrum of Dirac hamiltonian in presence of constant magnetic field. More precisely we study solutions of the following equation
\begin{equation}\label{DiracH}
H \psi = \gamma^3(\vec{\gamma}.(\vec{\partial}-i \vec{a})+m_F)\psi=E \psi
\end{equation}
Where gauge field is given by
\begin{equation}
a_j =- b\frac{\epsilon_{ji} x^i} {2}. 
\end{equation}
This corresponds to constant magnetic field through $12$ plane. Working in polar coordinates \eqref{DiracH} can be rewritten as
\begin{equation}
H = H_0 + i \gamma^3 \vec{\gamma}.\vec{e}_r H_r+i \gamma^3 \vec{\gamma}.\vec{e}_\phi H_\phi
\end{equation}
Here unit vectors are
\begin{equation}
 \vec{e}_r=\frac{\vec{r}}{r}, ~ \vec{e}_\phi=-\sin(\phi) \vec{e}_1+\cos(\phi)\vec{e}_2
\end{equation}
and
\begin{equation}
 \begin{split}
  H_0=\gamma^3 m_F, ~ H_r=-i\partial_r, ~ H_\phi=-i\frac{1}{r}\partial_\phi -a_\phi
 \end{split}
\end{equation}
At this point we make a similarity transformation $\Lambda$ so that $\vec{\gamma}.\vec{e}_r$ and $\vec{\gamma}.\vec{e}_\phi$ gets mapped to $\gamma^1$ and $\gamma^2$ respectively, i.e., 
\begin{equation}
 \Lambda \vec{\gamma}.\vec{e}_r \Lambda^{-1}=\gamma^1, ~ \Lambda \vec{\gamma}.\vec{e}_\phi \Lambda^{-1}=\gamma^2, ~ \Lambda=\sqrt{r} e^{\frac{i}{2}\gamma^3 \phi}
\end{equation}
Correspondingly we have
\begin{equation}
 \tilde{H}=\Lambda H \Lambda^{-1}, ~  \tilde{\psi}=\Lambda \psi
\end{equation}
Angular part is easily diagonalised by 
\begin{equation}
 \tilde{\psi}_{\pm}=\frac{e^{i J \phi}}{\sqrt{2 \pi}} \chi(r)_{\pm}
\end{equation}
Where $\pm$ refers to upper and lower component of the spinor. At this point we know angular dependence of the field completely. We impose the condition that under rotation the field behaves as a fermion
\begin{equation}
	\begin{split}
		e^{-i\frac{1}{2} \gamma^3 2 \pi}\psi(r,2 \pi)=-\psi(r,0)
	\end{split}
\end{equation} 
This implies $j=\pm \frac{1}{2},\pm \frac{3}{2},...$.

Radial dependance of the field is determined from two first order coupled equations given below
\begin{equation}\label{1storder}
 \chi_{\mp}(r)=\frac{1}{E \pm m_F}
\left( \mp \frac{d}{dr}+\frac{J}{r}-\frac{b}{2}r  \right)\chi_{\pm}(r)
\end{equation}
Below we first consider the case of $|E| \neq |m_F|$ and consider this special of  $|E|=|m_F|$ carefully later.

For $|E| \neq |m_F|$, above first order equations can be put in the following form (Kummer's equation) by eliminating one of the variables
\begin{equation}\label{2ndorder}
	\begin{split}
		u \tilde{\chi}''_{\pm}(u)+(1-u+\alpha_{\pm})\tilde{\chi}'_{\pm}(u)+n \tilde{\chi}_{\pm}(u)=0
	\end{split}
\end{equation}
where $$\alpha_{\pm}=|J \mp \frac{1}{2}|, ~ u = \frac{|b|}{2}r^2, ~ E^2=m^2-b(J \pm \frac{1}{2})+|b|(2n+1+\alpha_{\pm})$$ and $$ \chi_{\pm}(r)=u^{\alpha_{\pm}+1/2}e^{-u/2}\tilde{\chi}_{\pm}(u)$$
Among two linearly independent solution of \eqref{2ndorder}, only solution that is regular at the origin and normalisable according to 
\begin{equation}\label{inner product}
(\chi, \psi )= \int d^2x \chi^\dagger(\vec{x}) \psi(\vec{x})
\end{equation}
 are given by $$ \tilde{\chi}_{\pm}(u) \propto L_n^{\alpha_{\pm}}(u)$$ Further $n$ is restricted to take values $0,1,2,...$ only. Given the value of $J, E$ we choose one sign from above equation and determine the other component by pugging it into RHS of \eqref{1storder}. 

Solution for $|E|=|m_F|$ are obtained by setting one of $\chi_+,\chi_-$  to zero. In that case when the component that vanishes is plugged into RHS of \eqref{1storder}, $\rm{sgn}(E)$ is determined in terms of $\rm{sgn}(m_F)$ for consistency (the denominator must vanish for the other component to be non-zero).

Finally the undetermined proportionality constant is fixed by demanding orthonormality of various solutions. Below we summarize the results.

As is usual, the  Dirac Hamiltonian has eigenfunctions of both positive and negative energy. We use the 
label $\eta$ to distinguish the two kinds 
of solutions; $\eta=1$ for positive energy 
solutions and $\eta=-1$ for negative energy
solutions. Positive and negative energy 
solutions are both further labelled by two discrete parameters, which we call $n$ and $J$. These parameters range over the values
(together with exceptional values listed below) \footnote{Note that condition on range of parameters corresponds to $n_\pm \geq-1$ (defined in \eqref{coefficients} ) excluding the particular case of $n_+=n_-=-1$. With further restriction that $n_+=-1$ or $n_-=-1$ is allowed only if $ \eta b m_F >0$. }
\begin{equation}\label{rangenj}
\begin{split}
&n= 1, 2, \ldots  ~~~J= \pm \frac{1}{2}, \pm \frac{3}{2}, \pm \frac{5}{2}, 
\ldots \\
& n=0, J= \pm \frac{1}{2}, \pm \frac{3}{2}, \pm \frac{5}{2}, \ldots ~~~{\rm when} ~bJ<0 \\
& n=0, J= - \frac{1}{2}, - \frac{3}{2}, - \frac{5}{2}, \ldots ~~~{\rm when} ~b<0
\end{split}
\end{equation}
In addition we have the exceptional range 
of parameters 
\begin{equation}\label{rangenjex}
\begin{split}
&n= 0  ~~~J=  \frac{1}{2}, \frac{3}{2},  \frac{5}{2}, 
\ldots ~~~{\rm when} ~~b>0 ~~{\rm and} 
~~ \eta m_F>0\\
& n= -1  ~~~J= - \frac{1}{2}, - \frac{3}{2}, -\frac{5}{2}, 
\ldots ~~~{\rm when} ~~b<0 ~~{\rm and} 
~~ \eta m_F<0\\
\end{split}
\end{equation}
As we will see below, the exceptional range of parameters correspond to states with 
$|E|=|m_F|$. 
The quantity $J$ just labels the angular 
momentum of states; $J$ runs over half integer values because the total angular 
momentum is the sum of an orbital and a spin 
component.

We denote the most general solution by 
$\psi^{\eta n J}$. Recall that 
$\psi^{\eta n J}$ is a two component spinor, 
whose components we denote by 
\begin{equation}\label{psisgA}
\psi^{\eta n J}=\begin{bmatrix}
    \psi_+^{\eta n J} \\
   \psi_-^{\eta n J}
  \end{bmatrix}
\end{equation}
After solving for the eigenfunctions we find following explicit results \footnote{$L_n^l$ is associated Laguerre polynomial. For non-negative integer $n$, $L_n^l(x)$ is 
a polynomial in $x$ of degree $n$  normalised as $ L_n^l(0)=\frac{\Gamma(n+l+1)}{\Gamma(n+1) \Gamma(l+1)}$. For negative integer $n$,  $ L_n^l(x)$ is taken to be zero by definition. In the 
exceptional range of parameters, the 
expressions \eqref{psisg} involve the 
quantity $L_{-1}^{l_\pm}$.
Consequently the final expressions for $\psi_\pm$ listed in \eqref{psexp} only involve the functions $L_n^l(x)$ for $n \geq 0$. }

\begin{equation}\label{psexpA}
\begin{split}
\psi_{\pm}^{\eta n J}(u,\phi)= c_{\pm}\frac{ e^{i  \rm{sgn}(J) l_{\pm}  \phi}}{\sqrt{2 \pi}} u^{\frac{l_{\pm}}{2}} e^{-\frac{u}{2}} L_{n_{\pm}}^{l_{\pm}}(u)
\end{split} 
\end{equation} 
where $(u = \frac{b}{2}x_s^2$, $\phi)$ are coordinates on space and \footnote{Here $|x|$ stands for absolute value of $x$. }
\begin{equation}\label{coefficients}
\begin{split}
& \ l_+=|J-\frac{1}{2}|, \ l_-=|J-\frac{1}{2}|+\rm{sgn}(J), \  n_+=n, \ n_-=n-\frac{1}{2}\rm{sgn}(J)(1+\rm{sgn}(bJ))\\
 &  c_-=c_+ \left(\frac{\sqrt{2|b|}}{m+\rm{sgn}(J)E^{\eta n J}}\right)^{\rm{sgn}(J)}
\\
& c_+=\left(\frac{1}{|b|} \left(\frac{ \Gamma(1+n_++l_+)}{\Gamma(1+n_+)}+
\left(\frac{c_-}{c_+}\right)^2\frac{ \Gamma(1+n_-+l_-)}{\Gamma(1+n_-)}
 \right)   \right)^{-1/2}
\end{split}
\end{equation}
At this point we note an important property of the eigenfunctions given in \eqref{psexpA} - $\psi^{\eta n J}_\pm$ is non-zero at the origin only if $l_\pm=0$, i.e., precisely when they don't depend on azimuthal angle $\phi$.\footnote{This property ensures that wave-functions are not singular at the origin.} 

The energy $E^{\eta n J}$ of the eigensolution $\psi^{\eta n J}$ is given by \footnote{Note that expression of energy given in \eqref{enoso} is invariant under following simultaneous replacements: $b \to -b, J \to -J, m_F\to -m_F, n\to n-\frac{1}{2}\rm{sgn}(J)(1+\rm{sgn}(bJ))$. This is reminiscent of parity invariance of the theory under consideration. }
\begin{equation} \label{enoso}
\begin{split}
 E^{\eta n J}=& \eta 
\sqrt{ m_F^2 + |b|\left( 2n+1    + 
	| \left( J-\frac{1}{2} \right) | -
	\rm{sgn}(b) \left( J+ \frac{1}{2} \right)  \right)}\\
	=& \eta 
\sqrt{ m_F^2 + |b|\left( 2n+1 - \frac{1}{2}\rm{sgn}(J)(1+\rm{sgn}(bJ))   + |J|(1-\rm{sgn}(b J))  \right)}
\end{split}
\end{equation}

Notice that while the wave functions 
\eqref{psexpA} depend separately on $J$ and 
$n$, the energy $E^{\eta n J}$ depends on 
these discrete parameters only through the 
combination 
\begin{equation}\label{enu} 
2\nu=2n+1 - \frac{1}{2}\rm{sgn}(J)(1+\rm{sgn}(bJ))   + |J|(1-\rm{sgn}(b J))
\end{equation}
The formula for the eigenenergies, 
\eqref{enoso}, takes the following simplified 
form when re-expressed in terms of $\nu$
\begin{equation}\label{enosonu}
E^{\eta\nu}= \eta 
\sqrt{ m_F^2 + 2|b|\nu}
\end{equation}
Below we discuss the range over which $\nu$ takes values.
Let us first suppose that $bJ <0$. In this case $\nu$ simplifies to 
\begin{equation}\label{nuint}
2\nu=2(n+ |J|)+ 1.
\end{equation}
When $bJ>0$, on the other hand, 
$\nu$ is independent of $|J|$ and we have 
\begin{equation}\label{nupos}
2\nu= 2n+1-\rm{sgn}(J)
\end{equation}
Notice that in both cases \eqref{nuint} 
and \eqref{nupos} $\nu$ ranges over integer 
values. The specific range of values is 
as follows: 
\begin{equation}\label{rangeM}
\begin{split} 
&\nu = 1, 2, 3 \ldots ~~~ b ~\eta ~ m_F <0\\
&\nu = 0, 1, 2 , 3 \ldots ~~~ b ~\eta ~m_F >0\\
\end{split} 
\end{equation}

If we fix $\nu$ at some value - say $\nu_0$. 
Let us first take the case $b>0$. For 
$\nu_0>0$ the set of energy eigenstates with any particular value of $\nu_0$ (and so with eigen energy determined by $\nu_0$) is given by 
$$(0, -\nu_0+\frac{1}{2}), ~~(1, -\nu_0+\frac{3}{2}) \ldots (\nu_0-1, -\frac{1}{2}) , ~~\sum_{l=0}^{\infty} (\nu_0, l+ \frac{1}{2})   $$
(where we have denoted the labels of states 
by the notation $(n,J)$). As we have remarked above, eigenstates with $\nu_0=0$ exist only when $\eta m_F>0$, in which case 
the set of $(n, J)$ values that lead to 
$\nu_0=0$ are given by 
 $$\sum_{l=0}^{\infty} (\nu_0, l+ \frac{1}{2})$$
 When $b<0$, on the other hand, the set 
 of values that lead to any value of 
 $\nu_0 >0$ are given by 
$$(0, \nu_0-\frac{1}{2}), ~~(1, \nu_0-\frac{3}{2}) \ldots (\nu_0-1, \frac{1}{2}) , ~~\sum_{l=0}^{\infty} (\nu_0-1, -(l+ \frac{1}{2}))   $$
The particular case of $\nu_0=0$ is obtained when $\eta m_F<0$ and that corresponds to the set
$$ \sum_{l=0}^{\infty} (\nu_0-1, -(l+ \frac{1}{2})) $$.

\section{Star product and twisted convolution}

In this appendix we review standard properties 
of the Moyal star product (see, for example,  \cite{Gopakumar:2000zd} and references therein for more details). For two given functions $\phi_{1,2}$ on $R^d$, (Moyal-) star product is defined to be the following function 
\begin{equation}
	\begin{split}
		(\phi_1 \star_\Theta \phi_2)(q)=\lim_{p \to q} e^{\frac{i}{2}\Theta^{\mu \nu } \frac{\partial}{ \partial p_\mu} \frac{\partial}{\partial q_\mu} } \phi_1(p) \phi_2(q)
	\end{split}
\end{equation}
Where $\Theta$ is an anti-symmetric non-degenerate 2 form. Another equivalent definition is given by
\begin{equation}
	\begin{split}
		(\phi_1 \star_\Theta \phi_2)(q)=\int d^d q_1 d^d q_2  \ \phi_1(q_1) \phi_2(q_2)K(q_1,q_2,q)
	\end{split}
\end{equation}
where
\begin{equation}
	K(q_1,q_2,q)=\frac{1}{det(\Theta)} e^{2i (\Theta^{-1})^{\mu \nu}(q-q_1)_\mu (q-q_2)_\nu}
\end{equation}
It follows from the definition that, star product is not commutative. In Fourier space star product becomes twisted convolution
\begin{equation} \label{spfourier}
	\begin{split}
		\int \frac{d^dq}{(2\pi)^d} (\phi_1 \star_\Theta \phi_2 )(q)e^{i x q}=\int d^d y \ \phi_1(x-y)\phi_2(y)e^{-\frac{i}{2} \Theta_{\mu \nu}(x-y)^\mu y^\nu} \equiv (\phi_1 *_\Theta \phi_2)(x)
	\end{split} 
\end{equation}

The Moyal star product is isomorphic to product 
of operators acting on the Hilbert space of 
square integrable wave functions in one dimension 
(i.e. the Hilbert space relevant to the quantum mechanics for a non relativistic 
particle on a line). We pause to describe this 
isomorphism. 

\textit{
 If we define anti-commuting operators by demanding $$[\hat{q}_\mu,\hat{q}_\nu]=i \Theta_{\mu \nu} $$
and with any function $f$ of $q_\mu$ we associated a Weyl ordered operator $\hat{f} $ by \footnote{Here we denote functions in position and momentum space both by same symbol, but they are in general different mathematical functions in strict sense.}
\begin{equation}
 \begin{split}
  f(x)=\int \frac{d^dq}{(2 \pi)^3} e^{i q.x}f(q), \ ~ \hat{f}(\hat{q})=\int d^dx  e^{-i \hat{q}.x}f(x)
 \end{split}
\end{equation}
Then these operators enjoy the following identity
\begin{equation}
\begin{split}
\hat{f}(\hat{q}) \hat{g}(\hat{q})=\hat{h}(\hat{q}), ~ (f \star_\Theta g)(q)=h(q)
\end{split}
\end{equation}
}
With the help of above theorem, we can convert star products to operator products and vice-versa. 

Comparing \eqref{spfourier} with \eqref{stp} we see that for the purpose of this paper we have 
\begin{equation}
 \begin{split}
  \Theta_{\mu \nu}=b \ \epsilon_{\mu \nu 3}
 \end{split}
\end{equation}

\subsection{Star product basis}

In this subsection we follow \cite{Gopakumar:2000zd} to construct a basis of functions on $R^2$ which has simple behaviour under star product (twisted convolution). To this end we restrict our attention to $b>0$ and define \footnote{Note that in any dimension we can always put $\Theta$ into canonical form to define set of such creation, annihilation operators. }
\begin{equation}
 \begin{split}
  \hat{a}=\frac{\hat{q}^+}{\sqrt{b}}, ~ \hat{a}^\dagger=\frac{\hat{q}^-}{\sqrt{b}} \implies [\hat{a},\hat{a}^\dagger]=1
 \end{split}
\end{equation}
In terms of these operators we consider following basis of operators \footnote{Here we are using standard normal ordering of $a,a^\dagger$,i.e., $a^\dagger$ on the left of $a$.}
\begin{equation}
 \begin{split}
  \hat{e}_{m,n}(\hat{q}) \equiv | m \rangle \langle n| =:\frac{(\hat{a}^\dagger)^m}{\sqrt{m!}} e^{-\hat{a}^\dagger \hat{a}} \frac{(\hat{a})^n}{\sqrt{n!}}: , ~~~~~ n,m=0,1,2,...
 \end{split}
\end{equation}
These define corresponding functions in position space by
\begin{equation}
\begin{split}
\hat{e}_{m,n}(\hat{q})=\int d^2x \ e^{-i \hat{q}.x}e_{m,n}(x) 
\end{split}
\end{equation}
More explicitly these functions are given by
\begin{equation}\label{enm} 
e_{ m,n}(u, \phi)=  i^{m-n}\frac{b}{2 \pi} \left(\frac{n!}{m!} \right)^{1/2} e^{-i(m-n)\phi}
 u^{\frac{m-n}{2}}e^{- \frac{u}{2}} L_n^{m-n}(u)
\end{equation}
where $u = \frac{b}{2}r^2$,  $(r, \phi)$ are polar coordinates on space.
The algebra of these operators turns into identities 
\begin{equation}
\begin{split}
& (e_{n,p} ~*_b ~e_{p', m})(\vec{x}) = \delta_{p p'} e_{n, m}(\vec{x}) \\
&\sum_{n=0}^{\infty}  e_{n,n}({\vec x})= \delta({\vec x})\\
\end{split}
\end{equation}
Further the functions 
$e_{n,m}$ are orthogonal to each other
in the usual quantum mechanical sense 
\begin{equation}\label{orthog} 
\int d^2x \frac{2 \pi}{b} e^*_{n,m}({\vec x } )
 e_{n',m'}({\vec x } ) = 
  \delta_{n n'} \delta_{m m'}
 \end{equation}

\subsection{Expanding functions in star product basis}

In addition to \eqref{orthog}, in this paper 
we will need to evaluate the integral of 
$e_{m,n}$ and $e^*_{m',n'}$ wighted by $1/x^+$ 
(the last factor comes from the Chern Simons 
gauge boson propagator in lightcone gauge). 
The relevant integrals can be done by 
using the explicit expressions in \eqref{enm}, 
first performing all angular integrals and then 
doing the radial integrals. At this stage 
the integrand has a product of associated Laguerre 
polynomials, and the needed integrals can be obtained by manipulating the integrand into a 
standard form (i.e. a form whose integrals is
presented in Tables of Integrals) using recursion 
relations and other simple identities of associated
Laguerre Polynomials. In this brief subsection 
we only present our final results.
\begin{equation} \label{decomp}
 \begin{split}
 \int d^2 x \frac{2 \pi}{b}e_{m',n'}(\vec{x})^* e_{m,n}(\vec{x}) \frac{i}{\sqrt{b}x^+}=\delta_{(m'-n')-(m-n),1}\sqrt{\frac{m!n'!}{n!m'!}} \times
 \begin{cases}
  +U(n'-n) ~~~~ ~~~~~\text{For } m\geq n\\
  -U(n-(n'+1)) ~ \text{For } m < n\\
 \end{cases}
 \end{split}
\end{equation}
 \footnote{Even though the integrand on the 
	LHS of \eqref{decomp} is singular at 
	$|x|=0$, the singularity is integrable
	and so the integral on the LHS of 
	\eqref{decomp} is well defined.} 
where $U(m)$ is the unit step function, i.e.
\begin{equation}\label{usf}
U(m)=1~~~m\geq 0, ~~~U(m)=0, ~~~{\rm otherwise}
\end{equation}
(The $\delta$ function on RHS of 
\eqref{decomp} enforces momentum conservation). 
It follows from \eqref{decomp}, \eqref{orthog} that 
\begin{equation}\label{enmz}
\begin{split}
\frac{i}{\sqrt{b}} \frac{e_{m,n}(\vec{x})}{x^+}
= 
\begin{cases} 
&+\sum_{p=0}^\infty \sqrt{\frac{m!(n+p)!}{n!(m+p+1)!}}
e_{m+p+1, n+p}(\vec{x})~~~~~~~\text{For } m\geq n\\
&- \sum_{p=1}^{m+1} \sqrt{\frac{m!(n-p)!}{n!(m-p+1)!}}
e_{m-p+1, n-p}(\vec{x})~~~~~~~\text{For } m < n\\
\end{cases}
\end{split}\end{equation}

\section{Fermionic gap equations}

In this appendix we reproduce the derivation 
of the gap equations of the Fermionic theory. 
The analysis of this section closely follows 
the work out presented in \cite{Jain:2012qi}. 

Consider the action given in (\ref{eucact}) and work in momentum space. We work in the gauge $A_{-}=0$. In this gauge self-interaction of gauge field $A$ becomes trivial and action for remaining components of $A$ becomes quadratic. Using equations of motion of $A_{+}$, $A_3$ can easily be solved in terms of fermionic  fields
\begin{equation}
 \begin{split}
 A_3^C(P)  =4 \pi \lambda_F \frac{1}{P_{-}}\int \frac{d^3q}{(2\pi)^3} \frac{1}{N} \bar{\psi}\left(-q+\frac{P}{2}\right)\gamma^+T^C\psi \left(q+\frac{P}{2} \right)
 \end{split}
\end{equation}
Putting it back in the action we get an effective (purely imaginary) non-local four-fermion interaction
\begin{equation}
 \begin{split}
 S=&  \int \frac{d^3q}{(2\pi)^3}\frac{d^3P}{(2\pi)^3}  \bar{\psi}\left(-q+\frac{P}{2}\right) \left((i q_\mu \gamma^\mu+m_F) (2 \pi)^3 \delta^3(P)  -i\gamma^\mu a_{\mu}(-P)\right) \psi \left(q+\frac{P}{2} \right)\\
 &  ~~~~~~~~~~~~ ~~~~~~~~~ -N \ 2 \pi i \lambda_F\int \frac{d^3q_1}{(2\pi)^3}  \frac{d^3q_2}{(2\pi)^3} \frac{d^3P}{(2\pi)^3} \frac{1}{(q_1-q_2)_-} \alpha(P,q_1) \gamma^+ \alpha( -P,q_2)\gamma^3
 \end{split}
\end{equation}
Here we have introduced fermion bilinear
\begin{equation}
 \alpha(P,q)=\frac{1}{N}\bar{\psi}\left(-q+\frac{P}{2}\right) \psi \left(q+\frac{P}{2} \right)
\end{equation}
where gauge indices are contracted.

To proceed further we use following standard trick
\begin{equation}
 \begin{split}
1=& \int d\alpha \ \delta  \left(\alpha(P,q)-\frac{1}{N}\bar{\psi}\left(-q+\frac{P}{2}\right) \psi \left(q+\frac{P}{2} \right)\right)\\
=&  \int d\alpha d \Sigma \ e^{N \int \frac{d^3q}{(2\pi)^3}\frac{d^3P}{(2\pi)^3} \ \Sigma(-P,q)\left(\alpha(P,q)-\frac{1}{N}\bar{\psi}\left(-q+\frac{P}{2}\right) \psi \left(q+\frac{P}{2} \right)\right) }
 \end{split}
\end{equation}
Plugging it in the path integration we can integrate out $\psi, \bar{\psi}$ to give an effective action for $\alpha, \Sigma $
\begin{equation}\label{effective action}
 \begin{split}
 S_{eff}=&  -N \int \frac{d^3q}{(2\pi)^3}\frac{d^3P}{(2\pi)^3}   log \ det\left((i q_\mu \gamma^\mu+m_F) (2 \pi)^3 \delta^3(P)  -i\gamma^\mu a_{\mu}(-P)+\Sigma(-P,q)\right)\\
 &  -N \ 2 \pi i \lambda_F\int \frac{d^3q_1}{(2\pi)^3}  \frac{d^3q_2}{(2\pi)^3} \frac{d^3P}{(2\pi)^3} \frac{1}{(q_1-q_2)_-} \alpha(P,q_1) \gamma^+ \alpha( -P,q_2)\gamma^3\\
 &-N \int \frac{d^3q}{(2\pi)^3}\frac{d^3P}{(2\pi)^3}  \alpha(P,q)\Sigma(-P,q) 
 \end{split}
\end{equation}
Varying above action with respect to $\Sigma$ we get
\begin{equation}
 \begin{split}
\left((i q_\mu \gamma^\mu+m_F) (2 \pi)^3 \delta^3(P)  -i\gamma^\mu a_{\mu}(-P)+\Sigma(-P,q)\right) \alpha(P,q)=-(2 \pi)^3 \delta^3(P)
 \end{split}
\end{equation}
Varying above action with respect to $\alpha$ we get
\begin{equation}
 \begin{split}
\Sigma(P,q)=-2 \pi i \lambda_F \int \frac{d^3q'}{(2\pi)^3} \frac{1}{(q-q')_{-}} \left(\gamma^+ \alpha(P,q') \gamma^3-\gamma^3 \alpha(P,q') \gamma^+ \right)
 \end{split}
\end{equation}
These two equations when translated to position space becomes (\ref{sde}) as presented in main text.

\section{Details of fermionic recursion relations}\label{recrel}

In this appendix, we consider fermionic gap equations for chemical potential in positive or negative exceptional band. We will show that the quantity
\begin{equation}\label{defi}
 \begin{split}
  I_n= \lambda_F^2  (\sqrt{2b}\chi_n)^2+\frac{1}{(\sqrt{2b}\chi_n)^2}
 \end{split}
\end{equation}
satisfies, for $n \geq 1$
\begin{equation}\label{recursionI}
 \begin{split}
  I_n=I_{n-1}+4
 \end{split}
\end{equation}
To see this we simplify the following quantity by writing it entirely in terms of differences. \footnote{We have used the fact that $K_{R,I,n}^F$ is independent of $n$.}
\begin{equation}
\begin{split}
\frac{1}{\chi_n^2}-\frac{1}{\chi_{n-1}^2}&=(\zeta_+^F(n)+\zeta_-^F(n)  )^2-(\zeta_+^F(n-1)+\zeta_-^F(n-1)  )^2\\
&=\left( (K_{R,I,n+1} +K_{R,I,n})^2 -8K_{R,+,n+1}K_{R,-,n}\right)\\
&~~~~~~ -\left( (K_{R,I,n} +K_{R,I,n-1})^2 -8K_{R,+,n}K_{R,-,n-1}\right)\\
&=(K_{R,I,n+1} +K_{R,I,n-1}+2K_{R,I,n})(K_{R,I,n+1} -K_{R,I,n-1})\\  &~~~~~~ -8i\sqrt{b}\left(\sqrt{n+1} (i\sqrt{b}\sqrt{n+1}+\Sigma_{R,+,n+1})-
\sqrt{n}  (i\sqrt{b}\sqrt{n}+\Sigma_{R,+,n})\right)\\
&=((K_{R,I,n+1}-K_{R,I,n}) -(K_{R,I,n}-K_{R,I,n-1})+4K_{R,I,n})\\&~~~~~~~~~~~~((K_{R,I,n+1}-K_{R,I,n}) +(K_{R,I,n}-K_{R,I,n-1}))\\  &~~~~~~~~~~~~~~~ -8i\sqrt{b}\left(\sqrt{n+1} (i\sqrt{b}\sqrt{n+1}+\Sigma_{R,+,n+1})-
\sqrt{n}  (i\sqrt{b}\sqrt{n}+\Sigma_{R,+,n})\right)\\
&=((\Sigma_{R,I,n+1}-\Sigma_{R,I,n}) -(\Sigma_{R,I,n}-\Sigma_{R,I,n-1})+4K_{R,I,n})\\&~~~~~~~~~~~~~~~~~~~~((\Sigma_{R,I,n+1}-\Sigma_{R,I,n}) +(\Sigma_{R,I,n}-\Sigma_{R,I,n-1}))\\  &~~~~~~~~~~~~~~~~~~~~~~~~~~ -8i\sqrt{b}\left(i\sqrt{b}+\sqrt{n+1} \Sigma_{R,+,n+1}-
\sqrt{n}  \Sigma_{R,+,n}\right)\\
\end{split}
\end{equation}
Now we use \eqref{gapeqsimplified} to replace differences in terms of $\Sigma_{R,I}$ and $\chi$. Next goal in algebraic manipulations below is to show that explicit  $\Sigma_{R,I}$ dependences cancel each other, leaving behind a recursion relation involving $\chi$ alone.
\begin{equation}
\begin{split}
\frac{1}{\chi_n^2}-\frac{1}{\chi_{n-1}^2}&=(2 \lambda_F b(\chi_n -\chi_{n-1})+4K_{R,I,n})(2 \lambda_F b(\chi_n +\chi_{n-1}))+8b
\\  &-8b\lambda_F \left( (K_{R,I,n-1}\chi_{n-1}+K_{R,I,n+1}\chi_{n})-\lambda_F( (\sqrt{b}\chi_n)^2-(\sqrt{b}\chi_{n-1})^2)  \right)\\
&=(2 \lambda_F b(\chi_n -\chi_{n-1})+4K_{R,I,n})(2 \lambda_F b(\chi_n +\chi_{n-1}))+8b
\\  & ~~~~~~~~~~~~~~~~~~~~~~~~~~ -8b\lambda_F ((K_{R,I,n-1}-K_{R,I,n}+K_{R,I,n})\chi_{n-1} \\ &~~~~~~~~~~~ +(K_{R,I,n+1}-K_{R,I,n}+K_{R,I,n})\chi_{n})+8b \lambda_F^2( (\sqrt{b}\chi_n)^2-(\sqrt{b}\chi_{n-1})^2) \\
&=(2 \lambda_F b(\chi_n -\chi_{n-1})+4K_{R,I,n})(2 \lambda_F b(\chi_n +\chi_{n-1}))+8b
\\  & ~~~~~~~-8b\lambda_F ((-2 \lambda_F b \chi_{n-1}  +K_{R,I,n})\chi_{n-1} +(2 \lambda_F b \chi_{n} +K_{R,I,n})\chi_{n})\\ &~~~~~~~~~~~~~~~~~~~~~~~~~~~~~~~~~~~~~~~~~~~~~~~~~~~ +8b \lambda_F^2( (\sqrt{b}\chi_n)^2-(\sqrt{b}\chi_{n-1})^2) \\
&=(2 \lambda_F b)^2(\chi_n^2 -\chi_{n-1}^2)-4 (2 \lambda_F b)^2(\chi_{n}^2-\chi_{n-1}^2)+2(2 \lambda_Fb)^2( \chi_n^2-\chi_{n-1}^2) +8b\\
&=-(2 \lambda_F b)^2(\chi_n^2 -\chi_{n-1}^2) +8b
\end{split}
\end{equation}
This can be rewritten as 
\begin{equation}\label{recursionI}
\begin{split}
I_n=I_{n-1}+4
\end{split}
\end{equation}

\section{Bosonic gap equations}\label{bosonicgapeq}
We consider the action given in \eqref{ActionScalar} and work in momentum space in gauge $A_-=0$. Once again our work out closely
follows the methodology laid out in \cite{Jain:2012qi}. 
For simplicity we set background field to zero, putting it back is trivial modification to what is derived below. Action in this gauge becomes
\begin{equation}\label{ac1}
	\begin{aligned}
		S_B & =\int \frac{d^3p}{(2 \pi)^3}\left(\frac{\kappa_B}{4 \pi}A_+^C(-p)ip_-A_3^C(p)+ \bar{\phi}(-p)(p^2+M_B^2)\phi(p) \right)\\
		& +\int \frac{d^3p}{(2 \pi)^3}\left(A_+^C(-p)J_-^C(p)+A_3^C(-p)J_3^C(p) \right)\\
		& +\int \frac{d^3q_1}{(2 \pi)^3}\frac{d^3q_2}{(2 \pi)^3}\frac{d^3q}{(2 \pi)^3}\left(A_3^C(q_1)A_3^D(q_2)\bar{\phi}(- \frac{q_1+q_2}{2}-q ) T^C T^D\phi(- \frac{q_1+q_2}{2}+q)\right)\\
		& + 4 \pi b_4 \lambda_B N \int \frac{d^3q_1}{(2 \pi)^3}\frac{d^3q_2}{(2 \pi)^3}\frac{d^3P}{(2 \pi)^3}\chi(P,q_1)\chi(-P,q_2)\\
		& + 4 \pi^2 x_6 \lambda_B^2 N \int \frac{d^3q_1}{(2 \pi)^3}\frac{d^3q_2}{(2 \pi)^3}\frac{d^3q_3}{(2 \pi)^3} \frac{d^3P_1}{(2 \pi)^3}\frac{d^3P_2}{(2 \pi)^3} \chi(P_1,q_1)\chi(P_2,q_2)\chi(-P_1-P_2,q_3)
	\end{aligned}
\end{equation}
where we have defined
\begin{equation}
	\begin{aligned}
		J_\mu^C(P)&=-\int \frac{d^3q}{(2 \pi)^3}\bar{\phi}(\frac{P}{2}-q )T^C\phi(\frac{P}{2}+q )2 q_\mu\\
		\chi(P,q)&=\frac{1}{N}\bar{\phi}(\frac{P}{2}-q )\phi(\frac{P}{2}+q )
	\end{aligned}
\end{equation}
$A_+^C(-p)$ equation of motion from the action above is
\begin{equation}\label{eqqq1}
	\begin{aligned}
		\frac{\kappa_B}{4 \pi}ip_-A_3^C(p)=-J_-^C(p)
	\end{aligned}
\end{equation}
Since $A_+$ appears only linearly in action \eqref{ac1} integrating it out sets its coefficient to zero. Only left over part of gauge field is $A_3$ which can be solved from \eqref{eqqq1} in terms of $J_-$. Plugging back $A_3$ into action \eqref{ac1} we get 
\begin{equation}\label{ac1}
	\begin{aligned}
		S_B & =\int \frac{d^3p}{(2 \pi)^3}\left( \bar{\phi}(-p)(p^2+m_B^2)\phi(p) \right)\\
				& + N \int \frac{d^3q_1}{(2 \pi)^3}\frac{d^3q_2}{(2 \pi)^3}\frac{d^3P}{(2 \pi)^3}C_1(P,q_1,q_2)\chi(P,q_1)\chi(-P,q_2)\\
		& +  N \int \frac{d^3q_1}{(2 \pi)^3}\frac{d^3q_2}{(2 \pi)^3}\frac{d^3q_3}{(2 \pi)^3} \frac{d^3P_1}{(2 \pi)^3}\frac{d^3P_2}{(2 \pi)^3} C_2(P_1,P_2,q_1,q_2,q_3)\chi(P_1,q_1)\chi(P_2,q_2)\chi(-P_1-P_2,q_3)
	\end{aligned}
\end{equation}
Where\footnote{Note that expressions given for $C_1, C_2$ is not unique as the integrand multiplying these has certain symmetries. For example $C_1(P,q_1,q_2)$ multiples $\chi(P,q_1)\chi(-P,q_2)$ which is invariant under simultaneous $q_1 \leftrightarrow q_2$, $P \to -P $ exchange and therefore addition of any expression to $C_1$ which is 'odd' under this exchange integrates to zero. }
\begin{equation}
	\begin{aligned}
		C_1(P,q_1,q_2)& =4 \pi b_4 \lambda_B -i 2 \pi \lambda_B \frac{(q_1+q_2)_-P_3-P_-(q_1+q_2)_3}{(q_1-q_2)_-} \\
		C_2(P_1,P_2,q_1,q_2,q_3)& =4\pi^2 x_6 \lambda_B^2+4 \pi^2 \lambda_B^2 \frac{(P_1-P_2+2(q_1+q_2))_-(P_1+2P_2+2(q_3+q_2))_-}{(P_1+P_2+2(q_1-q_2))_-(P_1+2(q_3-q_2))_-} 
	\end{aligned}
\end{equation}
To proceed further we use following standard trick
\begin{equation}
 \begin{split}
1=& \int d\alpha \ \delta  \left(\alpha(P,q)-\frac{1}{N}\bar{\phi}\left(-q+\frac{P}{2}\right) \phi \left(q+\frac{P}{2} \right)\right)\\
=&  \int d\alpha d \Sigma \ e^{N \int \frac{d^3q}{(2\pi)^3}\frac{d^3P}{(2\pi)^3} \ \Sigma(-P,q)\left(\alpha(P,q)-\frac{1}{N}\bar{\phi}\left(-q+\frac{P}{2}\right) \phi \left(q+\frac{P}{2} \right)\right) }
 \end{split}
\end{equation}
Plugging it in the path integration we can integrate out $\phi, \bar{\phi}$ to give an effective action for $\alpha, \Sigma $
\begin{equation}\label{ac2}
	\begin{aligned}
		S_B & =N \int \frac{d^3P}{(2 \pi)^3}\frac{d^3q}{(2 \pi)^3}log \ det\left((q^2+m_B^2) (2 \pi)^3 \delta^3(P) +\Sigma(-P,q)\right)\\
				& + N \int \frac{d^3q_1}{(2 \pi)^3}\frac{d^3q_2}{(2 \pi)^3}\frac{d^3P}{(2 \pi)^3}C_1(P,q_1,q_2)\alpha(P,q_1)\alpha(-P,q_2)\\
		& +  N \int \frac{d^3q_1}{(2 \pi)^3}\frac{d^3q_2}{(2 \pi)^3}\frac{d^3q_3}{(2 \pi)^3} \frac{d^3P_1}{(2 \pi)^3}\frac{d^3P_2}{(2 \pi)^3} C_2(P_1,P_2,q_1,q_2,q_3)\alpha(P_1,q_1)\alpha(P_2,q_2)\alpha(-P_1-P_2,q_3)\\
		&-N \int \frac{d^3q}{(2\pi)^3}\frac{d^3P}{(2\pi)^3}  \alpha(P,q)\Sigma(-P,q)
	\end{aligned}
\end{equation}
Varying above action with respect to $\Sigma$ we get
\begin{equation}
	\begin{aligned}
		\left((q^2+m_B^2) (2 \pi)^3 \delta^3(P) +\Sigma(-P,q)\right)\alpha(P,q)=(2 \pi)^3 \delta^3(P) 
	\end{aligned}
\end{equation}
Varying above action with respect to $\alpha$ we get
\begin{equation}
	\begin{aligned}
		\Sigma(P,q)&=\int \frac{d^3q'}{(2 \pi)^3} \left( C_1(P,q',q)+C_1(-P,q,q') \right) \alpha(P,q')\\
		&+\int \frac{d^3q_1}{(2 \pi)^3}\frac{d^3q_2}{(2 \pi)^3}\frac{d^3P'}{(2 \pi)^3}(C_2(-P,P',q,q_1,q_2)+ C_2(P',-P,q_1,q,q_2)\\
		& ~~~~~~~~~~~~~~ +C_2(P',P-P',q_1,q_2,q) )\alpha(P',q_1)\alpha(P-P',q_2)
	\end{aligned}
\end{equation}
To write this equations in position space we take following strategy: express every vertex factor in terms of momentums of two point function $\alpha(P,q)$ and gauge boson propagator
\begin{equation}
	\begin{aligned}
		G(P,q)=\frac{(2 \pi)^3 \delta^{(3)}(P)}{q_-} \implies G(x,y)=\frac{i \delta(x^3-y^3)}{2\pi (x-y)_-} \equiv G(x-y)
	\end{aligned}
\end{equation}
and use the mapping mentioned in Appendix \ref{conventions} to convert them to derivatives in position space. These manipulations lead to the equations presented in \eqref{bgapeq2}, \eqref{bgapeq21}, \eqref{bgapeq22A},\eqref{bgapeq22B}\eqref{bgapeq22B'}.

\section{Reduced set of bosonic gap equations }

In this appendix we simplify gap equation for bosons using specific form of the self-energy and propagators in presence of a uniform magnetic field. To do so we will need various properties of gauge field propagator given in \eqref{GaugeProp}  that we list below
\begin{equation}
 \begin{aligned}
  G(-x)=-G(x), ~~ \partial_{x^-}G(x)=i \delta^3(x),~~ G(x)(G(y)+G(x-y))=G(x-y)G(y)\\
 \end{aligned}
\end{equation}
Before proceeding, we introduce few notations
\begin{equation}
 \begin{aligned}
  (f_1f_2)(x)=f_1(x)f_2(x), ~~ (\partial f)(x)=\partial_{x^-}f(x)\\
 \end{aligned}
\end{equation}

We plug the form of $\alpha, \Sigma$ given in \eqref{ussm} into the gap equations presented in  \eqref{bgapeq2}, \eqref{bgapeq21}, \eqref{bgapeq22A},\eqref{bgapeq22B}\eqref{bgapeq22B'}. Below we present reduced part of each of the expressions.

\begin{equation}\label{graph2a}
	\begin{aligned}
    \Sigma_{2A}(x_T,x_R)|_{red} =&4 \pi^2 \lambda_B^2((x_6+1) \delta^3(x_R) \alpha_R(x_R)^2
    \\& \color{blue} +2 D_R(\alpha_R \partial G *_b \alpha_R G)(x_R)\\&
    \color{green} -(D_R+D_T)(D_R-D_T)(G \alpha_R *_b G \alpha_R)(x_R))
	\end{aligned}
\end{equation}

\begin{equation}\label{graph2b}
	\begin{aligned}
		\Sigma_{2B}(x_T,x_R) |_{red}&=4 \pi^2 \lambda_B^2(x_6\alpha_R(x_R)^2\delta^{(3)}(x_R)
		\\&
		\color{green}+(D_R+D_T)(D_R-D_T)(G(\alpha_R *_b \alpha_R G))(x_R))
		\\
		&  \color{violet}   +2(D_R+D_T)D_T(G(\alpha_R *_b \alpha_R G))(x_R)) 
		\\& \color{violet} -(D_R-D_T)2b \int  d^3z \ x_{R_1,-} \alpha_R(x_{1,R})G(x_{R})\alpha_R(x_{2,R})G(x_{2,R})e^{-\frac{ib}{2}\epsilon_{ij}x_{1,R}^ix_{2,R}^j}
		\\& -2 D_T 2b  \int  d^3z \ x_{R_1,-} \alpha_R(x_{1,R})G(x_{R})\alpha_R(x_{2,R})G(x_{2,R})e^{-\frac{ib}{2}\epsilon_{ij}x_{1,R}^ix_{2,R}^j}
		\\
		& \color{blue} -(D_R-D_T)(G \alpha_R *_b \alpha_R \partial G) (x_R)-2D_T(G \alpha_R *_b \alpha_R \partial G)(x_R)
		\\& \color{orange} +2\int  d^3z \ \alpha_R(x_{1,R})\partial^2_{x_R^-}G(x_{R})\alpha_R(x_{2,R})G(x_{2,R})e^{-\frac{ib}{2}\epsilon_{ij}x_{1,R}^ix_{2,R}^j}\\
		& \color{orange}  +\int  d^3z \ \alpha_R(x_{1,R})\partial_{x_R^-}G(x_{R})\alpha_R(x_{2,R}) \partial_{x_{2,R}^-}G(x_{2,R})e^{-\frac{ib}{2}\epsilon_{ij}x_{1,R}^ix_{2,R}^j}
		\\& \color{magenta} +2b\int  d^3z \ x_{R_1,-}\alpha_R(x_{1,R})\partial_{x_R^-}G(x_{R})\alpha_R(x_{2,R})G(x_{2,R}))e^{-\frac{ib}{2}\epsilon_{ij}x_{1,R}^ix_{2,R}^j}
	\end{aligned}
\end{equation}

\begin{equation}\label{graph2bp}
	\begin{aligned}
				\Sigma_{2B'}(x_T,x_R) =& 4 \pi^2 \lambda_B^2(x_6\alpha_R(x_R)^2\delta^{(3)}(x_R)
		\\&
		\color{green}+(D_R+D_T)(D_R-D_T)(G(\alpha_R G *_b \alpha_R ))(x_R))
		\\& \color{violet} -(D_R-D_T)2b \int  d^3z \ x_{R_1,-} \alpha_R(x_{1,R})G(x_{R})\alpha_R(x_{2,R})G(x_{1,R})e^{-\frac{ib}{2}\epsilon_{ij}x_{1,R}^ix_{2,R}^j}\\
		& \color{blue} -(D_R-D_T)(\partial G \alpha_R *_b \alpha_R  G) (x_R)
		\\& \color{orange} +2\int  d^3z \ \alpha_R(x_{1,R})\partial^2_{x_R^-}G(x_{R})\alpha_R(x_{2,R})G(x_{1,R})e^{-\frac{ib}{2}\epsilon_{ij}x_{1,R}^ix_{2,R}^j}\\
		& \color{orange}  +\int  d^3z \ \alpha_R(x_{1,R})\partial_{x_R^-}G(x_{R})\alpha_R(x_{2,R}) \partial_{x_{2,R}^-}G(x_{1,R})e^{-\frac{ib}{2}\epsilon_{ij}x_{1,R}^ix_{2,R}^j}
		\\& \color{magenta} +2b\int  d^3z \ x_{R_1,-}\alpha_R(x_{1,R})\partial_{x_R^-}G(x_{R})\alpha_R(x_{2,R})G(x_{1,R}))e^{-\frac{ib}{2}\epsilon_{ij}x_{1,R}^ix_{2,R}^j}
	\end{aligned}
\end{equation}
Where we have defined 
\begin{equation}
 D_R=2 \left(-\frac{b}{2}X_{T,-}+\partial_{X_R^-}\right), ~ D_T=2 \left(\frac{b}{2}X_{R,-}+\partial_{X_T^-}\right)
\end{equation}
Now we present few identities that will we will need to simplify above equations. These rely crucially on properties of twisted convolution and properties given in \eqref{identities} of gauge boson propagator . For any function $\alpha(x)$, it can be shown that
\begin{equation}\label{identities}
 \begin{aligned}
 & ((\alpha \partial G*_b \alpha )G)(x-y)=(\alpha \partial G *_b \alpha G)(x-y))\\
 & (G(\alpha *_b \alpha \partial G))(x-y)=(\alpha G*_b \alpha \partial G)(x-y)\\
 &(\partial G (\alpha *_b \alpha G))(x-y)=((\alpha G*_b \alpha) \partial G)(x-y)=0\\
 & (G(\alpha*_b\alpha G)+(\alpha G*_b\alpha)G)(x-y)=(\alpha G*_b\alpha G)(x-y)\\
& (\partial^2G(\alpha *_b \alpha G))(x-y)+\frac{1}{2}(\partial G \alpha *_b \alpha \partial G)(x-y)=+\frac{b}{2} \int \partial G(x-y) \alpha(x-y-z) \alpha(z) G(z) z_{-} d^3z\\
& (\partial^2G(\alpha G *_b \alpha ))(x-y)+\frac{1}{2}(\partial G \alpha *_b \alpha \partial G)(x-y)=-\frac{b}{2} \int \partial G(x-y) \alpha(x-y-z) \alpha(z) G(z) z_{-}d^3z\\
 & \int  d^3z \ x_{R_1,-}\alpha_R(x_{1,R})\partial_{x_R^-}G(x_{R})\alpha_R(x_{2,R})(G(x_{1,R})+G(x_{2,R}))=0\\
 & \int  d^3z \ x_{R_1,-}\alpha_R(x_{1,R})G(x_{R})\alpha_R(x_{2,R})(G(x_{1,R})+G(x_{2,R}))e^{-\frac{ib}{2}\epsilon_{ij}x_{1,R}^ix_{2,R}^j}=x_{R,-} (G(\alpha *_b \alpha G))(x_R)
 \end{aligned} 
\end{equation}

Due to 1st, 2nd identity in \eqref{identities}, 2nd line of \eqref{graph2a}, 6th line of \eqref{graph2b} and 4th line of \eqref{graph2bp} sums to zero. Due to 4th identity in \eqref{identities}, 3rd line of \eqref{graph2a}, 2nd line of \eqref{graph2b} and 2nd line of \eqref{graph2bp} sums to zero. Due to 5th, 6th identity in \eqref{identities}, 7th, 8th line of \eqref{graph2b} and 5th,6th line of \eqref{graph2bp} sums to zero.  Due to 7th identity in \eqref{identities}, last line of \eqref{graph2b} and last line of \eqref{graph2bp} sums to zero.

Now we use last identity in \eqref{identities} to simplify the following
\begin{equation}\label{complex1}
 \begin{aligned}
   &  \color{violet}   2(D_R+D_T)D_T(G(\alpha_R *_b \alpha_R G))(x_R)) 
		\\& \color{violet} -(D_R-D_T)2b \int  d^3z \ x_{R_1,-} \alpha_R(x_{1,R})G(x_{R})\alpha_R(x_{2,R})G(x_{2,R})e^{-\frac{ib}{2}\epsilon_{ij}x_{1,R}^ix_{2,R}^j}\\
		 &\color{violet} -(D_R-D_T)2b \int  d^3z \ x_{R_1,-} \alpha_R(x_{1,R})G(x_{R})\alpha_R(x_{2,R})G(x_{1,R})e^{-\frac{ib}{2}\epsilon_{ij}x_{1,R}^ix_{2,R}^j}\\
		 & \color{violet}  = (D_R+D_T)2bx_{R,-}(G(\alpha_R *_b \alpha_R G))(x_R))  -(D_R-D_T)2b x_{R,-} (G(\alpha_R *_b \alpha_R G))(x_R)\\
		  & \color{violet}  = 2D_T 2bx_{R,-}(G(\alpha_R *_b \alpha_R G))(x_R)) \\
		  & \color{violet}  = (2bx_{R,-})^2(G(\alpha_R *_b \alpha_R G))(x_R))
 \end{aligned}
\end{equation}
The only remaining complex term can be simplified 
as follows
\begin{equation}\label{complex2}
 \begin{aligned}
   & -2 D_T 2b  \int  d^3z \ x_{R_1,-} \alpha_R(x_{1,R})G(x_{R})\alpha_R(x_{2,R})G(x_{2,R})e^{-\frac{ib}{2}\epsilon_{ij}x_{1,R}^ix_{2,R}^j}
		\\
		&= -2 bx_{R,-} 2b \ G(x_{R})\int  d^3z \ (x_R-z)_{-} \alpha_R(x_R-z)\alpha_R(z)G(z)e^{-\frac{ib}{2}\epsilon_{ij}x_{1,R}^ix_{2,R}^j}
		\\
		&= -(2 bx_{R,-})^2  \ G(x_{R})\int  d^3z \ \alpha_R(x_R-z)\alpha_R(z)G(z)e^{-\frac{ib}{2}\epsilon_{ij}x_{1,R}^ix_{2,R}^j}\\
		& ~~~~~~~ -\left(\frac{b}{\pi}\right)^2 \delta(x_R^3)\int  d^3z \ \alpha_R(x_R-z)\alpha_R(z)\delta(z^3)e^{-\frac{ib}{2}\epsilon_{ij}x_{1,R}^ix_{2,R}^j}
		\\
		&= -(2 bx_{R,-})^2  (G(\alpha_R *_b \alpha_R G))(x_R))-\left(\frac{b}{\pi}\right)^2 \int  d^3z \   \delta(x_R^3-z^3)\alpha_R(x_R-z)\alpha_R(z)\delta(z^3)e^{-\frac{ib}{2}\epsilon_{ij}x_{1,R}^ix_{2,R}^j}
		\\
 \end{aligned}
\end{equation}
Adding the final results of \eqref{complex1} and 
\eqref{complex2} to the (uncanceled) expressions 
on the first lines  of the 
RHS of \eqref{graph2a}, \eqref{graph2b}, 
\eqref{graph2bp} we obtain two loop contribution to reduced self-energy to be
\begin{equation}
	\begin{aligned}
		4 \pi^2 \lambda_B^2(3x_6+1) \delta^3(x_R) \alpha_R(x_R)^2-\left(2 b \lambda_B\right)^2 \int  d^3z \   \delta(x_R^3-z^3)\alpha_R(x_R-z)\alpha_R(z)\delta(z^3)e^{-\frac{ib}{2}\epsilon_{ij}x_{1,R}^ix_{2,R}^j}
	\end{aligned}
\end{equation}

\section{Zero magnetic field limit}

In this appendix we discuss how to take $b \to 0$ limit carefully. Note that this limit is bit special - in absence of magnetic field energy levels of a free theory has a continuous spectrum, whereas for any non-zero magnetic field that becomes discrete. Therefore $b=0$ discussion is an averaged out version of many very closely spaced levels that are present for small values of $b$ as we will show below. We will discuss only the case zero chemical potential.

\textbf{Fermions:}

Note that significant contribution comes only from $n \to \infty$ part of summations, such that $w=bn$ is kept fixed.
In this limit $\chi_n$ as obtained in \eqref{chicolution} simplifies to
\begin{equation}\label{chiflatspace}
 \chi_n=\frac{1}{2(c_F^2+2 w)^{1/2}}
\end{equation}
Gap equations \eqref{gapeqsimplified} becomes
\begin{equation}\label{geqflat}
 \begin{split}
  \frac{\partial}{\partial w} (\Sigma_{R,I,n} )&= 2 \lambda_F \chi_n\\
  \frac{\partial }{\partial w} (\sqrt{w} \Sigma_{R,+,n})&= -2i \lambda_FK_{R,I,n}  \chi_n
 \end{split}
\end{equation}
First of these gap equations is solved by it's integrated form \eqref{sigmge}
\begin{equation}
 \Sigma_{R,I,n} (w)=-2 \lambda_F \int_w^\infty  \chi_n(w') dw'
\end{equation}
Where above equation is regularized by dimensional regularization on the quantity $c_F^2+2 w$, i.e., effectively on $\omega$ integration that we performed previously.
These gives
\begin{equation}\label{senflat}
 \begin{split}
  \Sigma_{R,I,n}(w)&=\lambda_F(c_F^2+2w)^{1/2} \\ \Sigma_{R,+,n}(w)&=-\frac{i}{2\sqrt{w}} \left(\Sigma_{R,I,n}(w)^2-\Sigma_{R,I,n}(w=0)^2 +2 m_F (\Sigma_{R,I,n}(w)-\Sigma_{R,I,n}(w=0))\right)  
 \end{split}
\end{equation}
Plugging these two into \eqref{chiflatspace} we get
\begin{equation}
 c_F^2=(m_F+\Sigma_{R,I,n}(w=0))^2
\end{equation}

Summing components given in \eqref{senflat} with corresponding basis functions listed in \eqref{enm} we get corresponding self-energies in position space
\begin{equation}
 \begin{split}
  \Sigma_{R,I}( \vec{x} )&= \sum_n \Sigma_{R,I,n}( \vec{x} )e_{n,n}(\vec{x})\\
  & \to \int \frac{dw}{2 \pi} \lim_{n\to \infty} \ L_n^0 \left(\frac{1}{n}\frac{wr^2}{2}\right) \Sigma_{R,I,n}\\
   & = \int \frac{p_s dp_s}{2 \pi}  \ J_0 \left(p_s r\right) \lambda_F(c_F^2+p_s^2)^{1/2} 
\end{split}
\end{equation}
Here we have changed variables $w=\frac{p_s^2}{2}$, and  used following identity of Laguerre polynomial 
\begin{equation}
 \lim_{n\to \infty} \frac{1}{n^l}\ L_n^l \left(\frac{x}{n}\right)=\frac{1}{x^l}\ J_l \left(2\sqrt{x}\right)
\end{equation}
To compare our results with previous works on the subject we switch to momentum space defined by 
\begin{equation}
 \tilde{F}(\vec{p})=\int d^2x F(\vec{x})e^{-i \vec{p}.\vec{x}}
\end{equation}
A basic property that relates position and momentum space is that the  function which has 
\footnote{We are using following notation $$ p^{\pm}=\frac{p_s}{\sqrt{2}}e^{\pm i \theta}, ~ x^{\pm}=\frac{r}{\sqrt{2}}e^{\pm i \phi}$$}
\begin{equation}
 \tilde{F}(\vec{p})= e^{i m\theta}f(p_s)
\end{equation}
in position space is given by
\begin{equation}
 F(\vec{x})=\int \frac{p_s dp_s}{2 \pi} f(p_s) \ J_{|m|} \left(p_s r\right) e^{i m \phi}  
\end{equation}
Comparing above obtained form of $\Sigma_{R,I}$ and similar considerations for $\Sigma_{R,+}$, dictates that solutions given in \eqref{senflat} has following form in momentum space 
\begin{equation}\label{Flatspacesol}
 \begin{split}
  \Sigma_{R,I}(\vec{p})&=\lambda_F(c_F^2+p_s^2)^{1/2} \\
  \Sigma_{R,+}(\vec{p})&=\frac{p_+}{p_s^2} \left(c_F^2-(m_F+\Sigma_{R,I}(\vec{p}))^2\right)\\
   c_F^2&=(m_F+\lambda_F|c_F|)^2
 \end{split}
\end{equation}
\eqref{Flatspacesol} agrees with the solution of 
the zero magnetic field gap equations presented 
in \cite{Giombi:2011kc, Aharony:2012ns, Jain:2013gza, Jain:2014nza}.

\textbf{Bosons:}

 First note that \eqref{eq1b}, \eqref{eq2b} suggests that 
 \begin{equation}
 	\zeta_+^B(n)=\zeta_-^B(n)=c_B^2+2bn
 \end{equation}
in the limit we are considering.
On the other hand \eqref{gapeqfinal} gives
\begin{equation}
	\int \frac{b dn}{\sqrt{c_B^2+2bn}}=-m_B^{cri} \implies |c_B|=m_B^{cri}
\end{equation}
Finally performing an analysis very similar to that for the fermions (see around  \eqref{sigiz}) gives
\begin{equation}
	\Sigma(\vec{p})=(m_B^{cri})^2-m_B^2
\end{equation}
in agreement with the gap equations presented 
in \cite{Aharony:2012nh, Aharony:2012ns, Jain:2013gza, Jain:2014nza}.

\newpage
\bigskip
\bibliography{cp}

\providecommand{\href}[2]{#2}\begingroup\raggedright



\end{document}